\documentclass[twocolumn,floats,showpacs,superscriptaddress,pre]{revtex4}
\usepackage{amssymb}
\usepackage{graphicx}
\usepackage{dcolumn}
\usepackage{amsmath}
\usepackage{bm}
\usepackage{epsfig}

\def\sech{\mathop{\rm sech}\nolimits}

\def\defi{{\buildrel \;def\; \over =}}
\newcommand{\be}{\begin{equation}}
\newcommand{\ee}{\end{equation}}

\newcommand{\media}[1]{\langle #1 \rangle}

\begin{document}

\title{Effective field theory for models 
defined over small-world networks. \\ 
First and second order phase transitions.}
\author{M. Ostilli }
\affiliation{Departamento de F{\'\i}sica da Universidade de Aveiro, 3810-193 Aveiro,
Portugal}
\affiliation{Center for Statistical Mechanics and Complexity, 
INFM-CNR SMC, Unit\`a di Roma 1, Roma, 00185, Italy.}
\author{J. F. F. Mendes}
\affiliation{Departamento de F{\'\i}sica da Universidade de Aveiro, 3810-193 Aveiro,
Portugal}

\begin{abstract}
 We present an effective field theory method to analyze, in a very
 general way, models defined over small-world networks.  Even if the
 exactness of the method is limited to the paramagnetic regions and
 to some special limits, it provides, yielding a clear and immediate
 (also in terms of calculation) physical insight, the exact critical
 behavior and the exact critical surfaces and percolation thresholds.
 The underlying structure of the non random part of the model,
 \textit{i.e.}, the set of spins filling up a given lattice
 $\mathcal{L}_0$ of dimension $d_0$ and interacting through a fixed
 coupling $J_0$, is exactly taken into account. When $J_0\geq 0$, the
 small-world effect gives rise, as is known, to a second-order phase
 transition that takes place independently of the dimension $d_0$ and
 of the added random connectivity $c$.  When $J_0<0$, a different
 and novel scenario emerges in which, besides a spin glass
 transition, multiple first- and second-order phase transitions may
 take place.  As immediate analytical applications we analyze the
 Viana-Bray model ($d_0=0$), the one dimensional chain ($d_0=1$), and
 the spherical model for arbitrary $d_0$.
\end{abstract}

\maketitle

\email{ostilli@roma1.infn.it}

\section{Introduction} \label{intro} Since the very beginning of the
pioneer work by Watts and Strogatz \cite{Watts}, the interest toward
small-world networks - an interplay between random and regular
networks - has been growing ``exponentially''.  Mainly, there are two
reasons that have caused such a ``diffusion''.

The first reason is due to the topological properties of the
small-world network. In synthesis, if $N$ is the size of the system,
for any finite probability $p$ of rewiring, or for any finite added
random connectivity $c$ (the two cases correspond to slightly
different procedures for building a small-world network) one has: a
``short-distance behavior'', implying that the average shortest
distance between two arbitrarily chosen sites grows as $l(N)\sim
\log(N)$, as in random networks, and a large clustering coefficient,
$C(N)\sim\mathop{O}(1)$, as in regular lattices.  The interplay
between these two features makes small-world networks representative
of many realistic situations including social networks, communications
networks, chemical reactions networks, protein networks, neuronal
networks, etc.

The second reason is due the fact that, in models defined over
small-world networks, despite the presence of an underlying finite
dimensional structure - a lattice $\mathcal{L}_0$ of dimension
$d_0<\infty$ - the existence of short-cut bonds makes such models
mean-field like and - hopefully - exactly solvable.  However, even if
such a claim sounds intuitively correct, the complexity of these
models turns out to be in general quite high and, compared to
numerical works, there are still few exact results for small-world
networks \cite{Barrat}-\cite{Bolle} (for the percolation and
synchronization problem see \cite{Newman} and \cite{Hastings3}).

In particular, for $d_0>1$ a mean-field critical behavior is expected
and has been also supported by Monte Carlo (MC) simulations
\cite{Herrero}.  Some natural questions then arise.  Are we able to
prove analytically such a behavior?  If for example $d_0=2$, does the
mean-field critical behavior hold for any situation?  Yet, does the
correlation length diverge at the critical temperature?

Furthermore, even if for $d_0=1$ an exact analytical treatment has
been developed at the level of replica symmetry breaking (RSB)
\cite{Niko} and one step replica symmetry breaking (1RSB)
\cite{Niko2}, the calculations are quite involved and the solutions of
the coupled equations to evaluate the order parameters require a
certain numerical work, which becomes rapidly hard in the 1RSB case.
In any case, even if
these methods are able to give in principle exact results at any
temperature, they are not in general suitable to provide a clear
simple and immediate physical picture of the model, even within some
approximations. The main problem in fact resides in the presence of
short loops: as soon as $d_0>1$ these loops cannot be neglected and
the ``traditional'' cavity and replica methods seem hardly applicable.
In particular, we are not able to predict what happens, for example,
if we set $J_0$ negative.  Should we still expect a second-order phase
transition?  And what about the phase diagram?

In this paper, we present a general method to study random Ising
models defined on small-world graphs built up by adding a random
connectivity $c$ over an underlying arbitrary lattice $\mathcal{L}_0$
having dimension $d_0$.  We will show that this method - in a very
simple and physically sound way - provides an answer to the above
questions as well as to many others.

Roughly speaking, as an effective field theory the method generalizes
the Curie-Weiss mean-field equation $m=\tanh(\beta J m)$ to take into account the
presence of the short-range couplings $J_0$ besides the long-range
ones $J$.  As we will show, the magnetization $m$ of the model defined
over the small-world network, shortly the \textit{random model},
behaves as the magnetization $m_0$ of the model defined over
$\mathcal{L}_0$, shortly the \textit{unperturbed model}, but
immersed in an effective external field to be determined
self-consistently.  Even if the exactness of this method is limited to
the paramagnetic regions (P), it provides the exact critical behavior
and the exact critical surfaces, as well as simple qualitative good
estimates of the correlation functions in the ferromagnetic (F) and
spin glass regions (SG).  Furthermore, in unfrustrated systems, the
method becomes exact at any temperature in the two limits $c\to 0^+$
and $c\to\infty$.

The consequences of such a general result are remarkable from both the
theoretical and the practical point of view.  Once the explicit form
of the magnetization of the unperturbed model, $m_0=m_0(\beta
J_0,\beta h)$, is known, analytically or numerically, as a function of
the couplings $J_0$ and of the external field $h$, we get an
approximation to the full solution of the random model, which is
analytical or numerical, respectively, and becomes exact in the P
region.  If we do not have $m_0=m_0(\beta J_0,\beta h)$ but we know at
least some of its properties, we can still use these properties to
derive certain exact relations and the critical behavior of the random
model.

In the first part of the paper, after presenting the self-consistent
equations, we focus on their application for a general study of the
critical surfaces and of the critical behavior.  In the second part,
we apply the method to study models of interest which can be solved
analytically (and very easily) as for them we know $m_0(\beta
J_0,\beta h)$: the Viana-Bray model (which can be seen as a $d_0=0$
dimensional small-world model), the one-dimensional chain small-world
model, and the spherical small-world model in arbitrary $d_0$
dimension.

We stress that the critical surfaces as well as the correlation
functions in the P region provided by the present method are exact and
not based on any special ansatz as the replica-symmetry and the
tree-like ansatz.  We prove in particular that: independently of the
added random connectivity $c$, of the underlying dimension $d_0$, of
the structure of the underlying lattice $\mathcal{L}_0$, and of the
nature of the phase transition present in the unperturbed model (if any),
for $J_0\geq 0$ we always have a second-order phase transition with
the classical mean-field critical indices but with a finite
correlation length if calculated along the ``Euclidean distance''
defined in $\mathcal{L}_0$; on the other hand, for $J_0<0$ we show
that, as soon as $c$ is sufficiently large, there exist at least two
critical temperatures which, depending on the behavior of
$\chi_0(\beta J_0,\beta h)$ - the susceptibility of the unperturbed
system - correspond to first- or second-order phase transitions.  This
phenomenon will be explicitly shown in the example of the
one-dimensional small-world model. Note that, as it will result from the detailed 
analysis of the self-consistent equation (Sec. IIIB), 
in any case, the critical behavior of the unperturbed model - if any - can never influence
the behavior of the random model.

The paper is organized as follows.  In Sec. II we introduce the class
of small-world networks over which we define the random Ising models,
stressing some important differences concerning the definition of the
correlation functions with respect to those usually considered in
``ordinary'' random models.  In Sec. III we present our method: in
Sec. IIIA we provide the self-consistent equations and their relations
with physical correlation functions, in Sec. IIIB we analyze the
stability of the solutions of the self consistent equations and the
critical surface and behavior of the system. We separate the Sec. IIIB
in the sub-cases $J_0\geq 0$ and $J_0<0$.
In Sec. IIIC we discuss the limits of the method.  In Sec. IIID we
study the stability between the F and the SG phases and the phase
diagram.  Finally, in Sec. IIIE we mention how to generalize the
method to cases with more different short-range
couplings $J_0$, and to analyze possible
disordered antiferromagnetic systems.  In Secs. IV, V and VI the theory is
applied to the three above mentioned example cases.  The successive Secs.
VII, VIII and IX are devoted to the derivation of the method.  The starting
point of the proof is given in Sec. VII and is based on a general
mapping between a random model and a non random one
\cite{MOI}-\cite{MOIII} suitably adapted to the present case.  The
self-consistent equations are then easily derived in Sec. VIII.  Note
that, apart from the equations concerning the stability between the
P-F and the P-SG transitions, which are derived in Sec. IX, the
derivations of the equations presented in Sec. IIIB are mostly left to
the reader, since they can be easily obtained by standard arguments of
statistical mechanics using the Landau free energy $\psi(m)$ that we
provide and that is derived in Sec. VIII too.  Finally, in Sec. X we
draw some conclusions.  In Appendix A we generalize the method to
inhomogeneous external fields to make clear the subtle behavior of the
correlation functions in small-world models.

%
%

\section{Random Ising models on small-world networks}
\label{models}
The family of models we shall consider are random Ising models
constructed by super-imposing random graphs with finite average connectivity $c$
onto some given lattice $\mathcal{L}_0$ whose set of bonds $(i,j)$ 
and dimension will be indicated by $\Gamma_0$ and $d_0$, respectively.
Given an Ising model (\textit{the unperturbed model}) 
of $N$ spins coupled over $\mathcal{L}_0$
through a coupling $J_0$  
with Hamiltonian
\begin{eqnarray}
H_0\defi -J_{0}\sum_{(i,j)\in \Gamma_0}\sigma_{i}\sigma_{j}-h\sum_i \sigma_i
\label{H0},
\end{eqnarray}
and given an ensemble $\mathcal{C}$ 
of unconstrained random graphs $\bm{c}$, $\bm{c}\in\mathcal{C}$,
whose bonds are determined by the adjacency matrix elements $c_{i,j}=0,1$,
we define the corresponding small-world model  
as described by the following Hamiltonian
\begin{eqnarray}
\label{H}
H_{\bm{c},\bm{J}}\defi H_0-\sum_{i<j} c_{ij}{J}_{ij}\sigma_{i}\sigma_{j},
\end{eqnarray}
the free energy $F$ and the averages $\overline{\media{\mathcal{O}}^l}$, with $l=1,2$,
being defined in the usual (quenched) way as ($\beta=1/T$)
\begin{eqnarray}
\label{logZ}
-\beta F\defi \sum_{\bm{c}\in\mathcal{C}} 
P(\bm{c})\int d\mathcal{P}\left(\bm{J}\right)
\log\left(Z_{\bm{c},\bm{J}}\right)
\end{eqnarray} 
and 
\begin{eqnarray}
\label{O}
\overline{\media{\mathcal{O}}^l}\defi 
\sum_{\bm{c}\in\mathcal{C}} P(\bm{c}) \int d\mathcal{P}\left(\bm{J}\right)
\media{\mathcal{O}}_{\bm{c},\bm{J}}^l, \quad l=1,2
\end{eqnarray} 
where $Z_{\bm{c},\bm{J}}$ 
is the partition function of the quenched system
\begin{eqnarray}
\label{Z}
Z_{\bm{c},\bm{J}}= \sum_{\{\sigma_{i}\}}
e^{-\beta H_{\bm{c},\bm{J}}\left(\{\sigma_i\}\right)}, 
\end{eqnarray} 
$\media{\mathcal{O}}_{\bm{c},\bm{J}}$ the Boltzmann-average 
of the quenched system ($\media{\mathcal{O}}$ depends on the
given realization of the ${J}$'s and of $\bm{c}$:
$\media{\mathcal{O}}=\media{\mathcal{O}}_{\bm{c};\bm{J}}$;
for shortness we later will omit to write these dependencies)
\begin{eqnarray}
\label{OO}
\media{\mathcal{O}}_{\bm{c},\bm{J}}\defi \frac{\sum_{\{\sigma_i\}}\mathcal{O}e^{-\beta 
H_{\bm{c},\bm{J}}\left(\{\sigma_i\}\right)}}{Z_{\bm{c},\bm{J}}}, 
\end{eqnarray} 
and $d\mathcal{P}\left(\bm{J}\right)$ and $P(\bm{c})$ 
are two product measures 
given in terms of two normalized measures $d\mu(J_{i,j})\geq 0$ and $p(c_{i,j})\geq 0$, respectively: 
\begin{eqnarray}
\label{dP}
d\mathcal{P}\left(\bm{J}\right)\defi \prod_{(i,j),i<j} 
d\mu\left( {J}_{i,j} \right),
\quad \int d\mu\left( {J}_{i,j} \right)=1,
\end{eqnarray}
\begin{eqnarray}
\label{Pg}
P(\bm{c})\defi \prod_{(i,j),i<j} p(c_{i,j}),
\quad \sum_{c_{i,j}=0,1} p(c_{i,j})=1.
\end{eqnarray}

The variables
$c_{i,j}\in\{0,1\}$ specify whether a ``long-range'' bond between the sites
$i$ and $j$ is present ($c_{i,j}=1$) or absent ($c_{i,j}=0$), whereas
the $J_{i,j}$'s are the random variables of the given bond $(i,j)$.
For the $J_{i,j}$'s we will not assume any particular distribution, 
while, to be specific, for the $c_{i,j}$'s we shall consider the following 
distribution
\begin{eqnarray}
\label{PP}
 p(c_{ij})=
\frac{c}{N}\delta_{c_{ij},1}+\left(1-\frac{c}{N}\right)\delta_{c_{ij},0}.
\end{eqnarray}
This choice leads in the thermodynamic limit $N\to\infty$ to a
number of long range connections per site distributed according
to a Poisson law with mean $c>0$ (so that in average there are in total $cN/2$ bonds). 
Note however that the main results we report
in the next section are easily generalizable to any case in which
Eq. (\ref{Pg}) holds, or holds only in the thermodynamic limit due
a sufficiently small number of constrains among the matrix elements
$c_{i,j}$. 

When we will need to be specific, 
for the $J_{i,j}$'s we will assume either the distribution
\begin{eqnarray}
\label{dPF}
\frac{d\mu\left( {J}_{i,j} \right)}{d{J}_{i,j}}=\delta\left( {J}_{i,j}-J\right),
\end{eqnarray} 
or 
\begin{eqnarray}
\label{dPSG}
\frac{d\mu\left( {J}_{i,j} \right)}{d{J}_{i,j}}=
p\delta\left({J}_{i,j}-J\right)d{J}_{i,j}+
(1-p)\delta\left( {J}_{i,j}+J\right),
\end{eqnarray} 
to consider ferromagnetism or glassy phases, respectively.
In Eq. (\ref{dPSG}) $p\in [0,1]$.

The quantities of major interest are the averages, and the quadratic
averages, of the correlation functions 
which for shortness will be indicated by ${{C}}^{(\mathrm{1})}$ and
${{C}}^{(\mathrm{2})}$. For example, the following are non connected correlation functions
of order $k$:
\begin{eqnarray}
\label{CF}
{{C}}^{(\mathrm{1})}&=&\overline{\media {\sigma_{i_1}\ldots \sigma_{i_k}} }, \\
\label{CG}
{{C}}^{(\mathrm{2})}&=&\overline{\media {\sigma_{i_1}\ldots \sigma_{i_k}}^2 },
\end{eqnarray} 
where $k\geq 1$ and the indices $i_1,\ldots,i_k$ are supposed all different. 
For shortness we will keep on to use the symbols ${{C}}^{(\mathrm{1})}$ 
and ${{C}}^{(\mathrm{2})}$
also for the connected correlation function since, as we shall see in the next
section, they obey to the same rules of transformations.
We point out that the set of indices $i_1,\ldots,i_k$ is
fixed along the process of the two averages. This implies in particular
that, if we consider the spin with index $i$ and the spin with index $j$, 
their distance remains undefined, or more precisely, 
the only meaningful distance between $i$ and $j$, is the distance  
defined over $\mathcal{L}_0$, \textit{i.e.}, the Euclidean distance
between $i$ and $j$, which we will indicate as $||i-j||_{_{_0}}$.
 
Therefore, throughout this paper, it must be kept in mind that, 
for example, ${{C}}^{(\mathrm{1})}(||i-j||_{_{_0}})=
\overline{\media{\sigma_{i}\sigma_{j}} }$ 
is very different from the correlation function ${{G}}^{(\mathrm{1})}(l)$ 
of two points at a fixed
distance $l$, $l$ being here the distance defined over both $\mathcal{L}_0$
and the random graph $\bm{c}$, \textit{i.e.}, the minimum number 
of bonds to join two points
among both the bonds of $\Gamma_0$ and the bonds of the random graph $\bm{c}$.
In fact, if, for $J_0=0$, one considers all the possible 
realizations of the Poisson graph, and then all the possible
distances $l$ between two given points $i$ and $j$, one has   
\begin{eqnarray}
\label{CF1}
{{C}}^{(\mathrm{1})}(||i-j||_{_{_0}})&=&
\overline{\media {\sigma_{i}\sigma_{j}} }-\overline{\media
  {\sigma_{i}}} \overline{\media {\sigma_{j}} }
\nonumber \\
&=&\sum_{l=1}^N P_N(l){{G}}^{(\mathrm{1})}(l)
\end{eqnarray} 
where here $P_N(l)$ is the probability that, in the system with $N$ spins, 
the shortest path between the vertices $i$ and $j$ has length $l$.
If we now use ${{G}}^{(\mathrm{1})}(l)$
$\sim (\tanh(\beta J))^l$ \cite{Corr} (in the P region holds the equality) and the fact that
the average of $l$ with respect to $P_N(l)$ 
is of the order $\log(N)$, we see that the two point connected
correlation function (\ref{CF1}) goes to 0 in the thermodynamic limit. 
Similarly, all the connected
correlation functions defined in this way are zero in this limit.
Note however, that this independence of the variables holds only if $J_0=0$.
This discussion will be more deeply analyzed along the proof by using another point of view,
based on mapping the random model to a suitable fully connected model.

\section{An effective field theory}
\subsection{The self-consistent equations}
Depending on the temperature T, and 
on the parameters of the probability distributions, $d\mu(\cdot)$ and $p(\cdot)$,
the random model may stably stay either in the paramagnetic (P), in the ferromagnetic (F), 
or in the spin-glass (SG) phase.
In our approach for the F and SG phases there are two natural order parameters
that will be indicated by $m^{(\mathrm{F})}$ and $m^{(\mathrm{SG})}$.
Similarly, for any correlation function, quadratic or not, there are
two natural quantities 
indicated by $C^{(\mathrm{F})}$ and $C^{(\mathrm{SG})}$, and that in turn
will be calculated in terms of $m^{(\mathrm{F})}$ and $m^{(\mathrm{SG})}$,
respectively. 
To avoid confusion, it should be kept in mind that
in our approach,
for any observable $\mathcal{O}$,
there are - in principle - always 
two solutions that we label as F and SG,
but, as we shall discuss in Sec. IIID, for any temperature,
only one of the two solutions is stable and useful 
in the thermodynamic limit.

In the following, we will use the label $\mathop{}_0$ 
to specify that we are referring
to the unperturbed model with Hamiltonian (\ref{H0}).
Note that all the equations presented in this paper 
have meaning and usefulness also for sufficiently large but
finite size $N$. For shortness we shall omit to write the dependence on $N$.
 
Let $m_0(\beta J_0,\beta h)$ be the stable magnetization 
of the unperturbed model with coupling $J_0$ and in the presence of a uniform
external 
field $h$ at inverse temperature $\beta$. 
Then, the order parameters 
$m^{(\Sigma)}$, $\Sigma$=F,SG, 
satisfy the following self-consistent decoupled equations
\begin{eqnarray}
\label{THEOa}
m^{(\Sigma)}=m_0(\beta J_0^{(\Sigma)},
\beta J^{(\Sigma)}m^{(\Sigma)}+\beta h),
\end{eqnarray} 
where the effective couplings $J^{(\mathrm{F})}$, $J^{(\mathrm{SG})}$,
$J_0^{(\mathrm{F})}$ and $J_0^{(\mathrm{SG})}$ are given by
\begin{eqnarray}
\label{THEOb}
\beta J^{(\mathrm{F})}= c\int d\mu(J_{i,j})\tanh(\beta J_{i,j}),
\end{eqnarray} 
\begin{eqnarray}
\label{THEOc}
\beta J^{(\mathrm{SG})}= c\int d\mu(J_{i,j})\tanh^2(\beta J_{i,j}),
\end{eqnarray} 
\begin{eqnarray}
\label{THEOd}
J_0^{(\mathrm{F})}= J_0,
\end{eqnarray} 
and
\begin{eqnarray}
\label{THEOe}
\beta J_0^{(\mathrm{SG})}= \tanh^{-1}(\tanh^2(\beta J_0)).
\end{eqnarray}
Note that $|J_0^{(\mathrm{F})}|>J_0^{(\mathrm{SG})}$.

For the correlation functions ${{C}}^{(\Sigma)}$, $\Sigma$=F,SG, for sufficiently 
large $N$ we have
\begin{eqnarray}
\label{THEOh}
{{C}}^{(\Sigma)}=
{{C}}_0(\beta J_0^{(\Sigma)},\beta J^{(\Sigma)} m^{(\Sigma)}+\beta h)
+\mathop{O}\left(\frac{1}{N}\right),
\end{eqnarray} 
where ${{C}}_0(\beta J_0,\beta h)$ is the correlation function of the unperturbed 
(non random) model.

Concerning the free energy density $f$ we have 
\begin{eqnarray}
\label{THEOl} 
&& \beta f^{(\Sigma)} = -\frac{c}{2}\int d\mu(J_{i,j})
\log\left[\cosh(\beta J_{i,j})\right]
\nonumber \\
&& - \lim_{N\to\infty}\frac{1}{N}\sum_{(i,j)\in\Gamma_0}\log
\left[\cosh(\beta J_0)\right]
-\log\left[2\cosh(\beta h)\right]
\nonumber \\
&& + \{\lim_{N\to\infty}\frac{1}{N}\sum_{(i,j)\in\Gamma_0}\log
\left[\cosh(\beta J_0^{(\Sigma)})\right] 
\nonumber \\
&& +\log\left[2\cosh(\beta h)\right]\}
\times \frac{1}{l} +\frac{1}{l}L^{(\Sigma)}(m^{(\Sigma)}),
\end{eqnarray}
where $l=1,2$ for $\Sigma$=F,SG, respectively, and the non trivial free energy term $L^{(\Sigma)}$ is given by 
\begin{eqnarray}
\label{THEOll}
L^{(\Sigma)}(m)\defi
\frac{\beta J^{(\Sigma)}\left(m\right)^2}{2}+
\beta f_0\left(\beta J_0^{(\Sigma)},\beta J^{(\Sigma)}m+\beta h\right),
\end{eqnarray} 
$f_0(\beta J_0,\beta h)$ being the free energy density in the thermodynamic
limit of the unperturbed model with coupling $J_0$ and in the presence
of an external field $h$, at inverse temperature $\beta$.

For given $\beta$, among all the possible solutions of Eqs. (\ref{THEOa}), in the thermodynamic
limit, for both $\Sigma$=F and $\Sigma$=SG, 
the true solution $\bar{m}^{(\Sigma)}$, or leading solution, 
is the one that minimizes $L^{(\Sigma)}$:
\begin{eqnarray}
\label{THEOlead}
L^{(\Sigma)}\left(\bar{m}^{(\Sigma)}\right)&=&
\min_{m\in [-1,1]} L^{(\Sigma)}\left(m\right).
\end{eqnarray} 

Finally, let $k$ be the order of a given correlation function
$C^{(\mathrm{1})}$ or $C^{(\mathrm{2})}$.
The averages and the quadratic averages over the disorder,
$C^{(\mathrm{1})}$ and $C^{(\mathrm{2})}$, are related to  
$C^{(\mathrm{F})}$ and $C^{(\mathrm{SG})}$, as follows
\begin{eqnarray}
\label{THEOa0}
C^{(\mathrm{1})}&=&C^{(\mathrm{F})}, \quad \mathrm{in~F}, \\
\label{THEOa01}
C^{(\mathrm{1})}&=& 0, \quad k ~ \mathrm{odd}, \quad \mathrm{in~SG}, \\
\label{THEOa02}
C^{(\mathrm{1})}&=&C^{(\mathrm{SG})}, \quad k ~ \mathrm{even}, \quad \mathrm{in~SG},
\end{eqnarray} 
and
\begin{eqnarray}
\label{THEOa03}
C^{(\mathrm{2})}&=&\left(C^{(\mathrm{F})}\right)^2, \quad \mathrm{in~F}, \\
\label{THEOa04}
C^{(\mathrm{2})}&=&\left(C^{(\mathrm{SG})}\right)^2, \quad \mathrm{in~SG}.
\end{eqnarray} 

From Eqs. (\ref{THEOa03}) and (\ref{THEOa04}) for $k=1$, we note that
the Edward-Anderson order parameter \cite{EA}
$C^{(\mathrm{2})}=\overline{\media{\sigma}^2}=q_{EA}$ is 
equal to $(C^{(\mathrm{SG})})^2=(m^{(\mathrm{SG})})^2$ only in the SG phase, whereas 
in the F phase we have $q_{EA}=(m^{(\mathrm{F})})^2$.
Therefore, since $m^{(\mathrm{SG})}\neq m^{(\mathrm{F})}$,
$m^{(\mathrm{SG})}$ is not equal to $\sqrt{q_{EA}}$;
in our approach $m^{(\mathrm{SG})}$ represents a sort of a spin glass order parameter \cite{Parisi}.

The localization and the reciprocal stability between the F and SG phases will be
discussed in Sec. IIID. Note however that, at least for lattices $\mathcal{L}_0$
having only loops of even length, the stable P region is always that 
corresponding to a P-F phase diagram, so that in the P region
the correlation functions must be calculated only
through Eqs. (\ref{THEOa0}) and (\ref{THEOa03}).

As an immediate consequence of Eq. (\ref{THEOa}) we get the susceptibility 
$\tilde{\chi}^{(\Sigma)}$ of the random model:
\begin{eqnarray}
\label{THEOchie}
\tilde{\chi}^{(\Sigma)}= 
\frac{\tilde{\chi}_0\left(\beta J_0^{(\Sigma)},\beta J^{(\Sigma)}m^{(\Sigma)}+\beta h\right)}
{1-\beta J^{(\Sigma)}\tilde{\chi}_0
\left(\beta J_0^{(\Sigma)},\beta J^{(\Sigma)}m^{(\Sigma)} +\beta h\right)},
\end{eqnarray}
where $\tilde{\chi}_0$ stands for the susceptibility $\chi_0$ of the
unperturbed model divided by $\beta$ (we will adopt throughout this dimensionless 
definition of the susceptibility):
\begin{eqnarray}
\label{THEOchiedef}
\tilde{\chi}_0 \left(\beta J_0,\beta h\right)\defi 
\frac{\partial m_0\left(\beta J_0,\beta h\right)}{\partial (\beta h)}
=\frac{1}{\beta}\frac{\partial m_0\left(\beta J_0,\beta h\right)}{\partial h},
\end{eqnarray}
and similarly for the random model.

For the case $\Sigma=$F without disorder ($d\mu(J')=\delta(J'-J)dJ'$), 
Eq. (\ref{THEOchie}) was already derived in \cite{Hastings2} by series expansion
techniques at zero field ($h=0$) in the P region (where $m=0$). 

Another remarkable consequence of our theory comes from Eq. (\ref{THEOh}).
We see in fact that in the thermodynamic limit any correlation function
of the random model fits with the correlation function of the unperturbed model
but immersed in an effective field that is exactly 
zero in the P region and zero external field ($h=0$).
In other words, in terms of correlation functions, 
in the P region, the random model and the unperturbed model are indistinguishable
(modulo the transformation $J_0\to J_0^{(\mathrm{SG})}$ for $\Sigma=$SG).
Note however that this assertion holds only for a given correlation function 
calculated in the thermodynamic limit. In fact, the corrective $\mathop{O}(1/N)$ term
appearing in the rhs of Eq. (\ref{THEOh}) cannot be neglected when
we sum the correlation functions over all the sites $i\in\mathcal{L}_0$,
as to calculate the susceptibility; yet it is just
this corrective $\mathop{O}(1/N)$ 
term that gives rise to the singularities in the random model.

More precisely, for the two point connected correlation function 
\begin{eqnarray}
\label{THEOC2}
\tilde{\chi}_{i,j}^{(\Sigma)}\defi 
\overline{\media{\sigma_i\sigma_j}^l-\media{\sigma_i}^l\media{\sigma_j}^l},
\end{eqnarray}
where $l=1,2$ for $\Sigma=$ F, SG, respectively, if 
\begin{eqnarray}
\label{THEOC2a}
\tilde{\chi}_{0;i,j}\defi 
\media{\sigma_i\sigma_j}_0-\media{\sigma_i}_0\media{\sigma_j}_0,
\end{eqnarray}
we have
\begin{eqnarray}
\label{THEOC2b}
\tilde{\chi}_{i,j}^{(\Sigma)}&=&
\tilde{\chi}_{0;i,j}(\beta J_0^{(\Sigma)},\beta J^{(\Sigma)} m^{(\Sigma)}+\beta h)+\frac{\beta J^{(\Sigma)}}{N}
\nonumber \\ &\times & \frac{\left[\tilde{\chi}_{o} 
(\beta J_0^{(\Sigma)},\beta J^{(\Sigma)}{{m}}^{(\Sigma)}+ \beta h)\right]^2}
{1-\beta J^{(\Sigma)} \tilde{\chi}_{o} 
(\beta J_0^{(\Sigma)},\beta J^{(\Sigma)}{{m}}^{(\Sigma)}+ \beta h)},
\end{eqnarray}
where the dependence on $N$ in $\tilde{\chi}_{i,j}^{(\Sigma)}$ and $\tilde{\chi}_0$ are understood. 
Eq. (\ref{THEOC2b}) clarifies the structure of the correlation
functions in small-world models. In the rhs we have two terms: the former
is a distance-dependent short-range term whose finite correlation length, for $T\neq T_{c0}^{(\Sigma)}$ 
($T_{c0}^{(\Sigma)}$ being the critical temperature of the unperturbed model with coupling $J_0^{(\Sigma)}$),
makes it normalizable, the latter is instead a distance-independent long-range term which turns out
to be normalizable thanks to the $1/N$ factor. Once summed, both the terms give a finite
contribution to the susceptibility.
It is immediate to verify that by summing $\tilde{\chi}_{i,j}^{(\Sigma)}$ 
over all the indices $i,j\in\mathcal{L}_0$ and dividing by $N$ we get back - as it must be -
Eq. (\ref{THEOchie}). Eq. (\ref{THEOC2b}) will be derived in Appendix A where we generalize
the theory to a non homogeneous external field.

\subsection{Stability: critical surfaces and critical behavior}
Note that, for $\beta$ sufficiently small (see later), 
Eq. (\ref{THEOa}) has always the solution $m^{(\Sigma)}=0$, and furthermore,
if $m^{(\Sigma)}$ is a solution, $-m^{(\Sigma)}$ is a solution as well. 
From now on, if not explicitly said, 
we will refer only to the positive (possibly zero) solution,
the negative one being understood.
A solution $m^{(\Sigma)}$ of Eq. (\ref{THEOa}) is stable (but in general not unique) if
\begin{eqnarray}
\label{THEOO}
1- \beta J^{(\Sigma)}\tilde{\chi}_0 
\left(\beta J_0^{(\Sigma)},\beta J^{(\Sigma)}m^{(\Sigma)} + \beta h\right)>0.
\end{eqnarray}

For what follows, we need to rewrite
the non trivial part of the free energy density
$L^{(\Sigma)}(m)$ as
\begin{eqnarray}
\label{THEOL}
L^{(\Sigma)}(m)&=&
\beta f_0 \left(\beta J_0^{(\Sigma)},0\right) 
- m_0\left(\beta J_0^{(\Sigma)},0\right)\beta h 
\nonumber \\
&& + \psi^{(\Sigma)}\left(m\right),
\end{eqnarray}
where the introduced term $\psi^{(\Sigma)}$ plays the role of a Landau free energy
density and is responsible for the critical behavior of the system. 
Around $m=0$, up to terms $\mathop{O}(h^2)$ and $\mathop{O}(m^3 h)$,
$\psi^{(\Sigma)}(m)$ can be expanded as follows
\begin{eqnarray}
\label{THEOL1}
\psi^{(\Sigma)}\left(m\right)
&=& \frac{1}{2}a^{(\Sigma)} m^2 + 
\frac{1}{4} b^{(\Sigma)} m^4
+\frac{1}{6} c^{(\Sigma)} m^6
\nonumber \\ &&
 - m\beta \tilde{h}^{(\Sigma)}
\nonumber \\ &&
+\Delta\left(\beta f_0\right)(\beta J_0^{(\Sigma)},\beta J^{(\Sigma)}m),
\end{eqnarray}
where 
\begin{eqnarray}
\label{THEOL2}
a^{(\Sigma)}=
\left[1-\beta J^{(\Sigma)}\tilde{\chi}_0 
\left(\beta J_0^{(\Sigma)},0\right)\right]
\beta J^{(\Sigma)},
\end{eqnarray}
\begin{eqnarray}
\label{THEOL3}
b^{(\Sigma)}=- \frac{\partial^2}{\partial(\beta h)^2}
{\left. {\tilde{\chi}_0\left(\beta J_0^{(\Sigma)},\beta h\right)}
\right|_{_{\beta h =0} }
\frac{\left(\beta J^{(\Sigma)}\right)^4}{3!}},
\end{eqnarray}
\begin{eqnarray}
\label{THEOL4}
c^{(\Sigma)}=- \frac{\partial^4}{\partial(\beta h)^4}
{\left. {\tilde{\chi}_0\left(\beta J_0^{(\Sigma)},\beta h\right)}
\right|_{_{\beta h =0} }
\frac{\left(\beta J^{(\Sigma)}\right)^6}{5!}},
\end{eqnarray}
\begin{eqnarray}
\label{THEOL5}
\tilde{h}^{(\Sigma)}=m_0\left(\beta J_0^{(\Sigma)},0\right)J^{(\Sigma)}+
\tilde{\chi}_0\left(\beta J_0^{(\Sigma)},0\right)\beta J^{(\Sigma)}\beta h,
\end{eqnarray}
finally, the last term 
$\Delta \left(\beta f_0\right) \left(\beta J_0,\beta J^{(\Sigma)}m\right)$
is defined implicitly to render Eqs. (\ref{THEOL}) and (\ref{THEOL1}) exact, 
but terms $\mathop{O}(h^2)$ and $\mathop{O}(m^3 h)$, explicitly:
\begin{eqnarray}
\label{THEOL6}
&&\Delta \left(\beta f_0\right) \left(\beta J_0^{(\Sigma)},\beta J^{(\Sigma)}m\right)=
\nonumber \\
&-&\sum_{k=4}^\infty \frac{\partial^{2k-2}}{\partial(\beta h)^{2k-2}}
{\left. {\tilde{\chi}_0\left(\beta J_0^{(\Sigma)},\beta h\right)}
\right|_{_{\beta h =0} }
\frac{\left(\beta J^{(\Sigma)}\right)^{2k}}{(2k)!}}.
\end{eqnarray}

We recall that the
$k-2$-th derivative of $\tilde{\chi}_0\left(\beta J_0^{(\Sigma)},\beta h\right)$
with respect to the second argument, 
calculated at $h=0$, gives the total sum of all the $k$-th
cumulants normalized to $N$:
$\partial_{\beta h}^{k-2}\tilde{\chi}_0
\left(\beta J_0^{(\Sigma)},\beta h\right)|_{h=0}=\sum_{i_1,\ldots,i_k}
\media{\sigma_{i_1}\cdots\sigma_{i_k}}_0^{(c)}/N$, 
where $\media{\sigma_{i_1}\cdots\sigma_{i_k}}_0^{(c)}$ stands for the
cumulant, or connected correlation function, of order $k$ of the unperturbed
model, $\media{\sigma_{i_1}\sigma_{i_2}}_0^{(c)}=\media{\sigma_{i_1}\sigma_{i_2}}_0
-\media{\sigma_{i_1}}_0\media{\sigma_{i_2}}_0$, etc..
Note that, apart from the sign, 
these terms are proportional to the Binder cumulants \cite{Binder}
(which are all zero above $T_{c0}$ for $k>2$) only for $N$ finite. In the thermodynamic
limit the terms $b^{(\Sigma)}$, $c^{(\Sigma)}$, $\ldots$, 
in general are non zero 
and take into account the large deviations of the block-spin
distribution functions from the Gaussian distribution.

Let $T_c^{(\Sigma)}=1/\beta_c^{(\Sigma)}$ be
the critical temperatures, if any, of the random model and 
let $t^{(\Sigma)}$ be the corresponding reduced temperatures:
\begin{eqnarray}
\label{THEOtt}
t^{(\Sigma)}\defi \frac{T-T_c^{(\Sigma)}}{T_c^{(\Sigma)}}=
\frac{\beta_c^{(\Sigma)}-\beta}{\beta_c^{(\Sigma)}}+\mathop{O}(t^{(\Sigma)})^2.
\end{eqnarray} 
Here, the term ``critical temperature'', stands for any
temperature where some singularity shows up. However, if we limit ourselves 
to consider only the critical temperatures crossing which the system passes from a P
region to a non P region, from Eq. (\ref{THEOO}) it is easy to see that,
independently on the sign of $J_0$ and on the nature of the phase transition,
we have the important inequalities
\begin{eqnarray}
\label{THEOf1}
\beta_c^{(\Sigma)}<\beta_{c0}^{(\Sigma)},
\end{eqnarray} 
where we have introduced $\beta_{c0}^{(\Sigma)}$, 
the inverse critical temperature of the unperturbed model
with coupling $J_0^{(\Sigma)}$ and zero external field.
If more than one critical temperature is present in the unperturbed model,
$\beta_{c0}^{(\Sigma)}$ is the value corresponding to the smallest value
of these critical temperatures (highest in terms of $\beta$).
Formally we set $\beta_{c0}^{(\Sigma)}=\infty$  
if no phase transition is present in the unperturbed model.
A consequence of Eq. (\ref{THEOL5}) is that, in studying the critical behavior
of the system for $h=0$, we can put $\tilde{h}^{(\Sigma)}=0$.
Throughout this paper, we shall reserve the name
critical temperature of the unperturbed model as a P-F
critical temperature through which the magnetization 
$m_0\left(\beta J_0,0\right)$ passes from a zero to a non zero value,
continuously or not. This implies, in particular, that for
$J_0<0$ we have - formally - $\beta_{c0}=\infty$. 

In this paper we shall study only the  
order parameters $m^{(\mathrm{F})}$ and $m^{(\mathrm{SG})}$,
whereas we will give only few remarks on how to
generalize the method for possible antiferromagnetic order parameters.
We point out however that the existence of possible antiferromagnetic transitions
of the unperturbed model does not affect the results we present
in this paper.

It is convenient to distinguish the cases $J_0\geq 0$ and $J_0<0$,
since they give rise to two strictly different scenarios.

\subsubsection{The case $J_0\geq 0$}
In this case $\beta J^{(\Sigma)}\tilde{\chi}_0 
\left(\beta J_0,\beta J^{(\Sigma)}m^{(\Sigma)} + \beta h\right)$
is an increasing function of $\beta$ 
for $\beta<\beta_{c0}^{(\Sigma)}$ (and for $h\geq 0$). As a consequence, 
we have that for sufficiently low temperatures,
the solution $m^{(\Sigma)}=0$ of Eq. (\ref{THEOa}) becomes 
unstable and two - and only two - non zero solutions 
$\pm m^{(\Sigma)}$ are instead favored.
The inverse critical temperatures 
$\beta_c^{(\mathrm{F})}$ and $\beta_c^{(\mathrm{SG})}$ can be determined
by developing - for $h=0$ - Eqs. (\ref{THEOa}) for small 
$m^{(\mathrm{F})}$ and $m^{(\mathrm{SG})}$, respectively, which, in terms of
$\tilde{\chi}_0$ gives the following exact equation
\begin{eqnarray}
\label{THEOg}
{\tilde{\chi}_0\left(\beta_c^{(\Sigma)}J_0^{(\Sigma)},0\right)}
\beta_c^{(\Sigma)}J^{(\Sigma)}=1, \quad \beta_{c}^{(\Sigma)}<\beta_{c0}^{(\Sigma)},
\end{eqnarray} 
where the constrain $\beta_{c}^{(\Sigma)}<\beta_{c0}^{(\Sigma)}$ excludes other
possible spurious solutions that may appear when $d_0\geq 2$ (since in this case $\beta_{c0}^{(\Sigma)}$ may be finite).

The critical behavior of the system can be derived by developing 
Eqs. (\ref{THEOa}) for small fields. Alternatively, one can study the
critical behavior by analyzing the Landau free energy density
$\psi^{(\Sigma)}(m^{(\Sigma)})$ given by Eq. (\ref{THEOL1}).

In the following we will suppose that for $J_0>0$, 
$b^{(\Sigma)}$ be positive. We have checked this hypothesis in all the models
we have until now considered and that will be analyzed in Secs. IV, V and VI.
Furthermore, even if the sign of $c^{(\Sigma)}$   
cannot be in general \textit{a priori} established, for the convexity
of the function $f_0$ with respect to $\beta h$, the sum of the six-th term with
$\Delta \left(\beta f_0\right) \left(\beta J_0^{(\Sigma)},\beta
J^{(\Sigma)}m^{(\Sigma)}\right)$, in Eq. (\ref{THEOL1}) must go 
necessarily to $+\infty$ for $m^{(\Sigma)}\to\infty$. 
In conclusion, when $J_0\geq 0$, for the  critical
behavior of the system, the only relevant parameters of $\psi^{(\Sigma)}$ are 
$a^{(\Sigma)}$, $b^{(\Sigma)}$ and
$h^{(\Sigma)}=\tilde{\chi}(\beta J_0^{(\Sigma)},0)J^{(\Sigma)}h$, 
so that the critical behavior
can be immediately derived as in the Landau theory for the so called \textit{$m^4$ model} \cite{Landau}.
On noting that
\begin{eqnarray}
\label{THEOAa}
\left\{
\begin{array}{l}
a^{(\Sigma)}\geq 0, \qquad \mathrm{for} ~ t^{(\Sigma)}\geq 0,\\
a^{(\Sigma)}<0, \qquad \mathrm{for} ~ t^{(\Sigma)}<0,\\
\end{array}
\right.
\end{eqnarray}
it is convenient to define
\begin{eqnarray}
\label{THEOA}
A^{(\Sigma)}\defi -\beta\frac{\partial}{\partial\beta}a^{(\Sigma)},
\end{eqnarray} 
so that we have
\begin{eqnarray}
\label{THEOA1}
a^{(\Sigma)}=A^{(\Sigma)}|_{_{\beta=\beta_c^{(\Sigma)}}}t^{(\Sigma)}
+\mathop{O}(t^{(\Sigma)})^2.
\end{eqnarray} 

Note that, due to the fact that $J_0\geq 0$, $A^{(\Sigma)}>0$,
and, as already mentioned, $b^{(\Sigma)}\geq 0$ as well. 
By using Eq. (\ref{THEOA1}) for $\beta<\beta_{c0}$ and near
$\beta_c^{(\Sigma)}$, we see that the minimum $\bar{m}^{(\Sigma)}$ 
of $\psi^{(\Sigma)}$, \textit{i.e.}, the solution
of Eq. (\ref{THEOa}) near the critical temperature, is given by 
\begin{eqnarray}
\label{THEOmag}
\bar{m}^{(\Sigma)}=\left\{
\begin{array}{l}
0,\quad \qquad \qquad \qquad \qquad \qquad \qquad t^{(\Sigma)}~\geq 0, \\
\sqrt{-\frac{A^{(\Sigma)}}
{b^{(\Sigma)}}
|_{_{\beta=\beta_c^{(\Sigma)}}}
t^{(\Sigma)}~}+\mathop{O}(t^{(\Sigma)}),
\quad t^{(\Sigma)}< 0.
\end{array}
\right.
\end{eqnarray} 

Similarly, we can write general formulas for the susceptibility and
the equation of state. We have
\begin{eqnarray}
\label{THEOchi}
\tilde{\chi}^{(\Sigma)}=\left\{
\begin{array}{l}
\frac{\beta J^{(\Sigma)}\tilde{\chi}_0\left( \beta J_0^{(\Sigma)},0\right)}
{A^{(\Sigma)}}|_{_{\beta=\beta_c^{(\Sigma)}}}
\frac{1}{t^{(\Sigma)}}+\mathop{O}(1),~ t^{(\Sigma)}\geq 0, \\
\frac{\beta J^{(\Sigma)}\tilde{\chi}_0\left( \beta J_0^{(\Sigma)},0\right)}
{-2A^{(\Sigma)}}|_{_{\beta=\beta_c^{(\Sigma)}}}
\frac{1}{t^{(\Sigma)}}+\mathop{O}(1),~ t^{(\Sigma)}<0,
\end{array}
\right.
\end{eqnarray} 
\begin{eqnarray}
\label{THEOstate}
\bar{m}^{(\Sigma)}(h)=\left[
\frac{\beta J^{(\Sigma)}\tilde{\chi}_0\left( \beta J_0^{(\Sigma)},0\right)}
{A^{(\Sigma)}}
\right]^{\frac{1}{3}}_{\beta=\beta_c^{(\Sigma)}} h^{\frac{1}{3}}
+\mathop{O}\left(h^{\frac{2}{3}}\right).
\end{eqnarray} 

Finally, on using Eqs. (\ref{THEOL1}) and (\ref{THEOmag}) we get
that the specific heat $\mathcal{C}^{(\Sigma)}$ 
has the following finite jump discontinuity at $\beta_c^{(\Sigma)}$
\begin{eqnarray}
\label{THEOheat}
\mathcal{C}^{(\Sigma)}=\left\{
\begin{array}{l}
\mathcal{C}_c^{(\Sigma)}, \qquad \qquad \qquad t^{(\Sigma)}~\geq 0, \\
\mathcal{C}_c^{(\Sigma)}+ \frac{\left(A^{(\Sigma)}\right)^2}{2b^{(\Sigma)}}
|_{_{\beta=\beta_c^{(\Sigma)}}},
~ t^{(\Sigma)}< 0,
\end{array}
\right.
\end{eqnarray} 
where $\mathcal{C}_c^{(\Sigma)}$ is the continuous part of the specific heat corresponding
to the part of the free energy density without $\psi^{(\Sigma)}$. 

Hence, as a very general result, 
independently of the structure of the underlying graph $\mathcal{L}_0$ and its dimension $d_0$, 
independently of the nature of the phase transition present in unperturbed model (if any),
and independently of the added random connectivity $c$, provided positive,
we recover that 
the random model has always a mean-field critical behavior 
with a second-order phase transition with the classical exponents
$\beta=1/2$, $\gamma=\gamma'=1$, $\delta=3$ and $\alpha=\alpha'=0$, and 
certain constant coefficients depending on the susceptibility $\tilde{\chi}_0$ 
and its derivatives calculated at $\beta=\beta_c^{(\Sigma)}$ and external field $h=0$. 
Note however, that the correlation length of the system 
calculated along the distance of $\mathcal{L}_0$, $||\cdot||_0$, remains finite also 
at $\beta_c^{(\Sigma)}$. In fact, from Eq. (\ref{THEOh}), for the two point correlation
function at distance $r\defi ||i-j||_{_{_0}}$ in $\mathcal{L}_0$ we have
\begin{eqnarray}
\label{THEOCORR}
{{C}}^{(\Sigma)}(r)=
{{C}}_0(\beta J_0^{(\Sigma)},\beta J^{(\Sigma)}m^{(\Sigma)}+\beta h ;r).
\end{eqnarray} 
If we now assume for ${{C}}_0(\beta J_0,0;r)$ 
the following general Ornstein-Zernike form 
\begin{eqnarray}
\label{THEOCORR1}
{{C}}_0(\beta J_0,0;r)=\frac{e^{-r/\xi_0}}{f_0(r)},
\end{eqnarray} 
$f_0(r)=f_0(\beta J_0;r)$ being
a smooth function of $r$ (which has not to be confused with the free energy density), 
and $\xi_0=\xi_0(\beta J_0)$ the correlation length,
which is supposed to diverge only at $\beta_{c0}$ (if any),
on comparing Eqs. (\ref{THEOCORR}) and (\ref{THEOCORR1}) 
for $\beta\geq \beta_c^{(\Sigma)}$ we have (notice that, as explained in Sec. IIIA,
at least for lattices $\mathcal{L}_0$ having only loops of even length, 
the physical correlation function is only that corresponding to $\Sigma=$F,
\textit{i.e.}, $C^{(1)}=C^{(\mathrm{F})}$)
\begin{eqnarray}
\label{THEOCORR2}
{{C}}^{(\Sigma)}(r)=\frac{e^{-r/\xi^{(\Sigma)}}}{f^{(\Sigma)}(r)},
\end{eqnarray} 
where 
\begin{eqnarray}
\label{THEOCORR3}
f^{(\Sigma)}(r)=f_0(\beta J_0^{(\Sigma)};r),
\end{eqnarray} 
and 
\begin{eqnarray}
\label{THEOCORR4}
\xi^{(\Sigma)}=\xi_0(\beta J_0^{(\Sigma)}).
\end{eqnarray} 
Therefore, due to the inequalities (\ref{THEOf1}), we see that
\begin{eqnarray}
\label{THEOCORR5}
\xi^{(\Sigma)}|_{\beta=\beta_c^{(\Sigma)}}=
\xi_0(\beta_c^{(\Sigma)} J_0^{(\Sigma)})<\infty.
\end{eqnarray} 
The knowledge of ${{C}}_0(\beta J_0,\beta h;r)$ also for $h\neq 0$ would allows
us to find the general expression for ${{C}}^{(\Sigma)}(r)$ through 
Eq. (\ref{THEOCORR}) also for 
$\beta> \beta_c^{(\Sigma)}$. However, since 
${{C}}_0(\beta J_0,\beta h;r)$ has no critical behavior for $h\neq 0$, 
it follows that ${{C}}^{(\Sigma)}(r)$ cannot have a critical behavior 
for $\beta> \beta_c^{(\Sigma)}$ either 
(and then also for $\beta\to \beta_c^{(\Sigma)}$ from the right).
This result is consistent with \cite{Barrat}.

\subsubsection{The case $J_0<0$}
In this case $J_0^{\mathrm{(F)}}<0$, so that - in general - $\beta J^{\mathrm{(F)}}\tilde{\chi}_0 
\left(\beta J_0,\beta J^{\mathrm{(F)}}m^{\mathrm{(F)}}\right)$
is no longer a monotonic function of $\beta$. However, it is easy to see that
that for $\beta=0$ and $\beta\to\infty$, this function goes to 0.
Therefore, for a sufficiently large connectivity $c$, 
from Eq. (\ref{THEOO}) we see that there may appear
at least two regions where the paramagnetic solution $m^{\mathrm{(F)}}=0$ is stable,
separated by a third region in which a non zero solution is instead stable.
However the situation is even more complicated since, unlike
the case $J_0\geq 0$, the non monotonicity of $\beta J^{\mathrm{(F)}}\tilde{\chi}_0 
\left(\beta J_0,\beta J^\mathrm{(F)}m^\mathrm{(F)}\right)$ reflects also
in the fact that the self-consistent Eq. (\ref{THEOa}) for $\Sigma=$F may have
more solutions of the kind $\pm m^\mathrm{(F)},\pm m'^\mathrm{(F)},\ldots$
which are still stable with respect to the stability condition (\ref{THEOO}),
for $h=0$. We face in fact here the problem to compare more stable solutions.
According to Eq. (\ref{THEOlead}), 
in the thermodynamic limit, among all the possible stable solutions,
only $\bar{m}^\mathrm{(F)}$, the solution that minimizes 
$L^\mathrm{(F)}$, survives, 
whereas the not leading ones play the role of metastable states. 
This kind of scenario, which includes also finite jump discontinuities,
has been besides observed in the context of 
small-world neural networks in \cite{Skantos} where
we even observe some analogy in the used formalism, at least 
for the simplest case of one binary pattern.

From Eqs. (\ref{THEOL3}) and (\ref{THEOL4}) we see that the signs of
the Landau coefficients $a^{(\Sigma)}$, $b^{(\Sigma)}$, 
$c^{(\Sigma)}$, $\ldots$, are functions of $\beta$ and $J_0$ only.
Given $J_0<0$, the most important quantity that features the non monotonicity of 
$\beta J^\mathrm{(F)}\tilde{\chi}_0 
\left(\beta J_0,\beta J^\mathrm{(F)}m^\mathrm{(F)}\right)$ 
is the minimum value of $\beta$ over which $b^\mathrm{(F)}$ becomes
negative:
\begin{eqnarray}
\label{THEOneg1}
b^\mathrm{(F)}\leq 0, \qquad \beta\geq\beta_*^\mathrm{(F)}.
\end{eqnarray} 

The equation for $\beta_*^\mathrm{(F)}$, as a function of $J_0$,
defines a point  where $b^\mathrm{(F)}=0$. 
If $J_0<0$, the most general equation for a generic
critical temperature is no longer given by Eq. (\ref{THEOg}).
In fact, in general, a critical temperature now is any temperature where
the stable and leading solution $\bar{m}^\mathrm{(F)}$ 
may have a singular behavior, also with finite jumps between two non zero values. 

There are some simplification when for the Landau coefficient $c^\mathrm{(F)}$,
we have $c^\mathrm{(F)}>0$,
or at least $c^\mathrm{(F)}>0$ out of the P region
In this situation in fact, from Eq. (\ref{THEOL1}) we see that $a^\mathrm{(F)}$,
$b^\mathrm{(F)}$ and $c^\mathrm{(F)}$ are the only relevant terms
for the critical behavior of the system and 
- for small values of $\bar{m}^\mathrm{(F)}$ - we can again apply 
the Landau theory, this time for the so called \textit{$m^6$ model} \cite{Landau}. 
In such a case, for the solution $\bar{m}^\mathrm{(F)}$ we have
\begin{eqnarray}
\label{THEOneg2}
\bar{m}^\mathrm{(F)}&=&
\sqrt{\frac{1}{2c^\mathrm{(F)}}\left(\sqrt{\left(b^\mathrm{(F)}\right)^2
-4a^\mathrm{(F)}c^\mathrm{(F)}}-b^\mathrm{(F)}\right)}, \quad \mathrm{if}, 
\nonumber \\
&& a^\mathrm{(F)}<0, \quad \mathrm{or} 
\nonumber \\
&& a^\mathrm{(F)}\geq 0 \quad \mathrm{and} \quad
b^\mathrm{(F)}\leq -4\sqrt{\frac{a^\mathrm{(F)}c^\mathrm{(F)}}{3}}, 
\end{eqnarray} 
whereas 
\begin{eqnarray}
\label{THEOneg3}
\bar{m}^\mathrm{(F)}&=& 0, \quad \mathrm{if},
\nonumber \\
&& a^\mathrm{(F)}\geq 0 \quad \mathrm{and} \quad 
b^\mathrm{(F)}> -4\sqrt{\frac{a^\mathrm{(F)}c^\mathrm{(F)}}{3}}.
\end{eqnarray} 
From Eqs. (\ref{THEOneg2}) and (\ref{THEOneg3}) 
we see that, if $b^\mathrm{(F)}>0$, 
we have a second-order phase transition and 
Eqs. (\ref{THEOg})-(\ref{THEOCORR5}) are recovered with
Eq. (\ref{THEOneg2}) becoming the second of Eqs. (\ref{THEOmag})
for small and negative values of $a^\mathrm{(F)}$. 
However, from Eq. (\ref{THEOneg2}) we see that,
if $b^\mathrm{(F)}$ is sufficiently negative, 
we have a first-order phase transition which, 
for small values of $a^\mathrm{(F)}$, gives
\begin{eqnarray}
\label{THEOneg4}
\bar{m}^\mathrm{(F)}&=& 
\sqrt{-\frac{b^\mathrm{(F)}}{c^\mathrm{(F)}}}
\left(1-\frac{a^\mathrm{(F)}c^\mathrm{(F)}}{2\left(b^\mathrm{(F)}\right)^2}\right),
\quad \mathrm{if},
\nonumber \\
&& a^\mathrm{(F)}<0 \quad \mathrm{and} \quad b^\mathrm{(F)}<0,\quad \mathrm{or} 
\nonumber \\
&& a^\mathrm{(F)}\geq 0 \quad \mathrm{and} \quad
b^\mathrm{(F)}\leq -4\sqrt{\frac{a^\mathrm{(F)}c^\mathrm{(F)}}{3}}.
\end{eqnarray} 
From Eq. (\ref{THEOneg2}) we see that the line $b^\mathrm{(F)}=-4\sqrt{a^\mathrm{(F)}c^\mathrm{(F)}/3}$
with $a^\mathrm{(F)}\geq 0$ establishes a line of first-order transitions through which 
$\bar{m}^\mathrm{(F)}$ changes discontinuously from zero to 
\begin{eqnarray}
\label{THEOneg5}
\Delta\bar{m}^\mathrm{(F)}= 
\left(\frac{3a^\mathrm{(F)}}{c^\mathrm{(F)}}\right)^{\frac{1}{4}}.
\end{eqnarray} 
The point $a^\mathrm{(F)}=b^\mathrm{(F)}=0$ is a tricritical point where 
the second and first-order transition lines meet. If we approach the
tricritical point along the line $b^\mathrm{(F)}=0$ we get 
the critical indices $\alpha=1/2$, $\alpha'=0$, $\beta=1/4$, $\gamma=\gamma'=1$ and
$\delta=5$. However, this critical behavior along the line
$b^\mathrm{(F)}=0$ has not a great practical interest since from
Eq. (\ref{THEOL3}) we see that it is not possible to keep $b^\mathrm{(F)}$
constant and zero as the temperature varies. 
Finally, we point out that, even if $c^\mathrm{(F)}>0$, 
when the transition is of the first-order,
Eqs. (\ref{THEOneg2}) and (\ref{THEOneg4}) hold only for $b^\mathrm{(F)}$,
and then $a^\mathrm{(F)}$, sufficiently small, since
only in such a case the finite discontinuity of $\bar{m}^\mathrm{(F)}$
is small and then the truncation of the Landau free energy term $\psi^\mathrm{(F)}$
to a finite order meaningful. Note that this question implies also
that we cannot establish a simple and general rule to determine
the critical temperature of a first-order phase transition (we will return
soon on this point).

When $c^\mathrm{(F)}<0$, the Landau theory of the $m^6$ model 
cannot be of course applied.
However, as in the case $J_0>0$, even if the sign of $c^\mathrm{(F)}$   
cannot be \textit{a priori} established, for the convexity
of the function $f_0$ with respect to $\beta h$, the sum of the six-th term with
$\Delta \left(\beta f_0\right) \left(\beta J_0^\mathrm{(F)},\beta
J^\mathrm{(F)}m^\mathrm{(F)}\right)$, in Eq. (\ref{THEOL1}) must go 
necessarily to $+\infty$ for $m^\mathrm{(F)}\to\infty$ 
and a qualitative similar behavior of the $m^6$ model
is expected. 
In general, when $J_0<0$, the exact results are limited to the following ones. 

From now on, if not otherwise explicitly said, we shall reserve the name critical temperature, 
whose inverse value of $\beta$ we still indicate with $\beta_c^\mathrm{(F)}$, 
to any temperature on the boundary of a P region (through which $\bar{m}^\mathrm{(F)}$
passes from 0 to a non zero value, continuously or not).
For each critical temperature, 
depending on the value of $\beta_*^\mathrm{(F)}$, we have three possible
scenario of phase transitions:

\begin{eqnarray}
\label{THEOneg7}
\left\{
\begin{array}{l}
\beta_{c}^\mathrm{(F)}>\beta_*^\mathrm{(F)} \quad \Leftarrow \mathrm{first~order},\\
\beta_{c}^\mathrm{(F)}=\beta_*^\mathrm{(F)} \quad \Leftrightarrow \mathrm{tricritical~point},\\
\beta_{c}^\mathrm{(F)}<\beta_*^\mathrm{(F)} \quad \Rightarrow \mathrm{second~order}.
\end{array}
\right.
\end{eqnarray} 

Note that, according to our definition of critical temperature, 
the critical behavior described by Eqs. (\ref{THEOneg2}) - (\ref{THEOneg5})
represents a particular case of the general scenario expressed by
Eqs. (\ref{THEOneg7}). We see also that, in general, when $b^\mathrm{(F)}\leq 0$, 
approaching the tricritical point, for the critical exponent $\beta$ we have $\beta\leq 1/4$.

In the case in which $\beta_{c}^\mathrm{(F)}$ corresponds to
a second-order phase transition, or in the case in which
$a^\mathrm{(F)}<0$ out of the P region (at least immediately near 
the critical temperature), $\beta_{c}^\mathrm{(F)}$ can be exactly
calculated by Eq. (\ref{THEOg}). When we are not in such cases, the only
exact way to determine the critical temperature is to find 
the full solution for $\bar{m}^\mathrm{(F)}$ which consists in
looking numerically 
for all the possible solutions of Eq. (\ref{THEOa}) and - among those satisfying
the stability condition (\ref{THEOO}) - selecting the one that gives the
minimum value of $L^\mathrm{(F)}$.

\subsection{Level of accuracy of the method}
In the P region, Eqs. (\ref{THEOa}-\ref{THEOC2b}) are exact, 
whereas in the other regions
provide an effective approximation whose level of accuracy depends  
on the details of the model. In particular, in the absence of frustration
the method becomes exact at any temperature in two important limits:
in the limit $c\to 0^+$, in the case of second-order phase transitions, due
to a simple continuity argument;
and in the limit $c\to\infty$, due to the fact that in this
case the system becomes a suitable fully connected model exactly described
by the self-consistent equations (\ref{THEOa}) (of course, 
when $c\to\infty$, to have a finite critical temperature one has to renormalize 
the average of the coupling by $c$).  

However, for any $c>0$, off of the P region
and infinitely near the critical temperature, Eqs. (\ref{THEOa}-\ref{THEOh}) 
are able to give the exact critical 
behavior in the sense of the critical indices and, in the limit
of low temperatures, Eqs. (\ref{THEOa}-\ref{THEOe}) provide the exact
percolation threshold. 
In general, as for the SK model, which can be seen as 
a particular model with $J_0=0$,
the level of accuracy is better for the F phase rather than for
the SG one and this is particularly true 
for the free energy density $f^{(\Sigma)}$, Eq. (\ref{THEOl}).
In fact, though the derivatives of $f^{(\Sigma)}$ are expected
to give a good qualitative and partly also a quantitative
description of the system,
$f^\mathrm{(SG)}$ itself can give wrong results when the SG phase
at low temperatures is considered. We warn the reader that
in a model with $J_0=0$, and a symmetrical distribution $d\mu(J_{i,j})$ with variance
$\tilde{J}$, the method gives a ground state energy per site $u^{(\mathrm{SG})}$,
which grows with $c$ as $u^{(\mathrm{SG})}\sim -\tilde{J}c$, whereas the correct
result is expected to be $u^{(\mathrm{SG})}\sim -\tilde{J} \sqrt{c}$ \cite{Derrida}.
As a consequence, in the SK model, 
in the limit $\beta\to\infty$, the method gives a completely
wrong result with an infinite energy. 
We stress however that the order parameters $m^{(\mathrm{F})}$ and $m^{(\mathrm{SG})}$,
and then also the correlation functions, by construction, 
are exact in the zero temperature limit. 

\subsection{Phase diagram}
The physical inverse critical temperature $\beta_c$ of the random model is in general
a non single-value function of $\bm{X}$: $\beta_c=\beta_c(\bm{X})$, where
$\bm{X}$ represents symbolically the parameters of the
probability $d\mu$ for the couplings $J_{i,j}$, 
and the parameter $c$, the average connectivity (which is also a parameter of the 
probability distribution of the short-cut bonds). The  
parameters of $d\mu$ can be expressed through the moments of $d\mu$, and as they vary the
probability $d\mu$ changes. For example, if $d\mu$ is a Gaussian 
distribution, as in the SK model, there are only two parameters given by
the first and second moment. A concrete example for the one dimensional
small-world model will be shown in Figs. (\ref{sw_pd1}) and (\ref{sw_pd2}).

In the thermodynamic limit, only one of the two solutions with label F or SG survives,
and it is the solution having minimum free energy.
In principle, were our method exact at all temperatures, we were able to derive
exactly all the phase diagram. 
However, in our method, the solution with label F or SG are exact only in their own P region, \textit{i.e.},
the region where $m^{(\mathrm{F})}=0$ or $m^{(\mathrm{SG})}=0$, respectively. 
Unfortunately, according to what we have seen in Sec. IIIC, 
whereas the solution with label F is still a good approximation also out of the P region,
in frustrated model (where the variance of $d\mu$ is large if compared to
its first moment) the free energy of the solution with label SG becomes 
completely wrong at low temperatures.
Therefore, we are not able to give in general the exact boundary between the solution
with label F and the solution with label SG, and in particular we are not able 
to give the physical frontier F/SG. 
However, within some limitations which we now prescribe, 
we are able to give the exact critical surface,
\textit{i.e.}, the boundary with the P phase,
establishing which one - in the thermodynamic limit - of the two
critical boundaries, P-F or P-SG, is stable (we will use here the more common expression ``stable''
instead of the expression ``leading''), and to localize
some regions of the phase diagram for which we can
say exactly whether the stable solution is P, F, or SG. 
We will prove the stability of these solutions in Sec. IX.
When for a region we are not able to discriminate between
the solution with label F and the solution with label SG
and they are both out of their own P region, we will indicate
such a region with the symbol ``SG and/or F'' (stressing in this way that in this region there may be also
mixed phases and re-entrance phenomena).

\begin{figure}[p]
\epsfxsize=75mm \centerline{\epsffile{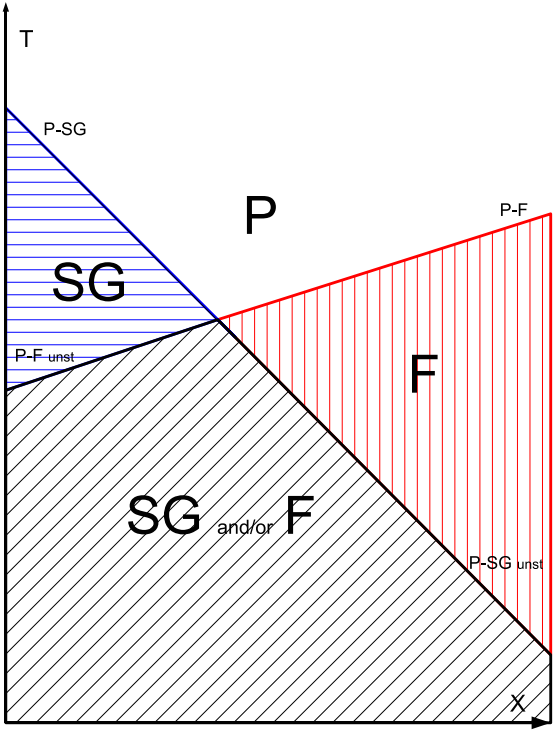}}
\caption{(Color online) Phase diagram for the \textbf{case (1)}: $J_0\geq 0$ and $d_0<2$ or $d_0=\infty$ in a broad sense.
Here T is the temperature while $\bm{X}$ represents symbolically the connectivity $c$ and the parameters of the probability distribution $d\mu$.} 
\label{phase_d1}
\end{figure}

\begin{figure}[p]
\epsfxsize=75mm \centerline{\epsffile{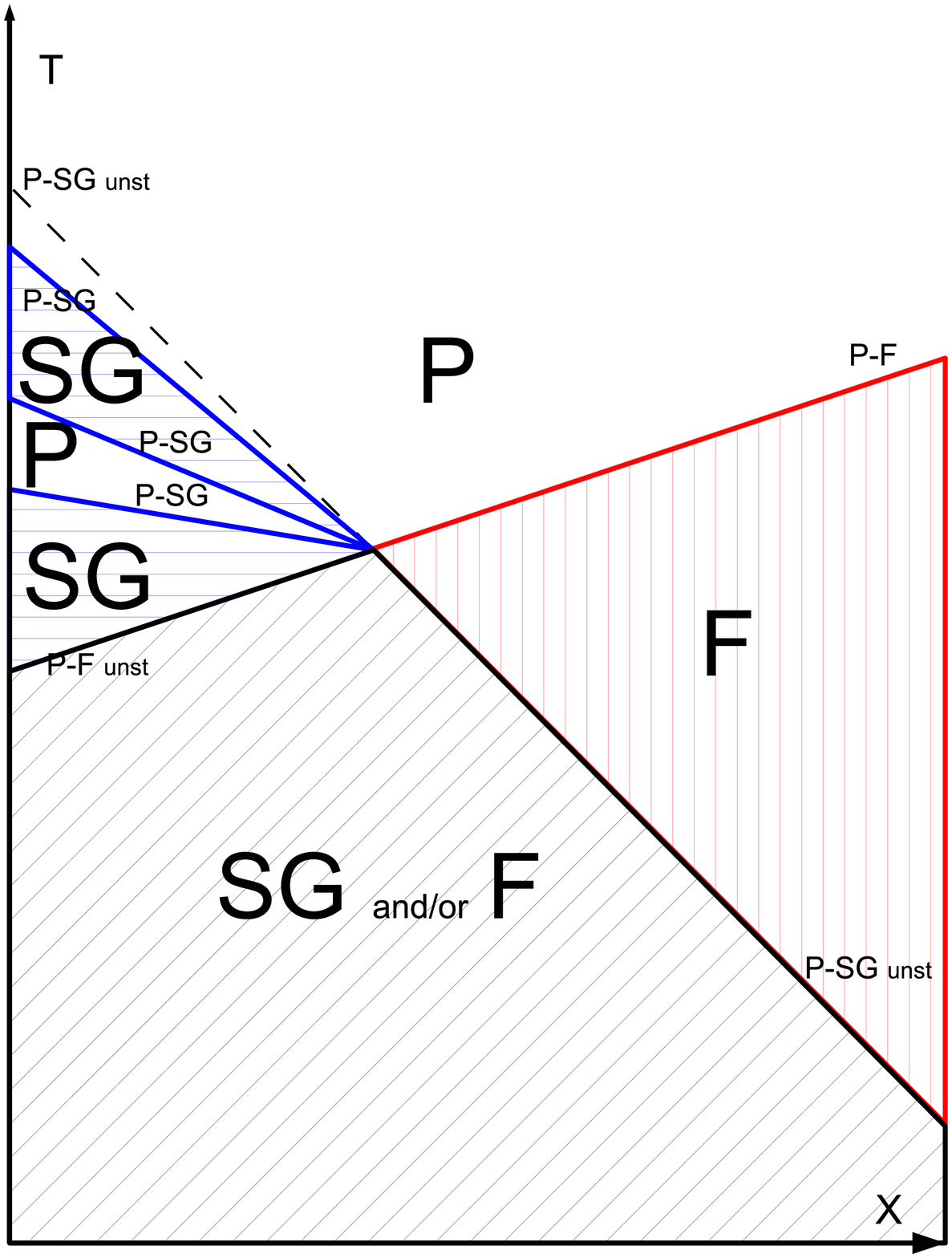}}
\caption{(Color online) Phase diagram for the \textbf{case (2)}: $J_0\geq 0$ and $2\leq d_0<\infty$. T and $\bm{X}$ as in Fig. \ref{phase_d1}} 
\label{phase_d2}
\end{figure}

\begin{figure}[p]
\epsfxsize=75mm \centerline{\epsffile{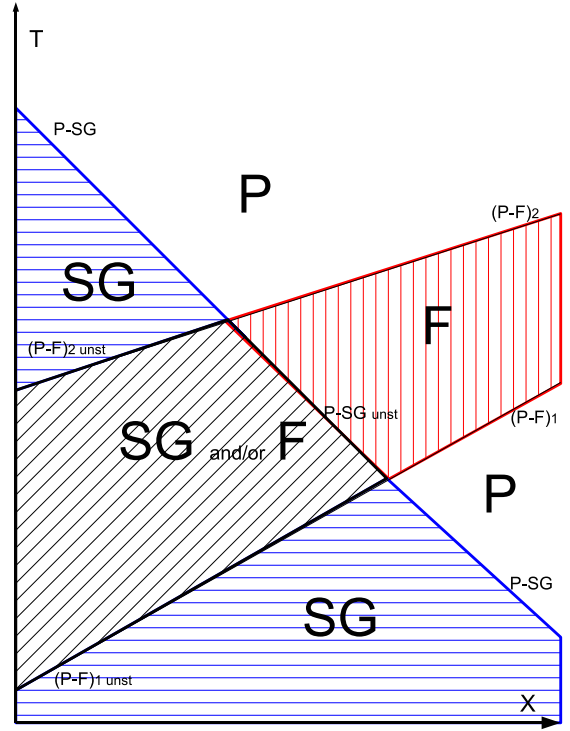}}
\caption{(Color online) Phase diagram for the \textbf{case (3)}: $J_0< 0$ and $d_0<2$ or $d_0=\infty$ in a broad sense. T and $\bm{X}$ as in Fig. \ref{phase_d1}} 
\label{phase_d3}
\end{figure}

\begin{figure}[p]
\epsfxsize=75mm \centerline{\epsffile{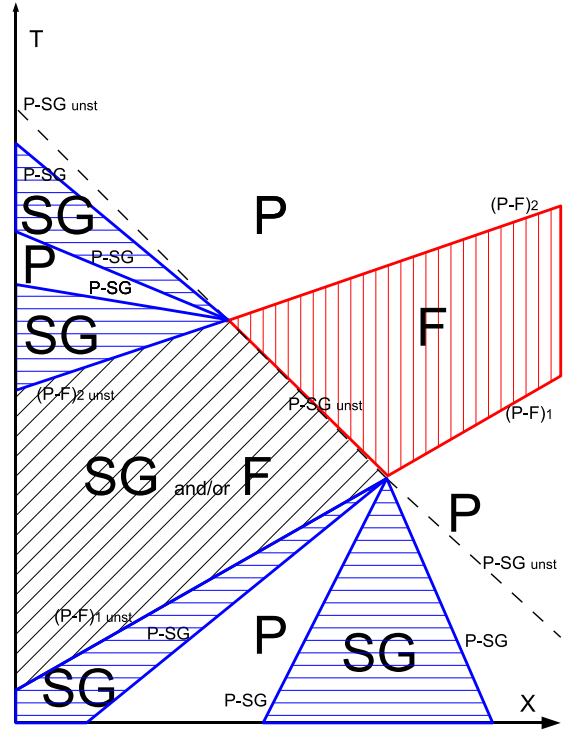}}
\caption{(Color online) Phase diagram for the \textbf{case (4)}: $J_0< 0$ and $2\leq d_0<\infty$. T and $\bm{X}$ as in Fig. \ref{phase_d1}} 
\label{phase_d4}
\end{figure}

In Sec. IX we prove that there are four possible kind of
phase diagrams that may occur according to the \textbf{cases}
\textbf{(1)} $(J_0\geq 0;~d_0<2,~\mathrm{or}~d_0=\infty)$, \textbf{(2)} $(J_0\geq 0;~2\leq d_0<\infty)$,
\textbf{(3)} $(J_0< 0;~d_0<2,~\mathrm{or}~d_0=\infty)$, and \textbf{(4)} $(J_0< 0;~2\leq d_0<\infty)$.  
The four kind of possible phase diagrams are schematically depicted 
in Figs. \ref{phase_d1}-\ref{phase_d4} in the plane $(T,~\bm{X})$.

\subsubsection{$J_0\geq 0$}
As we have seen in Sec. IIIB1, if $J_0\geq 0$, for both the solution with label F
and SG, we have one - and only one - critical temperature.
In the following, to avoid confusion, it should be kept in mind the distinction between the
physical $\beta_c=\beta_c(\bm{X})$ and $\beta_c^{(\Sigma)}$, with $\Sigma=$F or SG.
$\beta_c$ satisfies the following rules.

\textbf{Case (1)}: If $d_0<2$ and $J_0$ is a finite range coupling, or else $d_0=\infty$ at least
in a broad sense (see \cite{MOIII}), $\beta_c(\bm{X})$ is a single-value function of $\bm{X}$, and we have
\begin{eqnarray}
\label{THEOrule}
\beta_c=\mathrm{min}\{\beta_c^{(\mathrm{F})},\beta_c^{(\mathrm{SG})}\}.
\end{eqnarray} 
A schematic representation of this case is given in Fig.~1.

\textbf{Case (2)}: If instead $2\leq d_0<\infty$ we have 
\begin{eqnarray}
\label{THEOrule2}
\left\{
\begin{array}{l}
\beta_c=\beta_c^{(\mathrm{F})}, \quad \quad \quad \quad ~~ \mathrm{if} 
\quad \beta_c^{(\mathrm{SG})}\geq\beta_c^{(\mathrm{F})}, \\
\beta_c^{(\mathrm{F})}\geq \beta_c>\beta_c^{(\mathrm{SG})}, \quad \mathrm{if} 
\quad \beta_c^{(\mathrm{SG})}<\beta_c^{(\mathrm{F})}.
\end{array}
\right.
\end{eqnarray} 
Notice in particular that the second line of Eq. (\ref{THEOrule2}) 
does not exclude that $\beta_c(\bm{X})$ might be a non single-value function of $\bm{X}$. 
A schematic representation of this case is given in Fig.~2.

\subsubsection{$J_0< 0$}
As we have seen in Sec. IIIB2, if $J_0< 0$, for a sufficiently large connectivity $c$, 
the solution with label F has at least two separated P regions corresponding
to two critical temperatures. 
Here we assume that the underlying lattice $\mathcal{L}_0$
has only loops of even length so that, for example, triangular lattices
are here excluded. 
Let us suppose to have for the solution with label F only two critical temperatures 
(the minimum number, if $J_0<0$), and let be
\begin{eqnarray}
\label{THEOrule4}
\beta_{c1}^\mathrm{(F)}\geq \beta_{c2}^\mathrm{(F)}.
\end{eqnarray} 

In general we have the following scenario.

\textbf{Case (3)}: If $d_0<2$ and $J_0$ is a finite range coupling, or $d_0=\infty$ in a broad
sense (see \cite{MOIII}), $\beta_{c2}(\bm{X})$ is a single-value function of $\bm{X}$
and satisfies Eq. (\ref{THEOrule}). 
The other critical inverse temperature $\beta_{c1}(\bm{X})$ is instead: 
either a two-value function of $\bm{X}$ and we have 
\begin{eqnarray}
\label{THEOrule6}
\beta_{c1}=\left(
\begin{array}{l}
\beta_{c1}^\mathrm{(F)} \\
\beta_{c}^\mathrm{(SG)}
\end{array}
\right), \quad \mathrm{if} \quad \beta_{c1}^\mathrm{(F)}\leq \beta_{c}^\mathrm{(SG)},
\end{eqnarray} 
or  
\begin{eqnarray}
\label{THEOrule7}
\nexists\quad \beta_{c1}, \quad \mathrm{if} \quad \beta_{c1}^\mathrm{(F)}> \beta_{c}^\mathrm{(SG)},
\end{eqnarray} 
where $\nexists$ in Eq. (\ref{THEOrule7}) means that if $\beta_{c1}^\mathrm{(F)}> \beta_{c}^\mathrm{(SG)}$
there is no stable boundary with the P region.     
A schematic representation of this case is given in Fig.~3.

\textbf{Case (4)}: If $2\leq d_0<\infty$, $\beta_{c2}$ satisfies Eq. (\ref{THEOrule2});
whereas for $\beta_{c1}$ we have either
\begin{eqnarray}
\label{THEOrule8}
\beta_{c1}=\left(
\begin{array}{l}
\beta_{c1}^\mathrm{(F)} \\
\beta_{c}^{\mathrm{(SG>)}}
\end{array}
\right), \quad \mathrm{if} \quad \beta_{c1}^\mathrm{(F)}\leq \beta_{c}^\mathrm{(SG)},
\end{eqnarray} 
or
\begin{eqnarray}
\label{THEOrule9}
\mathrm{if} \quad \exists \beta_{c1}\Rightarrow 
\beta_{c1}>\beta_{c}^{\mathrm{(SG)}}, \quad \mathrm{if} \quad \beta_{c1}^\mathrm{(F)}> \beta_{c}^\mathrm{(SG)},
\end{eqnarray} 
where in Eq. (\ref{THEOrule8}) we have introduced the symbol SG$>$ to indicate
that in general the stable P-SG surface is above (or below in terms of temperatures) the surface coming from the solution
with label SG:~$\beta_{c}^{\mathrm{(SG>)}}>\beta_{c}^\mathrm{(SG)}$.
Notice that, similarly to the case \textbf{(3)}, we cannot exclude that $\beta_{c1}$ in Eq. (\ref{THEOrule9})
be a non single-value function of $\bm{X}$, as well as $\beta_{c}^{\mathrm{(SG>)}}$ in Eq. (\ref{THEOrule8}).  
A schematic representation of this case is given in Fig.~4.

If more than two critical temperatures are present, the above scheme
generalizes straightforwardly.

Keeping our definition for the introduced symbol ``SG and/or F'',
we stress that: in all the fours cases the phases F and ``SG and/or F''
are exactly localized; in the cases \textbf{(1)} and \textbf{(3)} the phases P and SG are
exactly localized; in the cases \textbf{(2)} and \textbf{(4)} the SG phase is always 
limited below (in terms of temperatures) 
by the unstable P-SG surface coming from the solution with label SG 
(indicated as P-SG unst in Figs. 2 and 4). 
Finally, we stress that - under the hypothesis that $\mathcal{L}_0$ has only 
loops of even length - the stable P regions correspond always to the solution
with label F.

For $2\leq d_0<\infty$, from the second line of Eqs. (\ref{THEOrule2}) and (\ref{THEOrule8})
and from Eq. (\ref{THEOrule9}),
we see that the method is not able to give the complete information 
about the P-SG boundary since we have only inequalities,
not equalities. Furthermore, in these regions of the phase diagram
the physical critical temperature in general may be a non single-value function of $\bm{X}$.
On the other hand, we have the important information 
that in these equations the inequalities between the physical $\beta_c$ and $\beta_c^{(\mathrm{SG})}$ 
(the solution with label SG) are always strict.
As a consequence, we see that, 
when $2\leq d_0<\infty$, in these regions
the SG ``magnetization'' $m^{(\mathrm{SG})}$
will always have a finite jump discontinuity in crossing the surface given by
$\beta_c$. In other words, along such a branch of the
critical surface corresponding to 
the second line of Eqs. (\ref{THEOrule2}) and (\ref{THEOrule8}) and Eq. (\ref{THEOrule9}),
we have a first-order phase transition, independently of the fact that
the phase transition corresponding  to the $\beta_c^{(\mathrm{SG})}$ surface
is second-order, and independently of the sign of $J_0$. 


\subsection{Generalizations}
The generalization to the cases in which the unperturbed model
has an Hamiltonian $H_0$ involving couplings depending
on the bond $b\in\Gamma_0$ is straightforward. In this case 
we have just to substitute everywhere in the formulae 
(\ref{THEOa})-(\ref{THEOL6}), $J_0^{(\Sigma)}$ 
with the set $\{J_{0b}^{(\Sigma)}\}$. However, 
the critical behavior will be in general different and more complicated 
than that depicted in the Subsections IIIB1 and IIIB2.  
In particular, even in the case in which all the couplings
$J_{0b}^{(\Sigma)}$ are positive,
we cannot assume that the Landau coefficient $b^{(\Sigma)}$
be positive so that, even in such a case, first-order phase
transitions are in principle possible, as has been seen via Monte Carlo simulations 
in directed small-world models \cite{Sanchez}.

As anticipated, our method can be generalized also to study possible
antiferromagnetic phase transitions in the random model.
There can be two kind of sources of antiferromagnetism:
one due to a negative coupling $J_0$ in the unperturbed model,
the other due to random shortcuts $J_{i,j}$ having a measure $d\mu$ 
with a negative average. 

In the first case, if for example the sublattice $\mathcal{L}_0$ is bipartite 
into two sublattices $\mathcal{L}_0^{(a)}$ and $\mathcal{L}_0^{(b)}$,
the unperturbed model will have an antiferromagnetism described
by two fields $m_0^{(a)}$ and $m_0^{(b)}$. Correspondingly,
in the random model we will have to analyze two effective fields
$m^{(a)}$ and $m^{(b)}$ which will satisfy a set of two
coupled self-consistent equations similar to Eqs. (\ref{THEOa}) and
involving the knowledge of
$m_0^{(a)}$ and $m_0^{(b)}$. More in general, we can introduce the site-dependent 
solution $m_{0i}$ to find correspondingly in a set 
of coupled equations (at most $N$), the effective fields $m_i$
of the random model.

In the second case, following \cite{Almeida} we consider a lattice 
$\mathcal{L}_0$ which is composed of, say, $p$ sublattices
$\mathcal{L}_0^{(\nu)}$, $\nu=1,\ldots,p$. Then, we build up
the random model with the rule that any shortcut may connect
only sites belonging to two different sublattices. 
Hence, as already done in \cite{MOI} for the generalized SK model, we 
introduce $p$ effective fields $m^{(\nu)}$ which
satisfy a system of $p$ self-consistent equations involving
the $p$ fields 
$m_0^{(\nu)}$ and calculated in the $p$ external fields $J^{\mathrm{(F)}} m^{(\nu)}$
(note that here the symbol F stresses only the fact that the 
effective coupling must be calculated through Eq. (\ref{THEOb})).

\section{Small World in $d_0=0$ Dimension}
\subsection{The Viana-Bray model}
As an immediate example, let us consider the Viana-Bray model \cite{VianaBray}.
It can be seen as the simplest small-world model in which
$N$ spins with no short-range couplings (here $J_0=0$) 
are randomly connected by long-range connections $J$ (possibly also random). 
Note that formally here $\mathcal{L}_0$ has dimension $d_0=0$.
Since $J_0=0$, for the unperturbed model we have
\begin{eqnarray}
\label{VBa}\nonumber
-\beta f_0 (0,\beta h)=
\log\left[2\cosh(\beta h)\right],
\end{eqnarray} 
\begin{eqnarray}
\label{VBb}
m_0(0,\beta h)=\tanh(\beta h),
\end{eqnarray} 
\begin{eqnarray}
\label{VBc}\nonumber
\tilde{\chi_0} (\beta J_0,\beta h)&=& 1-\tanh^2(\beta h)|_{\beta h=0}=1,
\end{eqnarray} 
It is interesting to check that the first and second derivatives of
$\tilde{\chi}_0$ in $h=0$ are null and negative, respectively. In fact we have
\begin{eqnarray}
\label{VBd}\nonumber
&&\frac{\partial}{\beta h}\tilde{\chi_0} (0,\beta h)=
-2\tanh^2(\beta h)
\nonumber \\
&& \times \left[1-\tanh^2(\beta h)\right]|_{\beta h=0}=0,
\end{eqnarray} 
and
\begin{eqnarray}
\label{VBe}\nonumber
&& \frac{\partial^2}{(\beta h)^2}\tilde{\chi_0} (0,\beta h)=
-2\left[1-\tanh^2(\beta h)\right]^2 +\nonumber \\
&& 4\tanh^2(\beta h)\left[1-\tanh^2(\beta h)\right]
|_{\beta h=0}=-2.
\end{eqnarray} 

Applying these results to Eqs. (\ref{THEOa}-\ref{THEOe}) we get immediately
the self-consistent equations for the F and the SG magnetizations
\begin{eqnarray}
\label{VBf}
m^{(\mathrm{F})}&=&\tanh\left[m^{(\mathrm{F})}c\int d\mu\tanh(\beta J)\right],\\
\label{VBg}
m^{(\mathrm{SG})}&=&\tanh\left[m^{(\mathrm{SG})}c\int d\mu\tanh^2(\beta J)\right], 
\end{eqnarray} 
and the Viana-Bray critical surface
\begin{eqnarray}
\label{VBh}
c\int d\mu\tanh(\beta_c^{(\mathrm{F})} J)=1,\\
\label{VBi}
c\int d\mu\tanh^2(\beta_c^{(\mathrm{SG})} J)=1.
\end{eqnarray} 

On choosing for $d\mu$ a measure having average 
and variance scaling as $\mathop{O}(1/c)$,
for $c\propto N$, we recover
the equations for the Sherrington-Kirkpatrick model \cite{SK} 
already derived in this form in \cite{MOI} and \cite{MOII}. 
In these papers Eqs. (\ref{VBf}-\ref{VBi}) were derived by mapping the
Viana-Bray model and, similarly, the Sherrington-Kirkpatrick model to
the non random fully connected Ising model. In this sense it should be also clear
that, at least for $\beta\leq \beta_c$ and zero external field, 
in the thermodynamic limit, 
the connected correlation functions (of order $k$ greater than 1) 
in the Sherrington-Kirkpatrick 
and in the Viana-Bray model are exactly zero. In fact, in the thermodynamic
limit, the non random fully connected model can be exactly reduced to a model 
of non interacting spins immersed in
an effective medium so that among any two spins there is no correlation.
Such a result is due to the fact that, in these models, 
all the $N$ spins interact 
through the same coupling $J/N$, no matter how far apart they are, and the net effect
of this is that in the thermodynamic limit the system becomes equivalent to
a collection of $N$ non interacting spins seeing only an effective external
field (the medium) like in Eq. (\ref{VBb}) with $\beta h$ replaced by $\beta J m$.

For the measure (\ref{dPF}) our approximated Eq. (\ref{VBf}) can be compared with the exact known equation
that can be derived by using the Bethe-Peierls or the replica approach and is given by
(see for example \cite{Review} and references therein)
\begin{equation}\label{BP1}\nonumber
m^{(\mathrm{F})}=\sum_{q=0}^{\infty}\frac{e^{-c}c^{q}}{q!}\int \tanh \Bigl(\beta
\sum_{m=1}^{q}H_{m}\Bigr)\prod_{m=1}^{q}\Psi (H_{m})dH_{m},  
\end{equation}
where the effective field $H$ is determined by the following integral equation
\begin{eqnarray}\label{BP2}
\Psi (H)
&=&\sum_{q=0}^{\infty}\frac{e^{-c}c^{q-1}}{(q-1)!}\int \delta \Bigl(H-T\tanh ^{-1}\Bigl[\tanh
\beta J\times  \nonumber \\
&&\tanh \Bigl(\beta \sum_{m=1}^{q-1}H_{m}\Bigr)\Bigr]\Bigr)\prod_{m=1}^{q-1}\Psi
(H_{m})dH_{m}. \nonumber
\end{eqnarray}

In the limit $\beta \to \infty$, Eqs. (\ref{VBf}) and (\ref{VBg})
give the following size (normalized to 1) of the giant connected component
\begin{eqnarray}
\label{VBff}
m^{(\mathrm{F})}&=&\tanh(m^{(\mathrm{F})}c),\\
\label{VBgg}
m^{(\mathrm{SG})}&=&\tanh(m^{(\mathrm{SG})}c). 
\end{eqnarray} 

These equations are not exact, however they succeed in giving the 
exact percolation threshold $c=1$. In fact, concerning the equation (\ref{VBff})
for the F phase, the exact equation for $m^{(\mathrm{F})}$ is (see for example 
\cite{Review} and references therein)
\begin{eqnarray}
\label{VBff1}
1-m^{(\mathrm{F})}&=&e^{-cm^{(\mathrm{F})}},
\end{eqnarray} 
which, in terms of the function $\tanh$, becomes
\begin{eqnarray}
\label{VBff2}\nonumber
\frac{2m^{(\mathrm{F})}+(m^{(\mathrm{F})})^2}{2-m^{(\mathrm{F})}+(m^{(\mathrm{F})})^2}
&=&\tanh(m^{(\mathrm{F})}c),
\end{eqnarray} 
so that Eqs. (\ref{VBff}) and (\ref{VBff1}) are equivalent at the order 
$\mathop{O}(m^{(\mathrm{F})})$. We see also that, as stated in the Sec. IIIC,
Eqs. (\ref{VBff}) and (\ref{VBff1}) become equal in the limits $c\to0$ and $c\to\infty$.

\subsection{Gas of Dimers}
Let us consider for $\mathcal{L}_0$ a set of $2N$ spins coupled through a coupling $J_0$
two by two. The expression ``gas of dimers'' stresses the fact that the dimers,
\textit{i.e.} the couples of spins, do not interact each other.
As a consequence, the free energy,
the magnetization, and the susceptibility of the unperturbed model
can be immediately calculated. We have
\begin{eqnarray}
\label{gd}\nonumber
-\beta f_0 (\beta J_0,\beta h)=\frac{1}{2}
\log\left[2e^{\beta J_0}\cosh(2\beta h)+2e^{-\beta J_0}\right],
\end{eqnarray} 
\begin{eqnarray}
\label{gd1}\nonumber
m_0 (\beta J_0,\beta h)=\frac{e^{\beta J_0}\sinh(2\beta h)}
{e^{\beta J_0}\cosh(2\beta h)+e^{-\beta J_0}},
\end{eqnarray} 
\begin{eqnarray}
\label{gd1b}\nonumber
\tilde{\chi_0} (\beta J_0,\beta h)&=&\frac{2e^{\beta J_0}+2\cosh(2\beta h)}
{\left[e^{\beta J_0}\cosh(2\beta h)+e^{-\beta J_0}\right]^2}|_{\beta h=0}\nonumber \\
&=& \frac{e^{\beta J_0}}{\cosh(\beta J_0)},
\end{eqnarray} 
Let us calculate also the second derivative of $\tilde{\chi}_0$. From
\begin{eqnarray}
\label{gd2}\nonumber
&&\frac{\partial}{\beta h}\tilde{\chi_0} (\beta J_0,\beta h)=
4\sinh(\beta h) \nonumber \\
&\times& \frac{e^{-\beta J_0}-2e^{3\beta J_0}
-e^{\beta J_0}\cosh(2\beta h)}
{\left[e^{\beta J_0}\cosh(2\beta h)+e^{-\beta J_0}\right]^3},
\end{eqnarray} 
we get
\begin{eqnarray}
\label{gd3}\nonumber
\frac{\partial^2}{(\beta h)^2}\tilde{\chi_0} (\beta J_0,\beta h)|_{\beta h=0}=
-2\frac{\sinh(\beta J_0)+e^{3\beta J_0}}{\left[\cosh(\beta J_0)\right]^3}.
\end{eqnarray} 
We note that, as expected, the second derivative of $\tilde{\chi}_0$ in $h=0$,
for $J_0\geq 0$ is always negative, whereas, for $J_0<0$ it becomes positive as soon as 
$\beta |J_0|>\log(\sqrt{2})$. 

By using the above equations, 
from Sec. III we get immediately the following self-consistent equation for the
magnetizations 
\begin{eqnarray}
\label{gd4}\nonumber
m^{(\Sigma)}=\frac{\tanh(2\beta J^{(\Sigma)}m^{(\Sigma)}+ 2\beta h)}
{1+e^{-2\beta J_0}\sech(2\beta J^{(\Sigma)}m^{(\Sigma)}+ 2\beta h)},
\end{eqnarray} 
and - at least for $J_0\geq 0$ - the equation for the critical temperature 
\begin{eqnarray}
\label{gd5}\nonumber
\frac{e^{\beta_c^{(\Sigma)}J_0^{(\Sigma)}}}
{\cosh(\beta_c^{(\Sigma)} J_0^{(\Sigma)})}
\beta_c^{(\Sigma)} J^{(\Sigma)}=1.
\end{eqnarray} 

As it will be clear soon, this model lies between the Viana-Bray 
model and the more complex $d_0=1$ dimensional chain small-world model, which will
be analyzed in detail in the next section.
Our major interest in this simpler gas of dimers small-world model 
is related to the fact that, 
in spite of its simplicity and $d_0$=0 dimensionality - since 
the second derivative of $\tilde{\chi}_0$ may be positive when $J_0$ is
negative -
according to the general result of Sec. IIIB, 
it is able to give rise to 
also multiple first- and second-order phase transitions. 

\section{Small world in $d_0$=1 dimension}
In this section we will analyze the case in which $\mathcal{L}_0$ is the 
$d_0$=1-dimensional chain with periodic boundary conditions (p.b.c.). 
The corresponding small-world model with Hamiltonian (\ref{H}) in zero field
has already been analyzed in \cite{Niko} by using the replica method. Here we will
recover the results found in \cite{Niko} for $\beta_c$ and will provide the
self-consistent equations for the magnetizations $m^{(\mathrm{F})}$ and
$m^{(\mathrm{SG})}$  whose solution, as expected,
turns out to be in good agreement with the corresponding solutions found in
\cite{Niko} for $c$ small and large (the latter when the frustration is relatively small). 
It will be however rather evident how much the two methods 
differ in terms of simplicity and intuitive meaning. 
Furthermore, we will derive also an explicit expression for the two-points
connected correlation function which, to the best of our knowledge, had not been
published yet. Finally, we will analyze in the detail the completely novel scenario
for the case $J_0<0$ which, as mentioned, produces
multiple first- and second-order phase transitions.

In order to apply the method of Sec.III we have to solve the one dimensional
Ising model with p.b.c. immersed in an external field. 
The solution of this non random model is easy and 
well known (see for example \cite{Baxter}).  
If we indicate with $\lambda_1$ and $\lambda_2$ the two
eigenvalues coming from the transfer matrix method, one has
\begin{eqnarray}
\label{1dd}\nonumber
\lambda_{1,2}=
{e^{\beta J_0}\cosh(\beta h)\pm
\left[e^{2\beta J_0}\sinh^2(\beta h)+e^{-2\beta J_0}\right]^{\frac{1}{2}}},
\end{eqnarray} 
from which it follows that, for the free energy density, the magnetization 
and the two-points connected correlation function, we have
\begin{eqnarray}
\label{1d}\nonumber
-\beta f_0 (\beta J_0,\beta h)=\log\left( \lambda_1\right),
\end{eqnarray} 
\begin{eqnarray}
\label{1da}
m_0 (\beta J_0,\beta h)=\frac{e^{\beta J_0}\sinh(\beta h)}
{\left[e^{2\beta J_0}\sinh^2(\beta h)+e^{-2\beta J_0}\right]^{\frac{1}{2}}},
\end{eqnarray} 
\begin{eqnarray}
\label{1db}
{{C}}_0 (\beta J_0,\beta h;{||i-j||}_0)&\defi&{\media{\sigma_i\sigma_j}}_0-{\media{\sigma}}_0^2
\nonumber \\
 &=& \sin^2(2\varphi) \left(\frac{\lambda_2}{\lambda_1}\right)^{{||i-j||}_0},
\end{eqnarray} 
where the phase $\varphi$ is defined by  
\begin{eqnarray}
\label{1dc}
\cot(2\varphi)=e^{2\beta J_0}\sinh(\beta h), \quad 0<\varphi<\frac{\pi}{2},
\end{eqnarray} 
and ${||i-j||}_0$ is the (euclidean) distance between $i$ and $j$.

Let us calculate $\tilde{\chi}_0$ and its first and second derivatives.
From Eq. (\ref{1da}) we have
\begin{eqnarray}
\label{1de}
\tilde{\chi}_0 (\beta J_0,\beta h)=\frac{e^{-\beta J_0}\cosh(\beta h)}
{\left[e^{2\beta J_0}\sinh^2(\beta h)+e^{-2\beta J_0}\right]^{\frac{3}{2}}},
\end{eqnarray} 

\begin{eqnarray}
\label{1df}\nonumber
&&\frac{\partial}{\partial \beta h}\tilde{\chi}_0 (\beta J_0,\beta h)=\sinh(\beta h)
\nonumber \\ &\times&
\frac{\left[e^{-3\beta J_0}-2e^{\beta J_0}\cosh^2(\beta h)-e^{\beta J_0}\right]}
{\left[e^{2\beta J_0}\sinh^2(\beta h)+e^{-2\beta J_0}\right]^{\frac{5}{2}}},
\end{eqnarray}
 
\begin{eqnarray}
\label{1dg}
&& \frac{\partial^2}{\partial (\beta h)^2}\tilde{\chi}_0 (\beta J_0,\beta h)=\cosh(\beta h)
\nonumber \\ &\times&
\frac{\left[e^{-3\beta J_0}-2e^{\beta J_0}\cosh^2(\beta h)-e^{\beta J_0}\right]}
{\left[e^{2\beta J_0}\sinh^2(\beta h)+e^{-2\beta J_0}\right]^{\frac{5}{2}}}
+\mathop{O}(\beta h)^2.
\end{eqnarray} 

From Eq. (\ref{1dg}) we see that for $J_0>0$ and any $\beta$ we have,
for sufficiently small $h$,
\begin{eqnarray}
\label{1dh}
&& \frac{\partial^2}{\partial (\beta h)^2}\tilde{\chi}_0 (\beta J_0,\beta h)<\cosh(\beta h)
\nonumber \\ &\times&
\frac{\left[1-3e^{\beta J_0}\right]}
{\left[e^{2\beta J_0}\sinh^2(\beta h)+e^{-2\beta J_0}\right]^{\frac{5}{2}}}
+\mathop{O}(\beta h)^2<0, 
\end{eqnarray} 
whereas for $J_0<0$ we have
\begin{eqnarray}
\label{1di}
&& \frac{\partial^2}{\partial (\beta h)^2}\tilde{\chi}_0 (\beta J_0,\beta h) \geq 0
\quad \mathrm{for} \quad e^{-4\beta J_0}>3.
\end{eqnarray} 

We see therefore that, according to Sec. IIIB,
when $J_0<0$ for $\beta |J_0|\geq \log(3)/4=0.1193...$
the Landau coefficient $b^{(\mathrm{F})}$ is negative and
we may have a first-order phase transition.

From Eqs. (\ref{THEOa}) and (\ref{1da}), for the magnetizations $m^{(\mathrm{F})}$ and
$m^{(\mathrm{SG})}$ at zero external field we have
\begin{eqnarray}
\label{1dl}
m^{(\Sigma)}=\frac{e^{\beta J_0^{(\Sigma)}}\sinh(\beta J^{(\Sigma)}m^{(\Sigma)})}
{\left[e^{2\beta J_0^{(\Sigma)}}\sinh^2(\beta J^{(\Sigma)} 
m^{(\Sigma)})+e^{-2\beta J_0^{(\Sigma)}}\right]^{\frac{1}{2}}}.
\end{eqnarray} 

From Eqs. (\ref{THEOg}) and (\ref{1de}) we see that a solution $m^{(\Sigma)}$
becomes unstable at the inverse temperature $\beta_c^{(\Sigma)}$ given by
\begin{eqnarray}
\label{1dm}
e^{2\beta_c^{(\Sigma)} J_0^{(\Sigma)}}\beta_c^{(\Sigma)}J^{(\Sigma)}=1.
\end{eqnarray} 
For $J_0\geq 0$ the above equation gives 
the exact P-F and P-SG critical temperatures in agreement with \cite{Niko}.
When $J_0<0$ - unless the transition be of second-order - Eq. (\ref{1dm}) for $\Sigma=$F
does not signal a phase transition. In general, as $J_0<0$ the P-F critical temperature
must be determined by looking at all the stable solutions $m^{(\mathrm{F})}$ of the self-consistent 
equation (\ref{1dl}) and by choosing the one
minimizing the effective free energy $L^{\mathrm{(F)}}(m)$ of Eq. (\ref{THEOll}).

Finally, for the two-point connected correlation function,
from Eqs. (\ref{THEOh}), (\ref{1db}) and (\ref{1dc}), we have
\begin{eqnarray}
\label{1dp}\nonumber
{{C}}^{(\Sigma)}({||i-j||}_0)&=& 
\sin^2(2\varphi^{(\Sigma)}) e^{-{||i-j||}_0/{\xi^{(\Sigma)}}}, 
\end{eqnarray} 
where
\begin{eqnarray}
\label{1dq}\nonumber
2\varphi^{(\Sigma)}= \cot^{-1} 
\left[e^{2\beta J_0^{(\Sigma)}}
\sinh(\beta J^{(\Sigma)} m^{(\Sigma)})\right], 
\end{eqnarray} 
and the correlation length $\xi^{(\Sigma)}$ is given by performing 
the effective substitutions $\beta J_0\to \beta J_0^{(\Sigma)}$ and
$\beta h\to \beta J^{(\Sigma)} m^{(\Sigma)}$ in $\log(\lambda_1/\lambda_2)$.

Note that ${{C}}_0 (\beta J_0,\beta h)$ is even in $\beta h$, so that $C(-m)=C(m)$.
Near the critical temperature we have
\begin{eqnarray}
\label{1dr}\nonumber
\sin(2\varphi^{(\Sigma)})=1-\frac{\left(e^{2\beta J_0^{(\Sigma)}}m^{(\Sigma)}\right)^2}{2}
+\mathop{O}(m^{(\Sigma)})^4
\end{eqnarray} 
and 
\begin{widetext}
\begin{eqnarray}
\label{1ds}\nonumber
\left(\xi^{(\Sigma)}\right)^{-1}=\left|\log[\tanh(\beta J_0^{(\Sigma)})]-
\frac{(\beta J^{(\Sigma)} m^{(\Sigma)})^2}{4\sinh(\beta J_0^{(\Sigma)})}
\left[\left(e^{\beta J_0^{(\Sigma)}}+e^{3\beta J_0^{(\Sigma)}}\right)
\tanh(\beta J_0^{(\Sigma)})
+e^{3\beta J_0^{(\Sigma)}}-e^{\beta J_0^{(\Sigma)}}
\right] +\mathop{O}(m^{(\Sigma)})^4\right| .
\end{eqnarray} 
\end{widetext}
According to the general result, Eqs. (\ref{THEOCORR1})-(\ref{THEOCORR5}),
we see that the correlation length remains finite at all temperatures.

\begin{figure}
\epsfxsize=65mm \centerline{\epsffile{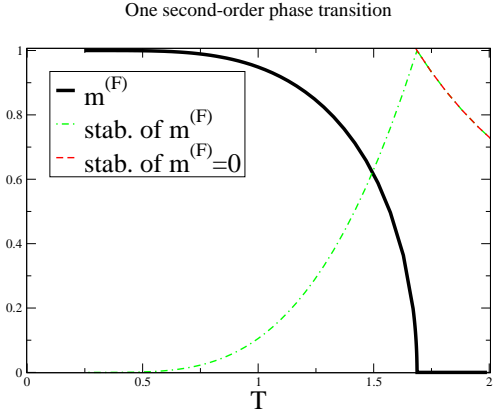}}
\caption{(Color online) Magnetization (thick solid line), and curves of stability 
(dashed and dot-dashed lines) for the case
$c=0.5$, $J_0=1$ and $J=3/5/$. Here $T_{c}=1.687$.} 
\label{sw_1_pos}
\end{figure}
\begin{figure}
\epsfxsize=65mm \centerline{\epsffile{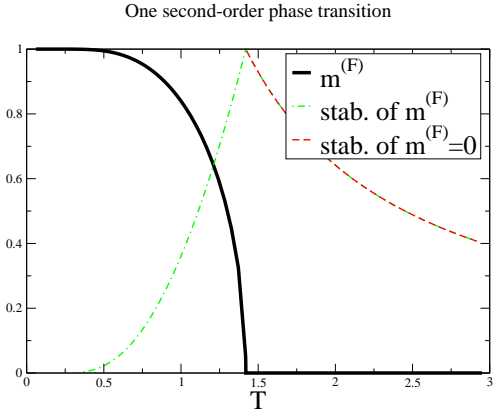}}
\caption{(Color online) Magnetization (thick solid line), and curves of stability 
(dashed and dot-dashed lines) for the case
$c=10$, $J_0=0.25$ and $J=1/c$. Here $T_{c}=1.419$.} 
\label{sw_2_pos}
\end{figure}
\begin{figure}
\epsfxsize=65mm \centerline{\epsffile{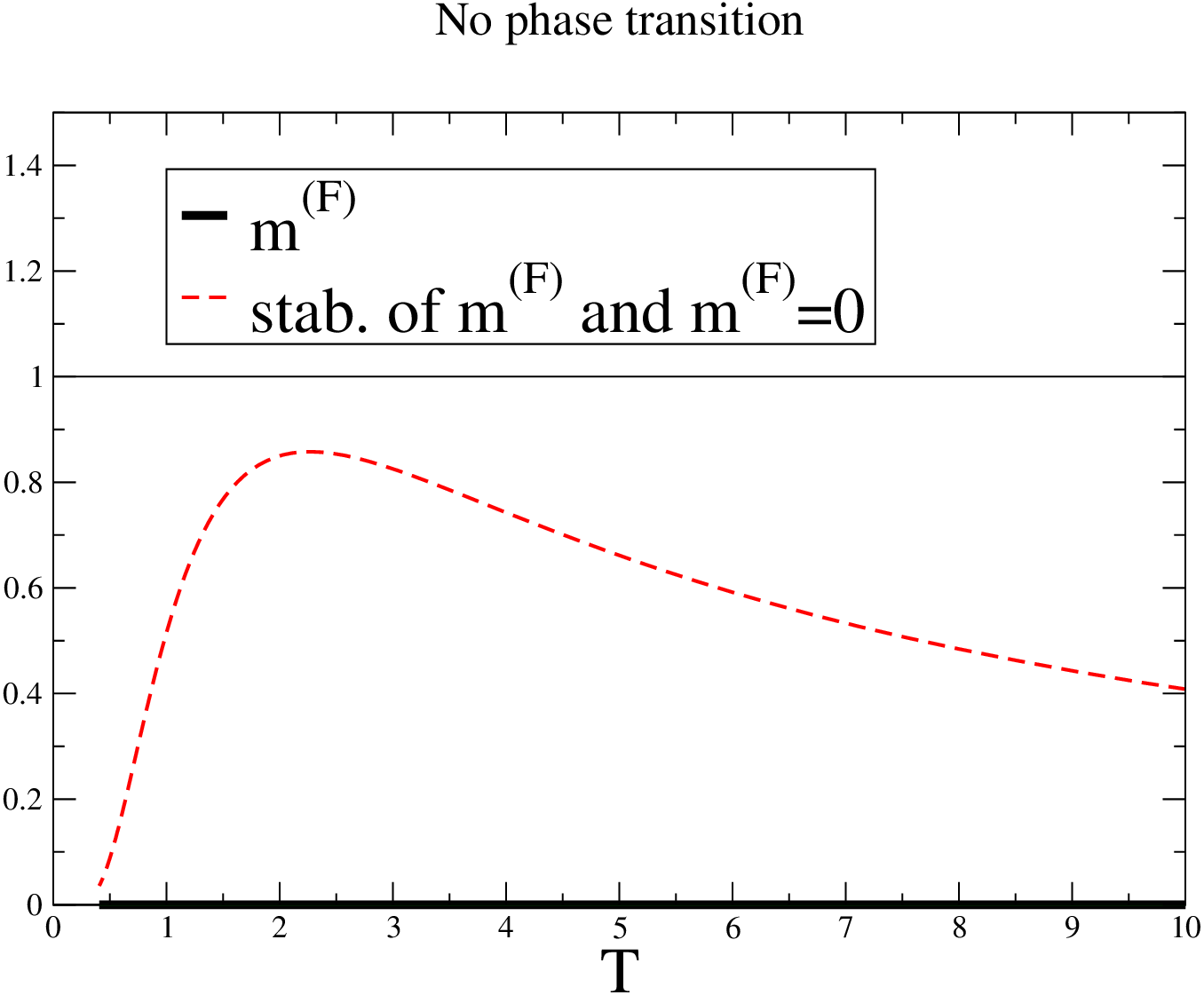}}
\caption{(Color online) Magnetization (thick solid line), and curves of stability 
(dashed and dot-dashed lines) for the case
$c=5$, $J_0=-1$ and $J=1$. 
Note that here $m^{(\mathrm{F})}=0$ and the two curves 
of stability coincide everywhere.} 
\label{sw_1}
\end{figure}
\begin{figure}
\epsfxsize=65mm \centerline{\epsffile{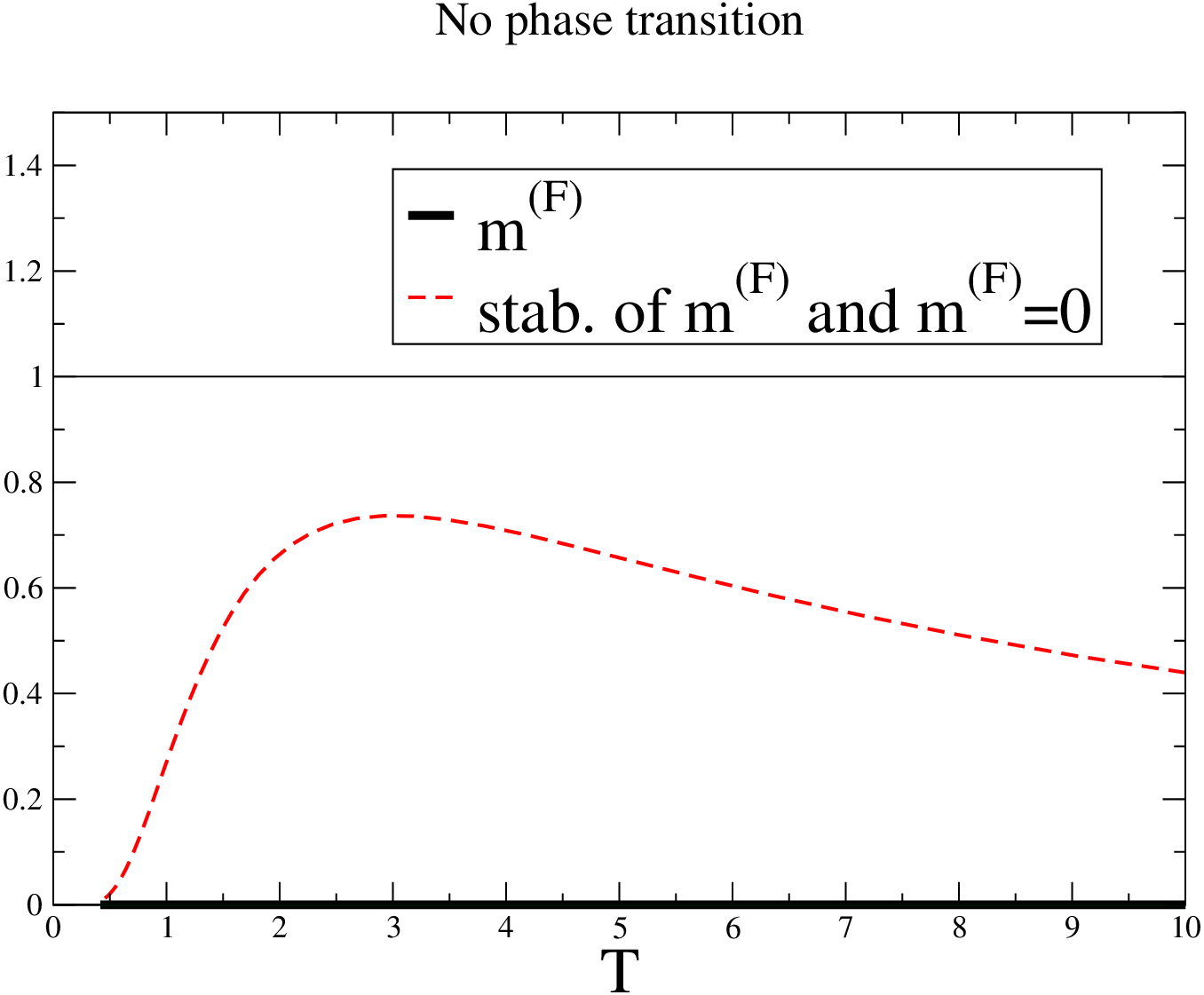}}
\caption{(Color online) Magnetization (thick solid line), and curves of stability 
(dashed and dot-dashed lines) for the case
$c=5.828$, $J_0=-1.4$ and $J=1$. 
Note that here $m^{(\mathrm{F})}=0$ and the two curves 
of stability coincide everywhere.} 
\label{sw_2}
\end{figure}
\begin{figure}[t]
\epsfxsize=65mm \centerline{\epsffile{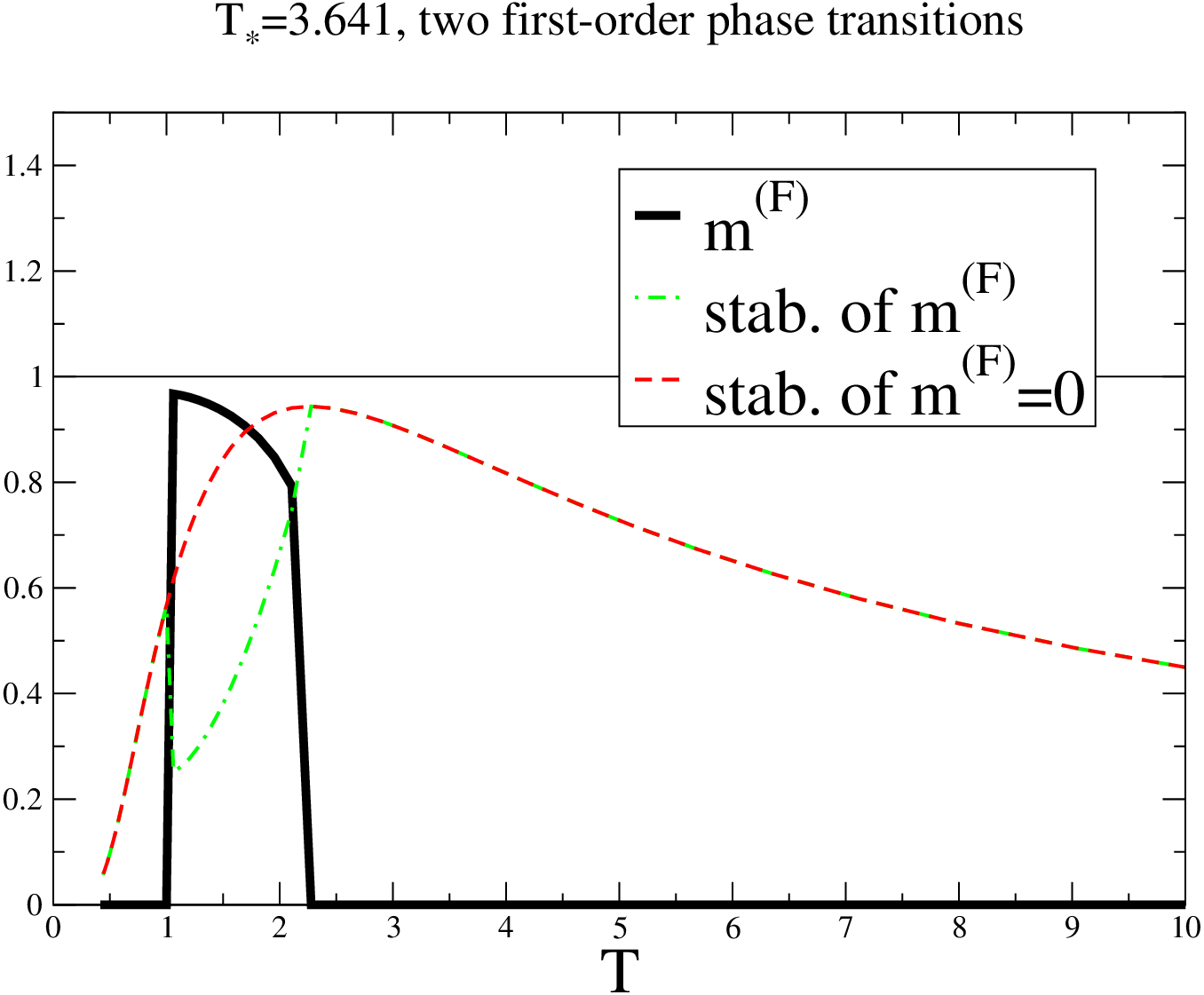}}
\caption{(Color online) Magnetization (thick solid line), and curves of stability 
(dashed and dot-dashed lines) for the case
$c=5.5$, $J_0=-1$, and $J=1$. Here $T_{c1}=1.02$ and $T_{c2}=2.27$.} 
\label{sw_3}
\end{figure}
\begin{figure}
\epsfxsize=65mm \centerline{\epsffile{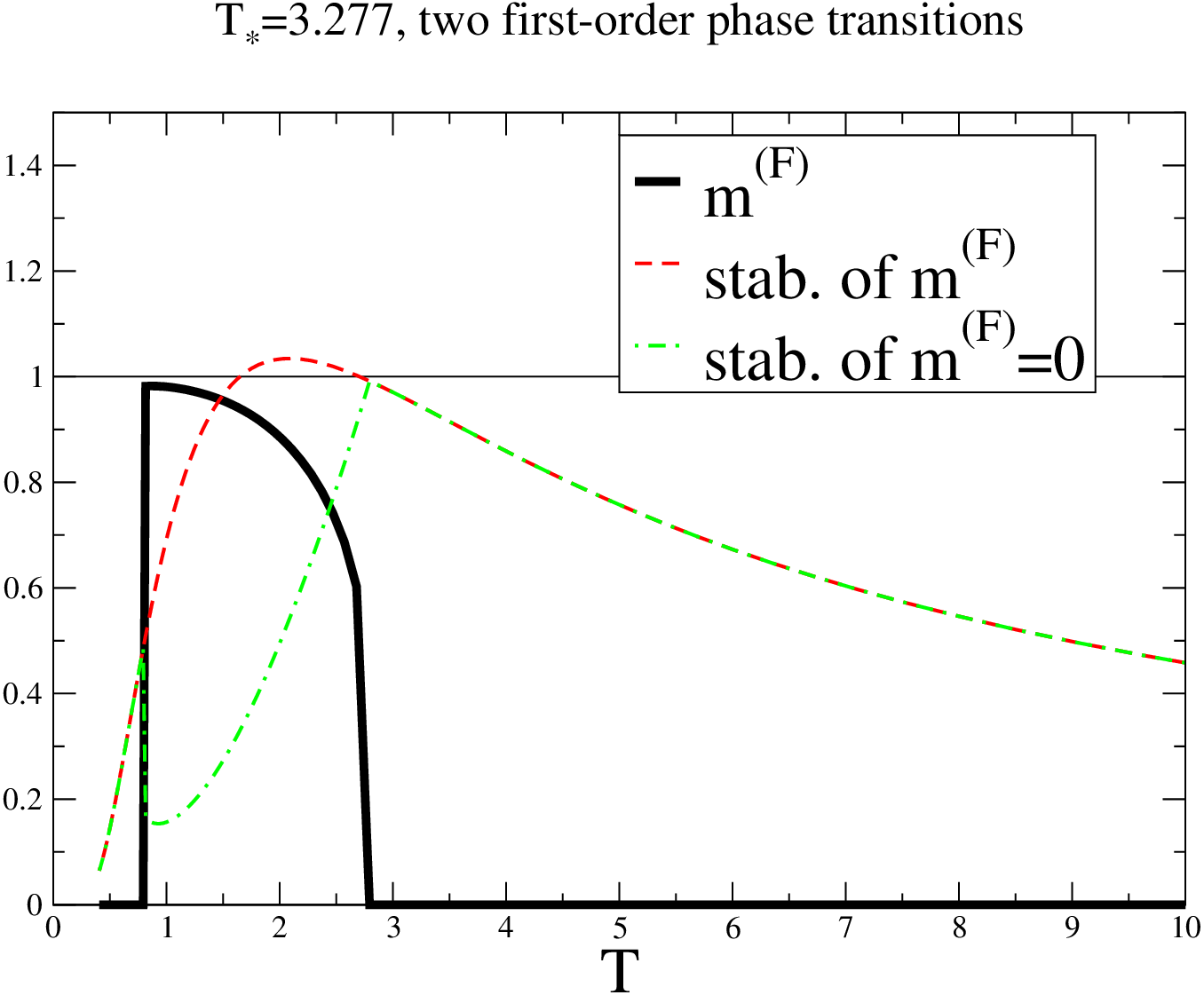}}
\caption{(Color online) Magnetization (thick solid line), and curves of stability 
(dashed and dot-dashed lines) for the case
$c=5.5$, $J_0=-0.9$ and $J=1$. Here $T_{c1}=0.85$ and $T_{c2}=2.78$.} 
\label{sw_4}
\end{figure}
\begin{figure}
\epsfxsize=65mm \centerline{\epsffile{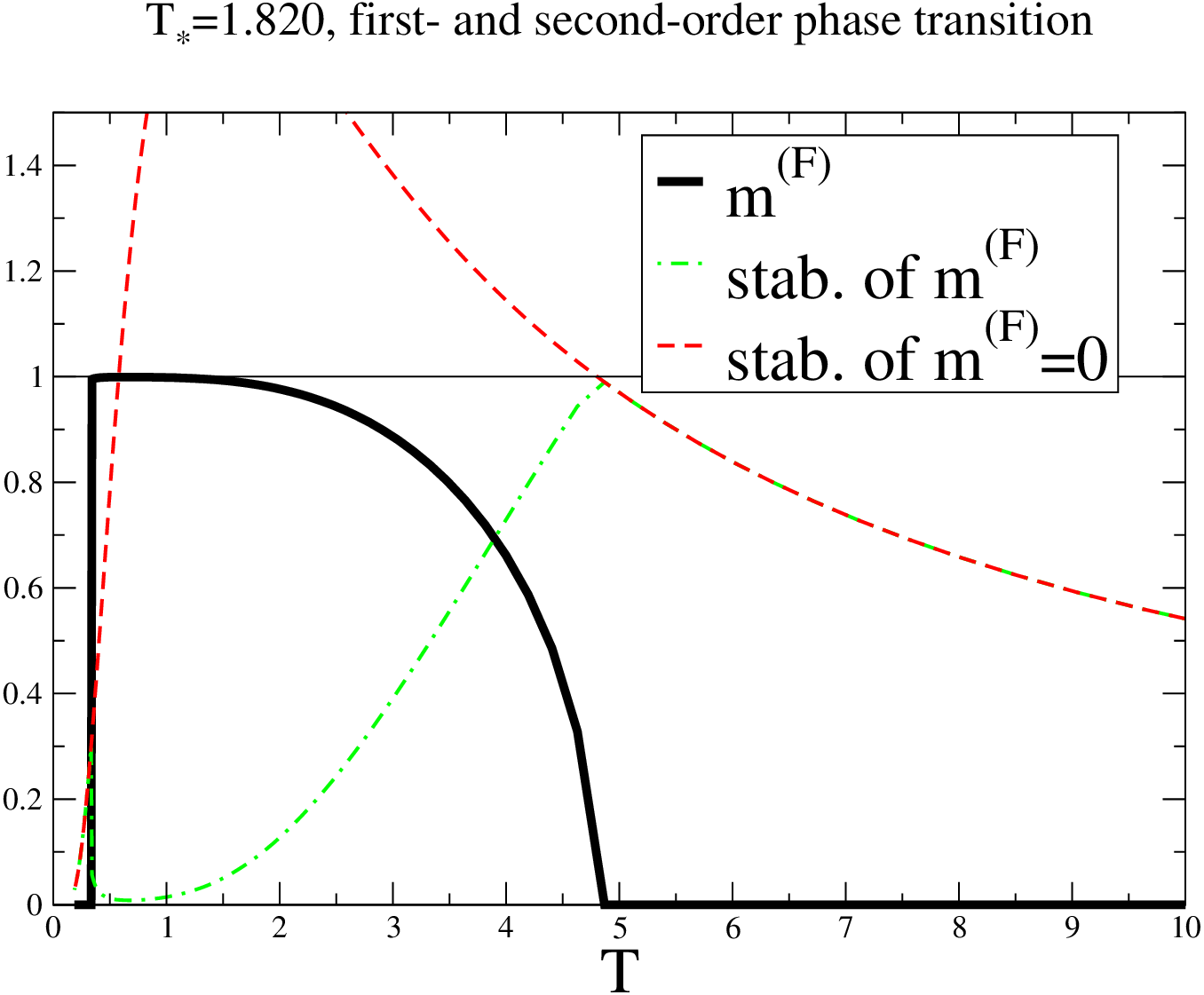}}
\caption{(Color online) Magnetization (thick solid line), and curves of stability 
(dashed and dot-dashed lines) for the case
$c=6$, $J_0=-0.5$ and $J=1$. Here $T_{c1}=0.35$ and $T_{c2}=4.87$.
$T_{c2}$ corresponds to a second-order phase transition.} 
\label{sw_5}
\end{figure}
\begin{figure}
\epsfxsize=65mm \centerline{\epsffile{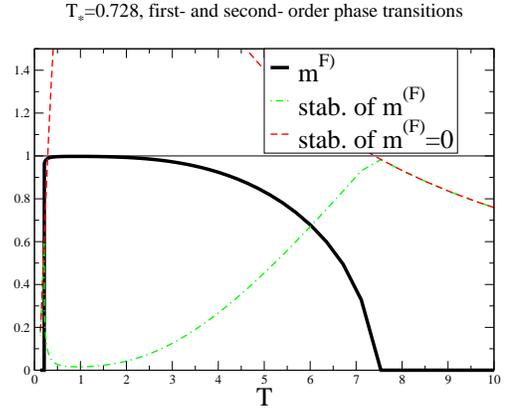}}
\caption{(Color online) Magnetization (thick solid line), and curves of stability 
(dashed and dot-dashed lines) for the case
$c=4$, $J_0=-0.2$ and $J=2$. Here $T_{c1}=0.22$ and $T_{c2}=7.55$.
$T_{c2}$ corresponds to a second-order phase transition.} 
\label{sw_6}
\end{figure}
\begin{figure}[t]
\epsfxsize=65mm \centerline{\epsffile{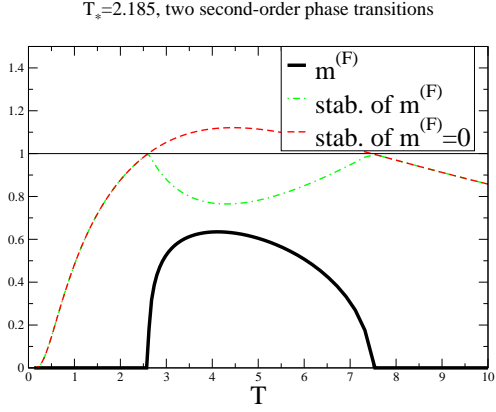}}
\caption{(Color online) Magnetization (thick solid line), and curves of stability 
(dashed and dot-dashed lines) for the case
$c=1.6$, $J_0=-0.6$ and $J=7$. Here $T_{c1}=2.58$ and $T_{c2}=7.55$.} 
\label{sw_7}
\end{figure}
\begin{figure}
\epsfxsize=65mm \centerline{\epsffile{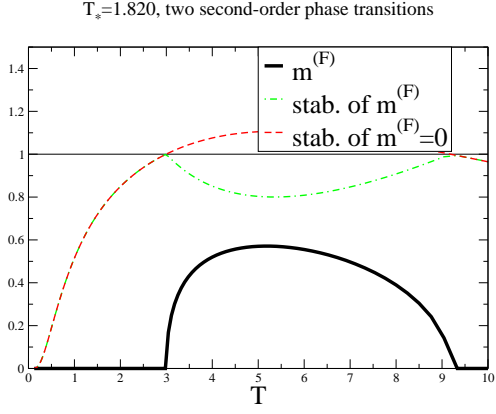}}
\caption{(Color online) Magnetization (thick solid line), and curves of stability 
(dashed and dot-dashed lines) for the case
$c=1.4$, $J_0=-0.5$ and $J=10$. Here $T_{c1}=3.00$ and $T_{c2}=9.34$.} 
\label{sw_8}
\end{figure}
\begin{figure}
\epsfxsize=65mm \centerline{\epsffile{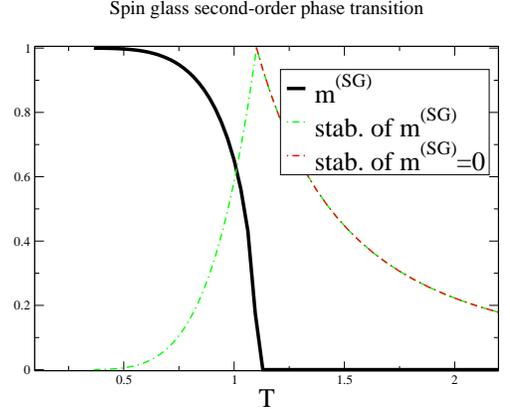}}
\caption{(Color online) Spin glass order parameter (solid line), and curves of stability 
(dashed and dot-dashed lines) for the case
$c=0.5$, $J_0=1$ and $J=3/5/c$. Here $T_{c}=1.130$.} 
\label{sw_sg_1}
\end{figure}
\begin{figure}
\epsfxsize=65mm \centerline{\epsffile{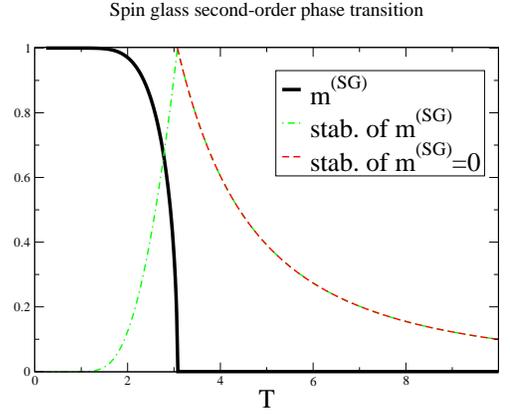}}
\caption{(Color online) Spin glass order parameter (solid line), and curves of stability 
(dashed and dot-dashed lines) for the case
$c=10$, $J_0=0.25$ and $J=1/c$. Here $T_{c}=0.424$.} 
\label{sw_sg_2}
\end{figure}
\begin{figure}[pt]
\epsfxsize=65mm \centerline{\epsffile{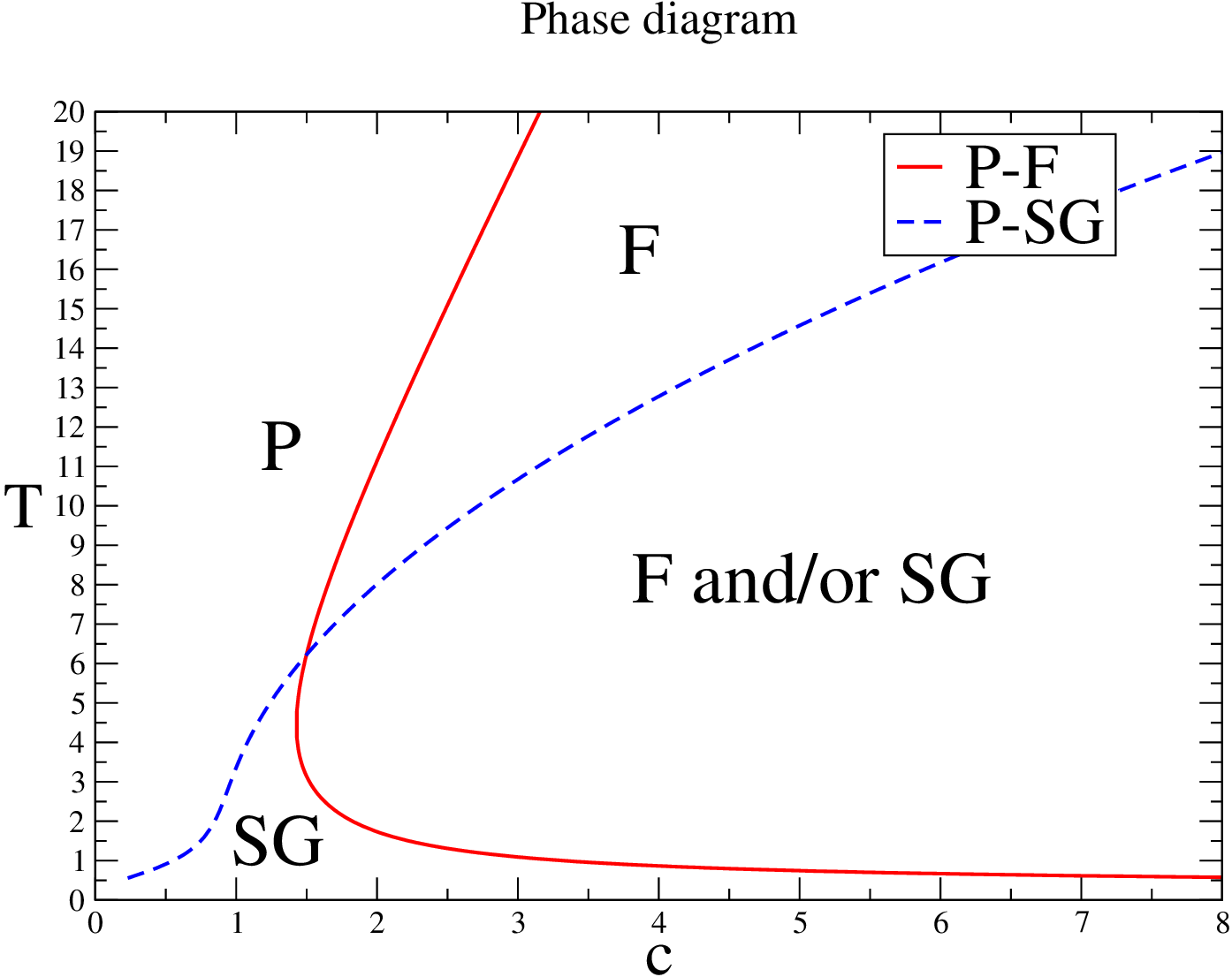}}
\caption{(Color online) Phase diagram for the case considered in Fig. \ref{sw_7}
with the measure of Eq. (\ref{dPF}).}
\label{sw_pd1}
\end{figure}
\begin{figure}[pt]
\epsfxsize=65mm \centerline{\epsffile{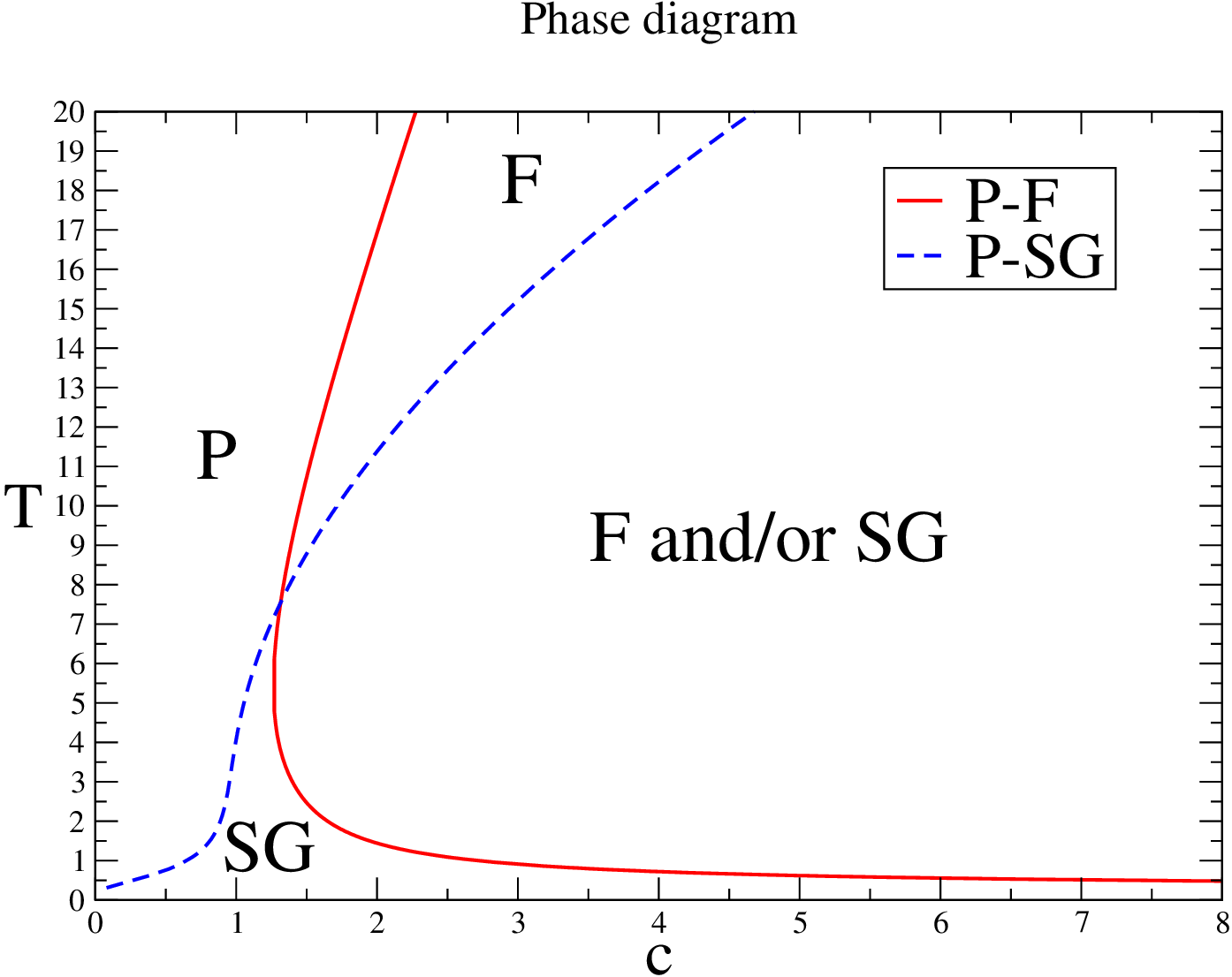}}
\caption{(Color online) Phase diagram for the case considered in Fig. \ref{sw_8}
with the measure of Eq. (\ref{dPF}).}
\label{sw_pd2}
\end{figure}

In Figs. \ref{sw_1_pos}-\ref{sw_8} we plot the 
stable and leading magnetization $m^{(\mathrm{F})}$ (thick solid line), 
${\tilde{\chi}_0\left(\beta^{(\mathrm{F})}J_0^{(\mathrm{F})},0\right)}
\beta^{(\mathrm{F})}J^{(\mathrm{F})}$ (dashed line), and 
${\tilde{\chi}_0\left(\beta^{(\mathrm{F})}J_0^{(\mathrm{F})},
\beta J^{(\mathrm{F})}m^{(\mathrm{F})}\right)}
\beta^{(\mathrm{F})}J^{(\mathrm{F})}$ (dot-dashed line) for several cases
obtained by solving Eq. (\ref{1dl}) numerically with $\mathrm{\Sigma}$=F.
The stable and leading solution (the only one drawn) corresponds to the solution that minimizes $L^{(\mathrm{F})}(m)$ (see Eq. (\ref{THEOlead})).
In all these examples we have chosen the measure (\ref{dPF}).
Figs. \ref{sw_1_pos} and \ref{sw_2_pos} concern two cases with $J_0>0$ so that one and only one
second-order phase transition is present. The input data of these two cases are
the same as those analyzed numerically in \cite{Niko} 
(note that in the model considered in \cite{Niko}, the long range
coupling $J$ is divided by $c$). As already stated in Sec. III, 
the self-consistent equations
become exact in the limit $c\to 0$, for second-order phase transitions, 
and in the limit $c\to\infty$. 
Therefore, for the magnetization, by comparison with \cite{Niko}, 
in Figs. \ref{sw_1_pos} and \ref{sw_2_pos}, where $c$ is relatively small and big, respectively, 
we see good agreement also below the critical temperature. 
 
Figs. \ref{sw_1}-\ref{sw_8} concern eight cases with $J_0<0$.
In these figures we plot also the line $y=1$, to make evident when the stability conditions for
the solutions $m^{(\mathrm{F})}=0$ and $m^{(\mathrm{F})}\neq 0$, which are given by 
${\tilde{\chi}_0\left(\beta^{(\mathrm{F})}J_0^{(\mathrm{F})},0\right)}
\beta^{(\mathrm{F})}J^{(\mathrm{F})}<1$ (dashed line), and 
${\tilde{\chi}_0\left(\beta^{(\mathrm{F})}J_0^{(\mathrm{F})},
\beta J^{(\mathrm{F})}m^{(\mathrm{F})}\right)}
\beta^{(\mathrm{F})}J^{(\mathrm{F})}<1$ (dot-dashed line), are satisfied, respectively.
As explained above, the critical behavior and the localization of
the critical temperatures is more complicated when $J_0<0$. In particular,
given $|J_0|$, if $c$ is not sufficiently high the solution $m^{(\mathrm{F})}=0$
remains stable at all temperatures and if it is also a leading solution,
no phase transition occurs.
Let us consider Eq. (\ref{1dm}). 
For $J_0<0$ the lhs of this equation has some maximum at a finite
value $\bar{\beta}$ given by
\begin{eqnarray}
\label{1dt}\nonumber
\bar{\beta} J=\frac{1}{2}\log\left[\frac{1+\delta(r)}{1-\delta(r)}\right],
\end{eqnarray} 
where 
$r\defi |J_0|/J$, and we have introduced
\begin{eqnarray}
\label{1du}\nonumber
\delta(r)\defi \sqrt{1+r^2}-r.
\end{eqnarray} 
Hence, we see that a sufficient condition 
for the solution $m^{(\mathrm{F})}=0$ to become unstable is that be
\begin{eqnarray}
\label{1dv}
c\left(\frac{1+\delta(r)}{1-\delta(r)}\right)^{r}\delta(r)\geq 1.
\end{eqnarray} 
Note that the above represents only a condition for the instability
of the solution $m^{(\mathrm{F})}=0$, but the true solution is the one
that is both stable and leading. In fact, when $J_0<0$,
a phase transition in general may be present also when Eq. (\ref{1dv}) 
is not satisfied and, correspondingly 
the possible critical temperatures will be not determined by Eq. (\ref{1dm}).
 
In Fig. \ref{sw_1} we report a case with $J=1$, $J_0=-1$ ($r=1$) and a relatively low value of $c$,
$c=5$, so that no phase transition is present. Similarly, in Fig. \ref{sw_2}
we report again a case in which no phase transition is present due to the fact
that here $r$ is relatively big, $r=1.1$.
It is interesting to observe that for $r=1$ Eq. (\ref{1dv}) requires a value
of $c$ greater than the limit value $c=3+2\sqrt{2}=5.8284...$.
In both Figs. \ref{sw_3} and \ref{sw_4} 
we report a case in which Eq. (\ref{1dv}) is still not satisfied,
but nevertheless two first-order phase transitions are present.    
In both Figs. \ref{sw_5} and \ref{sw_6} we have one first- and one second-order phase transition. 
In both Figs. \ref{sw_7} and \ref{sw_8} we have two second-order phase transitions. 
As anticipated in Sec. IIIB, we note that in Figs. \ref{sw_3}-\ref{sw_6}, \textit{i.e.}, the cases in which there is at least
one first-order phase transition, there are always regions where both the solutions, $m^{(\mathrm{F})}=0$
and $m^{(\mathrm{F})}\neq 0$, are simultaneously stable but only one solution is leading (the one drawn); 
whereas in Figs. \ref{sw_7} and \ref{sw_8}, as it was in Figs. \ref{sw_1_pos} and \ref{sw_2_pos},
since we have only second-order phase transitions,
the stability condition turns out to be a necessary and sufficient condition for determining the leading solution and
the critical temperature can be determined also by Eq. (\ref{1dm}) with $\Sigma=$F. 

In the top of Figs. \ref{sw_3}-\ref{sw_8} we write the discriminant 
temperature $T_*=4|J_0|/\log(3)$ below which
a phase transition (if any) may be first-order [see Eq. (\ref{THEOneg7}) and Eqs. (\ref{1dg})-(\ref{1di})].

Finally in Figs. \ref{sw_sg_1} and \ref{sw_sg_2} we plot the 
spin glass order parameter $m^{(\mathrm{SG})}$ (solid line), 
${\tilde{\chi}_0\left(\beta^{(\mathrm{SG})}J_0^{(\mathrm{SG})},0\right)}
\beta^{(\mathrm{SG})}J^{(\mathrm{SG})}$ (dashed line), and 
${\tilde{\chi}_0\left(\beta^{(\mathrm{SG})}J_0^{(\mathrm{SG})},
\beta J^{(\mathrm{SG})}m^{(\mathrm{SG})}\right)}
\beta^{(\mathrm{SG})}J^{(\mathrm{SG})}$ (dot-dashed line) 
obtained by solving Eq. (\ref{1dl}) numerically with $\mathrm{\Sigma}$=SG.
In these two examples we have chosen the measure (\ref{dPSG}) and,
for $c$, $|J|$, and $J_0$, we have considered the same parameters 
of Figs. \ref{sw_1_pos} and \ref{sw_2_pos} of the ferromagnetic case.
As anticipated in Sec. IIIB, due to the fact that the effective coupling $J_0^{(\mathrm{SG})}$
is positive, there is only a second-order phase transition, the stability condition turns out to be a necessary and sufficient condition 
for determining the leading solution, and
the critical temperature can be determined also by Eq. (\ref{1dm}) with $\Sigma=$SG. 

Note that, unlike the P-F critical surface, the P-SG critical
surface does not depend on the parameter $p$ entering in Eq. (\ref{dPSG}).
For the reciprocal stability between the P-F and the P-SG critical 
surface we remind the reader to the general rules of Sec. IIID 
(see cases \textbf{(1)} and \textbf{(3)}) which, for $J_0\geq 0$, 
reduce to the results reported in Sec. 6.1 of the Ref. \cite{Niko}. 
Here we stress just that, if $J_0\geq 0$, for $p\leq 0.5$, only the P-SG transition is possible.
However, when $J_0<0$ and $c$ is not sufficiently large, 
the SG phase may be the only stable phase even when $p=1$. 
In fact, although when $J_0<0$ the solution $m^{(\mathrm{F})}$
may have two P-F critical temperatures, in general, if the P-SG
temperature is between these, we cannot exclude that the solution 
$m^{(\mathrm{SG})}$ starts to be the leading solution at sufficiently
low temperatures. In Figs. \ref{sw_pd1} and \ref{sw_pd2},
on the plane $(T,c)$, we plot the phase diagrams
corresponding to the cases of Figs. \ref{sw_7} and \ref{sw_8}, respectively.
These phase diagrams are obtained by solving Eq. (\ref{THEOg}) supposing that
here, as in the cases of Figs. \ref{sw_7} and \ref{sw_8}, where
$c=1.4$ and $c=0.5$, respectively, the P-F transition is always second-order.
We plan to investigate in more detail the phase diagram in future works.

\section{Small-world spherical model in arbitrary dimension $d_0$}
In this section we will analyze the case in which the unperturbed model is
the spherical model built up over
a $d_0$-dimensional lattice $\mathcal{L}_0$ (see \cite{Baxter} and references
therein)~\footnote{
Note that from the point of view of the statistical mechanics the
spherical model is an infinite dimensional model, however we will keep on to
reserve the symbol $d_0$ for the dimension of $\mathcal{L}_0$.}.
In this case the $\sigma$'s are continuous ``spin'' variables ranging
in the interval $(-\infty,\infty)$ subjected to the sole constrain 
$\sum_{i\in\mathcal{L}_0}\sigma_i^2=N$, however our theorems and formalism can
be applied as well and give results that, within the same limitations
prescribed in Sec. III, are exact. 

Following \cite{Baxter}, for the unperturbed model we have
\begin{eqnarray}
\label{1s}\nonumber
-\beta f_0 (\beta J_0,\beta h)=
\frac{1}{2}\log\left(\frac{\pi}{\beta J_0}\right)+
\phi\left(\beta J_0,\beta h,\bar{z}\right),
\end{eqnarray} 
\begin{eqnarray}
\label{2s}
m_0 (\beta J_0,\beta h)=\frac{\beta h}{2\beta J_0 \bar{z}},
\end{eqnarray} 
where
\begin{eqnarray}
\label{3s}\nonumber
\phi\left(\beta J_0,\beta h,z\right)=\beta J_0 d_0 + \beta J_0 z - \frac{1}{2}g(z)+
\frac{\left(\beta h\right)^2}{4\beta J_0 z},
\end{eqnarray} 
\begin{eqnarray}
\label{4s}
g(z)&=&\frac{1}{(2\pi)^{d_0}}\int_0^{2\pi}\ldots \int_0^{2\pi}
d\omega_1\ldots d\omega_{d_0}
\nonumber \\
&\times& \log\left[d_0+z-\cos(\omega_1)-\ldots-\cos(\omega_{d_0})\right], \nonumber
\end{eqnarray} 
and $\bar{z}=\bar{z}(\beta J_0,\beta h)$ is the (unique) solution of 
the equation $\partial_{z}\phi\left(\beta J_0,\beta h,z\right)=0$:
\begin{eqnarray}
\label{5s}
\beta J_0-\frac{\left(\beta h\right)^2}{4\beta J_0 \bar{z}^2}=\frac{1}{2}g'(\bar{z}),
\end{eqnarray} 
from which it follows the equation for $m_0$ 
\begin{eqnarray}
\label{6s}
\beta J_0\left(1-m_0^2\right)=\frac{1}{2}g'\left(\frac{\beta h}{2\beta J_0 m_0}\right).
\end{eqnarray} 
The derivative $g'$ can in turn be expressed as
\begin{eqnarray}
\label{7s}
g'(z)=\int_0^{\infty}e^{-t(z+d_0)}\left[\mathcal{J}_0\left(it\right)\right]^{d_0} dt,
\end{eqnarray} 
$\mathcal{J}_0\left(it\right)$ being the usual Bessel function whose
behavior for large $t$ is given by
\begin{eqnarray}
\label{8s}\nonumber
\mathcal{J}_0\left(it\right) =
\frac{e^t}{(2\pi t)^{\frac{1}{2}}}\left(1+\mathop{O}\left(\frac{1}{t}\right)\right).
\end{eqnarray} 

The critical behavior of the unperturbed system depends on the values of
$g'(z)$ and $g''(z)$ near $z=0$. 
It turns out that for $d_0\leq 2$ one has $g'(0)=\infty $ and there is no
spontaneous magnetization, whereas for $d_0>2$ one has $g'(0)<\infty $ and at
$h=0$ the unperturbed system undergoes a second-order phase transition with
magnetization given by Eq. (\ref{6s}) which, for $\beta$ above
$\beta_{c0}$, becomes
\begin{eqnarray}
\label{9s}\nonumber
m_0(\beta J_0,0)=\sqrt{1-\frac{\beta_{c0}}{\beta}},
\end{eqnarray} 
where the inverse critical temperature $\beta_{c0}$ is given by
\begin{eqnarray}
\label{10s}\nonumber
\beta_{c0}J_0=\frac{1}{2}g'\left(0\right).
\end{eqnarray} 
Furthermore, it turns out that for $d_0\leq 4$ one has $g''(0)=\infty $, 
whereas for $d_0>4$ one has $g''(0)<\infty $. This reflects on the
critical exponents $\alpha$, $\gamma$ and $\delta$, which take the classical
mean-field values only for $d_0>4$.

According to Sec. III, to solve the random model - for simplicity - 
at zero external field, we have to perform the
effective substitutions $\beta J_0\to \beta J_0^{(\Sigma)}$ and
$\beta h\to \beta J^{(\Sigma)} m^{(\Sigma)}$ in the above equations.
From Eqs. (\ref{2s}), (\ref{5s}) and (\ref{6s}), we get immediately: 
\begin{eqnarray}
\label{11s}\nonumber
\bar{z}^{(\Sigma)}=
\frac{\beta J^{(\Sigma)}}
{2\beta J_0^{(\Sigma)}},
\end{eqnarray} 
the equations for
inverse critical temperature $\beta_c^{(\Sigma)}$  
\begin{eqnarray}
\label{12s}
\beta_c^{(\Sigma)} J_0^{(\Sigma)}=
\frac{1}{2}g'\left(
\frac{\beta_c^{(\Sigma)}J^{(\Sigma)}}
{2\beta_c^{(\Sigma)} J_0^{(\Sigma)}}\right),
\end{eqnarray} 
and the magnetizations 
$m^{(\Sigma)}$
\begin{eqnarray}
\label{13s}
m^{(\Sigma)}=\left\{
\begin{array}{l}
\sqrt{1-
\frac{1}{2\beta J_0^{(\Sigma)}}
g'\left(\frac{\beta J^{(\Sigma)}}
{2\beta J_0^{(\Sigma)}}\right)},\quad \beta>\beta_c^{(\Sigma)},
\\
0,\quad \beta<\beta_c^{(\Sigma)}\geq 0.
\end{array}
\right.
\end{eqnarray} 
Note that, as it must be from the general result of Sec. IIIB,
unlike the unperturbed model, as soon as the connectivity 
$c$ is not zero, Eq. (\ref{12s}) has always a finite
solution $\beta_c^{(\Sigma)}$, independently on the dimension $d_0$.
In fact, one has a finite temperature second-order 
phase transition even for $d_0\to 0^+$ where
from Eq. (\ref{7s}) we have  
\begin{eqnarray}
\label{7sb}\nonumber
g'(z)=\frac{1}{z}, \qquad (d_0=0)
\end{eqnarray} 
so that the equations for the critical temperature (\ref{12s}) become
\begin{eqnarray}
\label{12sb}\nonumber
\beta_c^{(\Sigma)} J^{(\Sigma)}=1, \qquad (d_0=0)
\end{eqnarray} 
which, as expected, coincide with Eqs. (\ref{VBh}) and (\ref{VBi}) 
of the Viana-Bray model.

Similarly, unlike the unperturbed model, in the random model
all the critical exponents take the classical mean-field values,
independently on the dimension $d_0$. In the specific case of
the spherical model, this behavior is due to the fact
that $g'(z)$ and $g''(z)$ can be singular only at $z=0$
but, as soon as the connectivity $c$ is not zero,
there is an effective external field $\beta J^{(\Sigma)} m^{(\Sigma)}$ 
so that $\bar{z}^{(\Sigma)}$ is not zero.
For the critical behavior, 
the dependence on the dimension $d_0$ reflects only in 
the coefficients, not on the critical exponents.
In particular, concerning the argument of the square root
of the rhs of Eq. (\ref{13s}), 
by expanding in the reduced temperature $t^{(\Sigma)}$, 
for $|t^{(\Sigma)}|\ll 1$ we have
\begin{eqnarray}
\label{14s}
&& 1-\frac{1}{2\beta_c^{(\Sigma)} J_0^{(\Sigma)}}
g'\left(\frac{\beta_c^{(\Sigma)}J^{(\Sigma)}}
{2\beta_c^{(\Sigma)} J_0^{(\Sigma)}}\right)=\nonumber \\ 
&& B^{(\Sigma)} t^{(\Sigma)}+\mathop{O}(t^{(\Sigma)})^2,
\end{eqnarray} 
where
\begin{widetext}
\begin{eqnarray}
\label{15s}
B^{\mathrm{(F)}}&=&-1+
\frac{1}{2\beta_c^{\mathrm{(F)}} J_0^{\mathrm{(F)}}}
g''\left(\frac{\beta_c^{\mathrm{(F)}}J^{\mathrm{(F)}}}
{(2\beta_c^{\mathrm{(F)}} J_0^{\mathrm{(F)}})^2}\right)
\nonumber \\ \times
&& \left(c\int d\mu(J_{i,j}) \left(1-
\tanh^2(\beta_c^{\mathrm{(F)}} J_{i,j})\right)\beta_c^{\mathrm{(F)}} J_{i,j}
- c\int d\mu(J_{i,j}) \tanh(\beta_c^{\mathrm{(F)}} J_{i,j})\right), \nonumber
\end{eqnarray} 
\begin{eqnarray}
\label{16s}\nonumber
B^{\mathrm{(SG)}}&=&\frac{1}{2\beta_c^{\mathrm{(SG)}}J_0^{\mathrm{(SG)}}}
\left[-4\frac{\tanh(\beta_c^{\mathrm{(SG)}}J_0)\beta_c^{\mathrm{(SG)}}J_0}
{1+\tanh^2(\beta_c^{\mathrm{(SG)}}J_0)}+
g''\left(\frac{\beta_c^{\mathrm{(SG)}}J^{\mathrm{(SG)}}}
{(2\beta_c^{\mathrm{(SG)}} J_0^{\mathrm{(SG)}})^2}\right) \right.
\nonumber \\ \times
&& \left(2c\int d\mu(J_{i,j}) \left(1-
\tanh^2(\beta_c^{\mathrm{(SG)}} J_{i,j})\right)
\tanh(\beta_c^{\mathrm{(SG)}} J_{i,j})\beta_c^{\mathrm{(SG)}} J_{i,j} \right.
\nonumber \\ 
&& \left. \left. - 4c\int d\mu(J_{i,j}) \tanh^2(\beta_c^{\mathrm{(SG)}} J_{i,j})
\frac{\tanh(\beta_c^{\mathrm{(SG)}}J_0)\beta_c^{\mathrm{(SG)}}J_0}
{\left(1+\tanh^2(\beta_c^{\mathrm{(SG)}}J_0)\right)
\beta_c^{\mathrm{(SG)}}J^{\mathrm{(SG)}}}
\right)\right], \nonumber
\end{eqnarray} 
\end{widetext}
so that from Eqs. (\ref{13s}) and (\ref{14s}) 
for the critical behavior of the magnetizations we get explicitly the
mean-field behavior:
\begin{eqnarray}
\label{17s}\nonumber
m^{(\Sigma)}=\left\{
\begin{array}{l}
\sqrt{B^{(\Sigma)} t^{(\Sigma)}~}+\mathop{O}(t^{(\Sigma)}),\quad t^{(\Sigma)}<0,
\\
0,\quad t^{(\Sigma)}\geq 0. \nonumber
\end{array}
\right.
\end{eqnarray} 

\section{Mapping to non random models}
In Sec. VIII we will derive the main result presented in
Sec. III. To this aim in the next Sec. VIIA we will recall the general mapping
between a random model built up over a given graph and a non random one
built up over the same graph, whereas in Sec. VIIB we will
generalize this mapping to random models built up over random graphs. 
We point out that the mapping does not consist in a sort of annealed approximation.

\subsection{Random Models defined on Quenched Graphs}
Let us consider the following random model.
Given a graph $\bm{g}$, which can be determined through the adjacency matrix
for shortness also indicate by $\bm{g}=\{g_b\}$, with $g_b=0,1$, $b$ being a bond,
let us indicate with $\Gamma_{\bm{g}}$ 
the set of the bonds $b$ of $\bm{g}$ and let us define over $\Gamma_{\bm{g}}$ 
the Hamiltonian 
\begin{eqnarray}
\label{HMD}
H\left(\{\sigma_i\};\{J_b\}\right)
\defi -\sum_{b\in\Gamma_{\bm{g}}} J_b \sigma_{i_b}\sigma_{j_b} - \sum_i h_i \sigma_i
\end{eqnarray} 
where $J_b$ is the random coupling at the bond $b$, and 
$\sigma_{i_b},\sigma_{i_b}$ are the Ising variables at the end-points of $b$. 
The free energy $F$ and the physics are defined as in Sec. II by
Eqs. (\ref{logZ})-(\ref{OO}):
\begin{eqnarray}
\label{logZD}
-\beta F\defi \int d\mathcal{P}\left(\{J_b\}\right)
\log\left(Z\left(\{J_b\}\right)\right),
\\
\label{logZD2}
\overline{\media{\mathcal{O}}^l}\defi\int d\mathcal{P}\left(\{J_b\}\right)
\media{\mathcal{O}}^l, \quad l=1,2
\end{eqnarray} 
%
where $d\mathcal{P}\left(\{J_b\}\right)$ 
is a product measure over all the possible bonds $b$ given 
in terms of normalized measures $d\mu_b\geq 0$ 
(we are considering a general measure $d\mu_b$ 
allowing also for a possible dependence on the bonds) 
\begin{eqnarray}
\label{dPM}
d\mathcal{P}\left(\{J_b\}\right)\defi \prod_{b\in\Gamma_{\mathrm{full}}} 
d\mu_b\left( J_b \right),
\quad \int d\mu_b\left( J_b \right) =1,
\end{eqnarray}
where $\Gamma_{\mathrm{full}}$ stands for the set of bonds of the fully connected graph.
As in Sec. II, we will indicate a generic correlation function, connected or not, by
${{C}}$ with understood indices $i_1,\ldots,i_k$ all different, see
Eqs. (\ref{CF}) and (\ref{CG}).

In the following, given an arbitrary vertex $i$ of $\bm{g}$, 
we will consider as first neighbors $j$ of $i$ only those vertices 
for which $\int d\mu_{i,j}(J_{i,j})J_{i,j}$ or $\int d\mu_{i,j}(J_{i,j})J_{i,j}^2$ are at least $\mathop{O}(1/N)$.
Note that we can always neglect couplings having lower averages.  
We will indicate with $D(\Gamma_{\bm{g}})$ the average number of first
neighbors of the graph $\bm{g}$. For a $d$-dimensional lattice,
$D(\Gamma_{\bm{g}})=2d-1$, for a Bethe lattice of coordination number q, 
$D(\Gamma_{\bm{g}})=q-1$, and for long range models, $D(\Gamma_{\bm{g}})\propto N$.
We will exploit in particular the fact that 
$D(\Gamma_{\mathcal{L}_0}\cup\Gamma_{\mathrm{full}})=
D(\Gamma_{\mathrm{full}})\propto N$. 

Given a random model defined trough Eqs. (\ref{HMD}-\ref{dPM}), 
we define, on the same set of bonds $\Gamma_{\bm{g}}$, its \textit{related Ising model} 
trough the following Ising Hamiltonian
\begin{eqnarray}
\label{HI}
H_{I}\left(\{\sigma_i\};\{J_b^{(I)}\}\right)\defi 
-\sum_{b\in\Gamma_{\bm{g}}} J_b^{(I)} \sigma_{i_b}\sigma{j_b}
-\sum_i h_i \sigma_i,
\end{eqnarray} 
where the Ising couplings $J_b^{(I)}$ have 
non random values such that $~\forall ~b,b'\in \Gamma_{\bm{g}}$
\begin{eqnarray}
\label{JI}
J_{b'}^{(I)}&=&J_b^{(I)} \quad \mathrm{if} \quad 
d\mu_{b'}\equiv d\mu_{b}, \\
\label{JIb}
J_b^{(I)}&\neq & 0 
\quad \mathrm{if} \left\{
\begin{array}{l}
\quad \int d\mu_b(J_b)J_b=\mathop{O}\left(\frac{1}{N}\right), \quad \mathrm{or} \\
\quad \int d\mu_b(J_b)J_b^2=\mathop{O}\left(\frac{1}{N}\right). 
\end{array}
\right.
\end{eqnarray}
In the following a suffix $I$ over quantities such as $H_{I}$,
$F_{I}$, $f_{I}$, $g_I$, etc\ldots, or $J_b^{(I)}$, $\beta_c^{(I)}$, etc\ldots,
will be referred to the related Ising system with Hamiltonian (\ref{HI}).

We can always split the free energy of the random system with $N$ spins 
as follows
\begin{eqnarray}
\label{0logZ2}
-\beta F&=&\sum_{b} \int d\mu_{b} \log\left[\cosh(\beta J_b)\right]+ \nonumber \\
&& \sum_i \log\left[2\cosh(\beta h_i)\right] +\phi,
\end{eqnarray}
$\phi$ being the high temperature part of the free energy.
Let $\varphi$ be the density of $\phi$ in the thermodynamic limit
\begin{eqnarray}
\label{varphi}
\varphi \defi \lim_{N\rightarrow \infty}\phi/N.
\end{eqnarray} 
Let us indicate with $\varphi_{I}$ the high temperature part
of the free energy density of the related Ising model defined through
Eqs. (\ref{HI}-\ref{JIb}). 
As is known, $\varphi_{I}$ can be expressed in terms of the
quantities $z_b=\tanh(\beta J_b^{(I)})$ and $z_i=\tanh(\beta h_i)$, \textit{i.e.}, 
the parameters of the high temperature expansion:
\begin{eqnarray}
\label{varphi1}
\varphi_{I} = \varphi_{I}\left(\{\tanh(\beta J_b^{(I)})\};\{\tanh(\beta h_i)\}\right).
\end{eqnarray} 

The related Ising model is defined by a set of,
typically few, independent couplings $\{J_b^{(I)}\}$, 
trough Eqs. (\ref{JI}-\ref{JIb})
and, for $h_i=0$, $i=1,\ldots,N$, its critical surface will be determined 
by the solutions of an equation, possibly vectorial, 
$G_I\left(\{\tanh(\beta J_b^{(I)})\}\right)=0$.

In \cite{MOI} we have proved the following mapping.

Let $\beta_c^{(\mathrm{SG})}$ and $\beta_c^{(\mathrm{F/AF})}$ be respectively 
solutions of the two equations 
\begin{eqnarray}
\label{mapp0g}
G_I\left(\{\int d\mu_b{\tanh^2(\beta_c^{(\mathrm{SG})} J_b)}\}\right)&=& 0, \\
\label{mapp01g}
G_I\left(\{\int d\mu_b{\tanh(\beta_c^{(\mathrm{F/AF})} J_b)}\}\right)&=& 0.
\end{eqnarray} 
Asymptotically, at sufficiently high dimensions $D(\Gamma_{\bm{g}})$,
the critical inverse temperature of the spin glass model $\beta_c$ is given by
\begin{eqnarray}
\label{mappg}
\beta_c=\mathrm{min}\{\beta_c^{(\mathrm{SG})},\beta_c^{(\mathrm{F/AF})}\};
\end{eqnarray} 
and in the paramagnetic phase for $D(\Gamma_{\bm{g}})>2$ the following mapping holds 
\begin{eqnarray}
\label{mapp1g}
\left|\frac{\varphi-\varphi_{eff}}{\varphi}\right|=
\left|\frac{{{C}}-{{C}}_{eff}}{{{C}}}\right|=
O\left(\frac{1}{D(\Gamma_{\bm{g}})}\right),
\end{eqnarray} 
\begin{eqnarray}
\label{mapp2g}
\varphi_{eff}=\frac{1}{l}\varphi^{(\Sigma)}\defi\frac{1}{l}
\varphi_{I}\left(\{\int d\mu_b{\tanh^{l}(\beta J_b)}\}\right),
\end{eqnarray}
and
\begin{eqnarray}
\label{mapp2gc}
{{C}}_{eff}=\frac{1}{l}{{C}}^{(\Sigma)}\defi\frac{1}{l}
{{C}}_{I}\left(\{\int d\mu_b{\tanh^{l}(\beta J_b)}\}\right),
\end{eqnarray}
where
\begin{eqnarray}
\label{mapp3g}
l=\left\{
\begin{array}{l}
2, \quad \mathrm{if}\quad
\varphi_{I}\left(\{\int d\mu_b{\tanh^{2}(\beta J_b)}\}\right)\geq \\
2|\varphi_{I}\left(\{\int d\mu_b{\tanh(\beta J_b)}\}\right)|, \\
1, \quad \mathrm{if}\quad
\varphi_{I}\left(\{\int d\mu_b{\tanh^{2}(\beta J_b)}\}\right)< \\
2|\varphi_{I}\left(\{\int d\mu_b{\tanh(\beta J_b)}\}\right)|,
\end{array}
\right.
\end{eqnarray} 
and $\Sigma$=F/AF or SG, for $l$=1 or 2, respectively.

In the limit $D(\Gamma_{\bm{g}})\rightarrow \infty$ and $h_i=0$, $i=1,\ldots,N$, 
Eqs. (\ref{mapp0g}-\ref{mapp3g}),
give the exact free energy and correlation functions 
in the paramagnetic phase (P); the
exact critical paramagnetic-spin glass (P-SG), $\beta_c^{(\mathrm{SG})}$, and
paramagnetic- F/AF (P-F/AF), 
$\beta_c^{(\mathrm{F/AF})}$, surfaces, whose reciprocal 
stability depends on which of the two ones has higher temperature. 
In the case of a measure $d\mu$ not depending on the bond $b$, 
the suffix F and AF stand
for ferromagnetic and antiferromagnetic, respectively.
In the general case, such a distinction is possible only
in the positive and negative sectors in the space of the 
parameters of the probability distribution, $\{\int d\mu_b J_b\geq 0\}$ and 
$\{\int d\mu_b J_b< 0\}$, respectively, whereas, for
the other sectors, we use the symbol
F/AF only to stress that the transition is not P-SG.

It is not difficult to see that, when the measure $d\mu$ 
does not depend on the specific bond $b$, \textit{i.e.}, if
$d\mu_b\equiv d\mu_{b'}~ \forall b,b'\in \Gamma_{\bm{g}}$, in the P region 
Eqs. (\ref{mapp0g}-\ref{mapp3g})
lead to the following exact limit for $\varphi$ and $C$ \cite{MOIII}
\begin{eqnarray}
\label{phi=0}
\lim_{D(\Gamma_{\bm{g}})\to \infty}\varphi=\lim_{D(\Gamma_{\bm{g}})\to \infty}{{C}}=0, 
\quad \mathrm{for}\quad \beta\leq\beta_c,
\end{eqnarray}
therefore, the basic role of Eqs. (\ref{mapp1g}-\ref{mapp3g}),
is to show how, in the limit $D(\Gamma_{\bm{g}})\to\infty$, $\varphi$ and ${{C}}$ 
approach zero and which are their singularities.
In particular this proves that for all the (random) infinite dimensional models
and any disorder non bond-dependent, the critical exponent $\alpha'$ for the specific
heat has the mean-field classical value, $\alpha'=0$, and that the correlation
functions (with different indices) above the critical temperature are exactly zero.
We point out however that, when the measure $d\mu_b$ depends explicitly on the
bond $b$, Eq. (\ref{phi=0}) in general does not hold~
\footnote{This was not strongly emphasized in \cite{MOIII}.}.
In fact, when the measure $d\mu_b$ is bond-dependent, 
the symmetry expressed by Eq. (\ref{phi=0}) is broken since the bonds are no
longer equivalent. 
As we will see in the next section, in small-world models with an underlying 
lattice $\mathcal{L}_0$ having $d_0<2$, 
even if Eq. (\ref{phi=0}) may still holds for $\varphi$, the symmetry is
broken for ${{C}}$ since the direction(s) of the axis(es) of 
$\mathcal{L}_0$ is(are) now favored direction(s). Yet, 
if $2\leq d_0< \infty$, 
the symmetry (\ref{phi=0}) for $\varphi$ is broken as well.  

The analytic continuation of Eqs. (\ref{mapp1g}-\ref{mapp3g}) to
$\beta>\beta_c$ and/or for $h\neq 0$ provide 
certain estimations which are expected to 
be qualitatively good. 
In general such estimations are not exact,
and this is particularly evident for the free energy
density of the SG phase.
However, the analytic continuation for
the other quantities gives a good qualitative result and
provide the exact critical behavior (in the sense of the
critical indices) and the exact percolation threshold.

For models defined over graphs satisfying a weak definition
of infinite dimensionality, as happens on a Bethe lattice with coordination number
$q>2$, a more general mapping has been established \cite{MOIII}. In this case, all the
above equations - along the critical surface (at least) - 
still hold exactly in the thermodynamic limit, 
where we can set effectively $D(\Gamma_{\bm{g}})=\infty$.
However, for the aims of this paper we do not need here to consider
this generalization of the mapping.

We have yet to make an important comment about Eqs. (\ref{THEOa01}),
(\ref{THEOa02}) and (\ref{THEOa04}), concerning the evaluation
of a correlation function in the SG phase 
here for a random system with $J_0=0$ (for the moment being). 
In fact Eq. (\ref{mapp2gc}), for both a normal   
and a quadratic correlation function $C^{(1)}$ or $C^{(2)}$,
has a factor 1/2 not entering in the physical Eqs. (\ref{THEOa01}),
(\ref{THEOa02}) and (\ref{THEOa04}).
The difference is just due to an artefact of the mapping
that separates the Gibbs state into two pure states \cite{Parisi}
not only in the F case, but also in the SG case. 
In fact, let us consider the correlation functions
of order $k=1$, that is, $C^{(1)}=\overline{\media{\sigma_1}}$ and
$C^{(2)}=q_{EA}=\overline{\media{\sigma_1}^2}$. We see that, for $C^{(1)}$,
Eq. (\ref{mapp2gc}) in the SG phase gives $C^{(1)}=m^{\mathrm{(SG)}}/2$.
On the other hand, for any non zero solution $m^{\mathrm{(SG)}}$
of the self-consistent Eq. (\ref{THEOa}), there exists another solution
$-m^{\mathrm{(SG)}}$, and both the solutions have 1/2 probability
to be realized in the random model. Since the SG phase is expected
to be the phase characterized by having $q_{EA}\neq 0$ and
$\overline{\media{\sigma_1}}=0$, we see that if we introduce both the
solutions $m^{\mathrm{(SG)}}$ and $-m^{\mathrm{(SG)}}$, 
we get $\overline{\media{\sigma_1}}=0$ in the SG phase.
Similarly, for $C^{(2)}$,
Eq. (\ref{mapp2gc}) in the SG phase gives $C^{(2)}=(m^{\mathrm{(SG)}})^2/2$,
which at zero temperature gives 1/2, whereas a completely frozen 
state with $q_{EA}=1$ is expected. Again, we recover the expected physical $q_{EA}$
by using both the
solutions $m^{\mathrm{(SG)}}$ and $-m^{\mathrm{(SG)}}$.
Repeating a similar argument for any correlation function of order $k$,
and recalling that for $k$ even (odd) the correlation function is
an even (odd) function of the external magnetic field $h$, we arrive at 
Eqs. (\ref{THEOa01}), (\ref{THEOa02}) and (\ref{THEOa04}).

\subsection{Random Models defined on Unconstrained Random Graphs}
Let us consider now more general random models in which
the source of the randomness comes from both
the randomness of the couplings and the randomness of the graph.
Given an ensemble of graphs $\bm{g}\in\mathcal{G}$ distributed with
some distribution $P(\bm{g})$, let us define 
\begin{eqnarray}
\label{HM}
H_{\bm{g}}\left(\{\sigma_i\};\{J_b\}\right)
&\defi& -\sum_{b\in\Gamma_{\bm{g}}} J_b \sigma_{i_b}\sigma_{j_b}
-h\sum_i\sigma_i\nonumber \\
&=& -\sum_{b\in\Gamma_{\mathrm{full}}} g_bJ_b \sigma_{i_b}\sigma_{j_b}
-h\sum_i\sigma_i.
\end{eqnarray} 
The free energy $F$ and the physics are now given by
\begin{eqnarray}
\label{logZDD}\nonumber
-\beta F\defi 
\sum_{{\bm{g}}\in\mathcal{G}}P({\bm{g}})\int d\mathcal{P}\left(\{J_b\}\right)
\log\left(Z_{\bm{g}}\left(\{J_b\}\right)\right),
\end{eqnarray} 
and similarly for $\overline{\media{O}^l},~l=1,2$.
Here $Z_g\left(\{J_b\}\right)$ 
is the partition function of the quenched system onto the graph realization 
$\bm{g}$ with bonds in $\Gamma_{\bm{g}}$ 
\begin{eqnarray}
\label{ZR}\nonumber
Z_{\bm{g}}\left(\{J_b\}\right)= \sum_{\{\sigma_i\}}e^{-\beta 
H_{\bm{g}}\left(\{\sigma_i\};\{J_b\}\right)}, 
\end{eqnarray} 
and $d\mathcal{P}\left(\{J_b\}\right)$ 
is again a product measure over all the possible bonds $b$ given 
as defined in Eq. (\ref{dPM}). Note that the bond-variables $\{g_b\}$ 
are independent from the coupling-variables $\{J_b\}$. 

For unconstrained random graphs, or for random graphs having a
number of constrains that grows sufficiently slowly with $N$, 
the probability $P(\bm{g})$, for large $N$, factorizes as
\begin{eqnarray}\nonumber
P(\bm{g})=\prod_{b\in\Gamma_{\mathrm{full}}} p_b(g_b).
\end{eqnarray} 
In such a case we can exploit the mapping we have previously seen 
for models over quenched graphs
as follows.
Let us define the effective coupling $\tilde{J}_b$:
\begin{eqnarray}
\label{sep}\nonumber
\tilde{J}_b\defi J_b \cdot g_b, 
\end{eqnarray}
correspondingly, since the random variables $J_b$ and $g_b$
are independent we have
\begin{eqnarray}
\label{dmutilde}\nonumber
d\tilde{\mu}_b(\tilde{J}_b)=d\mu_b(J_b) \cdot p_b(g_b), 
\end{eqnarray}
with the sum rule
\begin{eqnarray}\nonumber
\int d\tilde{\mu}_b(\tilde{J}_b)f(J_b;g_b)=\sum_{g_b=0,1}p_b(g_b)\int d\mu_b(J_b)f(J_b;g_b).
\end{eqnarray}
As a consequence, if we define the following global measure
\begin{eqnarray}\nonumber
d\tilde{\mathcal{P}}\left(\{\tilde{J}_b\}\right)=
P(\bm{g})\cdot d\mathcal{P}\left(\{J_b\}\right)=
\prod_{b\in\Gamma_{\mathrm{full}}}d\tilde{\mu}_b(\tilde{J}_b),
\end{eqnarray}
we see that
the mapping of the previous section can be applied as we had 
a single effective graph $\Gamma_p$ given by
\begin{eqnarray}
\label{GammaP}\nonumber
\Gamma_p\defi \{b\in\Gamma_{\mathrm{full}}:~p_b(g_b=1)\neq 0\},
\end{eqnarray} 
in fact we have
\begin{eqnarray}
\label{logZDD2}\nonumber
-\beta F= 
\int d\tilde{\mathcal{P}}\left(\{\tilde{J}_b\}\right)
\log\left(Z_p\left(\{\tilde{J}_b\}\right)\right),
\end{eqnarray}
where $Z_p$ is the partition function of the model with Hamiltonian $H_p$
given by
\begin{eqnarray}
\label{HM2}
H_p \left(\{\sigma_i\};\{\tilde{J}_b\}\right)
\defi -\sum_{b\in\Gamma_p} \tilde{J}_b \sigma_{i_b}\sigma_{j_b}
-h\sum_i\sigma_i.
\end{eqnarray} 
%

\section{Derivation of the self-consistent equations}
By using the above results, we are now able to derive easily
Eqs. (\ref{THEOa}-\ref{THEOlead}). Sometimes to indicate
a bond $b$ we will use the symbol $(i,j)$, or more shortly $ij$.

It is convenient to look formally at the
coupling $J_0$ also as a random coupling with distribution
\begin{eqnarray} 
\label{dmu0}
d\mu_0(J_0')/dJ_0'=\delta (J_0'-J_0).
\end{eqnarray}

Let us rewrite explicitly the Hamiltonian (\ref{H}) as follows
\begin{eqnarray}
\label{H1}\nonumber
H_{\bm{c}}&=&-\sum_{(i,j)\in\Gamma_0}\left(J_0+c_{ij}{J}_{ij}\right)\sigma_{i}\sigma_{j}
\nonumber \\
&& -\sum_{i<j,~(i,j)\notin \Gamma_0}c_{ij}{J}_{ij}\sigma_{i}\sigma_{j}
-h\sum_i\sigma_i,
\end{eqnarray}
and let us introduce the 
random variables $J_b'$, $g_b'$ and $\tilde{J}_b'$, where 
\begin{eqnarray}\nonumber
J_b'\defi  
\left\{
\begin{array}{l}
J_0+c_bJ_b, \quad b\in\Gamma_0,\\
J_b, \quad ~ b\notin\Gamma_0,
\end{array}
\right.
\end{eqnarray}
\begin{eqnarray}\nonumber
g_b'\defi  
\left\{
\begin{array}{l}
1, \quad b\in\Gamma_0,\\
c_b, \quad ~ b\notin\Gamma_0,
\end{array}
\right.
\end{eqnarray}
and 
\begin{eqnarray}\nonumber
\label{sep1}
\tilde{J}_b'\defi J_b' \cdot g_b'. 
\end{eqnarray}
Taking into account that 
the random variable $J_0+c_{ij}J_{ij}$, up to terms $\mathop{O}(1/N)$, is
distributed according to $d\mu_0(J_0)$,
the independent random variables $J_b'$ and $g_b'$ 
have distributions $d\mu_b'$ and $p_b'$ respectively given by
\begin{eqnarray}\nonumber
d\mu_b'(J_b') = 
\left\{
\begin{array}{l}
d\mu_0(J_b'), \quad b\in\Gamma_0,\\
d\mu(J_b'), \quad ~ b\notin\Gamma_0,
\end{array}
\right.
\end{eqnarray}
and
\begin{eqnarray}\nonumber
p_b'(g_b') = 
\left\{
\begin{array}{l}
\label{sep4}
\delta_{g_b',1}, \quad ~ b\in\Gamma_0 ,\\
\label{sep5}
p(g_b'), \quad b\notin\Gamma_0,
\end{array}
\right.
\end{eqnarray}
where the measures $d\mu$ and $p$ are those of the model introduced in Sec. II. 
As a consequence, Eq. (\ref{H1}) can be cast 
in the form of Eq. (\ref{HM2}) with the measure
\begin{eqnarray}
d\tilde{\mu}_b'(\tilde{J}_b') = 
\left\{
\begin{array}{l}
\label{sep2}
d\mu_0(J_b')\delta_{g_b',1}, \quad b\in\Gamma_0,\\
\label{sep3}
d\mu(J_b')p(g_b'), \quad ~ b\notin\Gamma_0.
\end{array}
\right.
\end{eqnarray}
Finally, since $p_b(g_b)\neq 0$ for any
$b\in\Gamma_{\mathrm{full}}$, we have also
\begin{eqnarray}
\Gamma_p=\Gamma_{\mathrm{full}},
\end{eqnarray} 
and due to the fact that $D(\Gamma_{\mathrm{full}})\propto N$, in the thermodynamic limit 
the mapping becomes exact. 

According to Eqs. (\ref{HI}-\ref{JIb}), the related Ising model
of our small-world model has the following Hamiltonian
with two free couplings: $J_0^{(I)}$, for $\Gamma_0$, and $J^{(I)}$, 
for $\Gamma_{\mathrm{full}}$
\begin{eqnarray}
\label{HR}
H_I&=&-J_0^{(I)}\sum_{(i,j)\in\Gamma_0}\sigma_{i}\sigma_{j}
-J^{(I)}\sum_{i<j,~(i,j)\notin \Gamma_0}\sigma_{i}\sigma_{j}
\nonumber \\
&&-h\sum_i\sigma_i.
\end{eqnarray}
After solving this Ising ($I$) model the mapping allows us to come back
to the random model by performing simultaneously for any $b\in \Gamma_{\mathrm{full}}$ 
the reverse substitutions
\begin{eqnarray}
\label{MT0}
\tanh\left(\beta J^{(I)}_b\right)\to \int
d\tilde{\mu}_b'(\tilde{J}_b')\tanh^l
\left(\beta \tilde{J}_b'\right), 
\end{eqnarray} 
where $l=1,2$ for $\Sigma=$ F or SG solution, respectively.
Since the couplings $J_0^{(I)}$ and $J^{(I)}$ are arbitrary, we find it convenient
to renormalize $J^{(I)}$ as $J^{(I)}/N$ and at the end of the 
calculation to put again $J^{(I)}$ instead of $J^{(I)}/N$.
Note that for the mapping 
nothing changes if we do not make this substitution;
the choice to use $J^{(I)}/N$ instead of $J^{(I)}$ is merely due
to a formal convenience, since in this way the calculations 
are presented in a more standard and physically understandable form. 
In fact, according to Eqs. (\ref{sep2}) and (\ref{MT0}) 
what matters after solving the related Ising model with 
$J^{(I)}/N$ instead of $J^{(I)}$
is that, once for $\Sigma$=F and once for $\Sigma$=SG,
we perform - simultaneously in the two couplings - 
the following reverse mapping transformations
($l=1,2$ for $\Sigma=$ F or SG, respectively):
\begin{eqnarray}
\label{MT}
&& \tanh\left(\beta J^{(I)}/N\right)\to \int
d\tilde{\mu}\left(\tilde{J}_{ij}\right)\tanh^l
\left(\beta \tilde{J}_{ij}\right),
\end{eqnarray}
for $(i,j)\notin \Gamma_0$, and
\begin{eqnarray}
\label{MT2}
&& \tanh\left(\beta J_0^{(I)}\right)\to \int
d\tilde{\mu}\left(\tilde{J}_{ij}\right)\tanh^l
\left(\beta \tilde{J}_{ij}\right),
\end{eqnarray}
for $(i,j)\in \Gamma_0$.
 
Explicitly, by applying Eqs. (\ref{sep2}), (\ref{PP}) and (\ref{dmu0}), the transformations 
(\ref{MT}) and (\ref{MT2}) become, respectively
\begin{eqnarray}
\label{MT3}
\beta J^{(I)} \to \beta J^{(\Sigma)}
\end{eqnarray}
and
\begin{eqnarray}
\label{MT4}
\beta J_0^{(I)} \to \beta J_0^{(\Sigma)},
\end{eqnarray}
where we have made use of the definitions (\ref{THEOb})-(\ref{THEOe})
introduced in Sec. III.

Let us now solve the related Ising model.
We have to evaluate the following partition function
\begin{eqnarray}
\label{ZI}\nonumber
Z_I=\sum_{\{\sigma_i\}}e^{\beta J_0^{(I)}\sum_{(i,j)\in\Gamma_0}\sigma_{i}\sigma_{j}+
\beta \frac{J^{(I)}}{2N}\sum_{i\neq j}\sigma_{i}\sigma_{j}
+\beta h\sum_i\sigma_i}.
\end{eqnarray}
In the following we will suppose that $J^{(I)}$ (and then $J^{(\Sigma)}$) is positive.
The derivation for $J^{(I)}$ (and then $J^{\mathrm{(F)}}$) negative differs from the other
derivation just for a rotation of $\pi/2$ in the complex $m$-plane, and leads to the
same result one can obtain by analytically continue the equations derived 
for $J^{(I)}>0$ to the region $J^{(I)}<0$. 

By using the Gaussian transformation we can rewrite $Z_I$ as
\begin{eqnarray}
\label{ZI1}
Z_I&=& c_N 
\sum_{\{\sigma_i\}}e^{\beta J_0^{(I)}\sum_{(i,j)\in\Gamma_0}\sigma_{i}\sigma_{j}}
\nonumber \\ 
&& \times \int_{-\infty}^{\infty} d{{m}} ~ e^{-\frac{\beta}{2} J^{(I)}{{m}}^2N +
\beta\left(J^{(I)}{{m}} + h\right)\sum_i\sigma_i},
\end{eqnarray}
where $c_N$ is a normalization constant
\begin{eqnarray}\nonumber
c_N = \sqrt{\frac{\beta J^{(I)}N}{2\pi}},
\end{eqnarray}
and, in the exponent of Eq. (\ref{ZI1}), 
we have again neglected terms of order $\mathop{O}(1)$.
For finite $N$ we can exchange the integral and the sum over the $\sigma$'s.
By using the definition of the unperturbed model with Hamiltonian $H_0$,
Eq. (\ref{H0}), whose free energy density, for given $\beta J_0$ and $\beta h$,
is indicated with $f_0(\beta J_0,\beta h)$, we arrive at
\begin{eqnarray}
\label{ZI2a}
Z_I&=& c_N \int_{-\infty}^{\infty} d{{m}} ~ e^{-N L({{m}})},
\end{eqnarray}
where we have introduced the function 
\begin{eqnarray}
\label{ZI2}
L({{m}})= \frac{\beta}{2} J^{(I)}{{m}}^2 
+\beta f_0 \left(\beta J_0^{(I)},\beta J^{(I)}{{m}} + \beta h\right).
\end{eqnarray}
By using 
$\partial_{\beta h}~\beta f_0(\beta J_0,\beta h)=-m_0(\beta J_0,\beta h)$, and
$\partial_{\beta h}~m_0(\beta J_0,\beta h)=\tilde{\chi}_0(\beta J_0,\beta h)$
we get  
\begin{eqnarray}
\label{ZI3}\nonumber
L'({{m}})= \beta J^{(I)}
\left[{{m}}-m_0 \left(\beta J_0^{(I)},\beta J^{(I)}{{m}} + \beta h\right)\right], 
\end{eqnarray}
\begin{eqnarray}
\label{ZI4}\nonumber
L''({{m}})= \beta J^{(I)}\left[1- \beta J^{(I)} 
\tilde{\chi}_0 \left(\beta J_0^{(I)},\beta J^{(I)}{{m}} + \beta h\right)\right]. 
\end{eqnarray}
If the integral in Eq. (\ref{ZI2a}) converges for any $N$,
by performing saddle point integration we see that the saddle point
${{m}}^{\mathrm{sp}}$ is solution of the equation
\begin{eqnarray}
\label{ZI5}
{{m}}^{\mathrm{sp}}=m_0 \left(\beta J_0^{(I)},
\beta J^{(I)}{{m}}^{\mathrm{sp}} + \beta h\right), 
\end{eqnarray}
so that, if the stability condition
\begin{eqnarray}
\label{ZI6}\nonumber
1- \beta J^{(I)}
\tilde{\chi}_0 \left(\beta J_0^{(I)},\beta J^{(I)}{{m}}^{\mathrm{sp}} + \beta h\right)>0, 
\end{eqnarray}
is satisfied, in the thermodynamic limit we arrive at the following
expression for the free energy density $f_I$ of the related Ising model 
\begin{eqnarray}
\label{ZI7}
\beta f_I = \left[\frac{\beta}{2} J^{(I)}{{m}}^2
+\beta f_0 \left(\beta J_0^{(I)},\beta J^{(I)}{{m}} 
+ \beta h\right)\right]_{{{m}}={{m}}^{\mathrm{sp}}}.
\end{eqnarray}
Similarly, any correlation function $C_I$ of the related Ising model 
is given in terms of the correlation function $C_0$ of the unperturbed
model by the following relation
\begin{eqnarray}
\label{ZI7b}
C_I =  C_0\left(\beta J_0^{(I)},\beta J^{(I)}{{m}} 
+ \beta h\right)|_{{{m}}={{m}}^{\mathrm{sp}}}.
\end{eqnarray}

Of course, the saddle point solution ${{m}}^{\mathrm{sp}}$ represents the magnetization
of the related Ising model, 
as can be checked directly by deriving
Eq. (\ref{ZI7}) with respect to $\beta h$ and by using Eq. (\ref{ZI5}).

If the saddle point equation (\ref{ZI5}) has more stable solutions,
the ``true'' free energy and the ``true'' observable of the related Ising model 
will be given by Eqs. (\ref{ZI7}) and (\ref{ZI7b}), respectively, calculated 
at the saddle point solution which minimizes Eq. (\ref{ZI7}) itself
and that we will indicate with $m_I$.

Let us call $\beta_{c0}^{(I)}$ the inverse critical temperature of the unperturbed
model with coupling $J_0^{(I)}$ and zero external field, 
possibly with $\beta_{c0}^{(I)}=\infty$ if no phase transition exists.
As stressed in Sec. IIIB, for the unperturbed model 
we use the expression ``critical temperature'' for 
any temperature where the magnetization $m_0$ at
zero external field passes from 0 to a non zero value, continuously or not.
Note that, as a consequence, if $J_0^{(I)}<0$,
we have formally $\beta_{c0}^{(I)}=\infty$, 
independently from the fact that
some antiferromagnetic order may be not zero.

Let us start to make the obvious observation that a necessary condition for the related Ising model to
have a phase transition at $h=0$ and for a finite temperature, 
is the existence of some paramagnetic region P$_I$ where
$m_I=0$. We see from the saddle point equation (\ref{ZI5}) that, for $h=0$,
a necessary condition for $m_I=0$ to be a solution is that be
$\beta\leq\beta_{c0}^{(I)}$ for any $\beta$ in P$_I$, 
from which we get also $\beta_c^{(I)}\leq\beta_{c0}^{(I)}$. 
In a few lines we will see however that the inequality must be strict 
if $\beta_{c0}^{(I)}$ is finite, which in particular excludes
the case $J_0<0$ (for which the inequality to be proved is trivial).

Let us suppose for the moment that be $\beta_c^{(I)}<\beta_{c0}^{(I)}$.
For $\beta<\beta_{c0}^{(I)}$
and $h=0$, the saddle point equation (\ref{ZI5}) has always the trivial 
solution $m_I=0$ which, according to the stability condition, is also a stable solution if
\begin{eqnarray}
\label{ZI8}
1- \beta J^{(I)}
\tilde{\chi}_0 \left(\beta J_0^{(I)},0\right)>0.
\end{eqnarray}
The solution $m_I=0$ starts to be unstable when 
\begin{eqnarray}
\label{ZI9}
1- \beta J^{(I)}
\tilde{\chi}_0 \left(\beta J_0^{(I)},0\right)=0.
\end{eqnarray}
Eq. (\ref{ZI9}), together with the constrain $\beta_c^{(I)}\leq\beta_{c0}^{(I)}$,
gives the critical temperature of the related Ising model $\beta_c^{(I)}$.
In the region of temperatures where Eq. (\ref{ZI8}) is violated, 
Eq. (\ref{ZI5}) gives two symmetrical stable 
solutions $\pm m_I\neq 0$. 
From Eq. (\ref{ZI9}) we see also that the case $\beta_c^{(I)}=\beta_{c0}^{(I)}$
is impossible unless be $J^{(I)}=0$, since the susceptibility 
$\tilde{\chi}_0(\beta J_0^{(I)},0)$ must diverge at $\beta_{c0}^{(I)}$.
We have therefore proved that $\beta_c^{(I)}<\beta_{c0}^{(I)}$.
Note that for $J_0^{(I)}\geq 0$ and $\beta<\beta_{c0}^{(I)}$ 
Eq. (\ref{ZI8}) is violated only for
$\beta>\beta_c^{(I)}$, whereas for $J_0^{(I)}<0$ Eq. (\ref{ZI8}) 
in general may be violated also in finite regions of the $\beta$ axis.

The critical behavior of the related Ising model 
can be studied by expanding Eq. (\ref{ZI5}) for small fields.
However, we find it more convenient to expand $L({{m}})$ in series around ${{m}}=0$
since in this way everything can be cast in the standard formalism of 
the Landau theory of phase transitions. From Eq. (\ref{ZI2}),
taking into account that the function $\tilde{\chi}_0 \left(\beta J_0,\beta h\right)$
is an even function of $\beta h$, we have the following general expression
valid for any $m$, $\beta$ and small $h$
\begin{eqnarray}
\label{ZI10}
L(m)= \beta f_0 \left(\beta J_0^{(I)},0\right) - m_0\left(\beta J_0^{(I)},0\right)\beta h +
\psi\left(m\right),
\end{eqnarray}
where we have introduced the Landau free energy density $\psi(m)$ given by
\begin{eqnarray}
\label{ZI11}
\psi\left(m\right)&=& \frac{1}{2}a m^2 + \frac{1}{4} b m^4+\frac{1}{6} c m^6
\nonumber \\ &&
 - m \beta \tilde{h}
+\Delta \left(\beta f_0\right)\left(\beta J_0^{(I)},\beta J^{(I)}m \right),
\end{eqnarray}
where 
\begin{eqnarray}
\label{ZI12}
a=\left[1-\beta J^{(I)}\tilde{\chi}_0 \left(\beta J_0^{(I)},0\right)\right]\beta J^{(I)},
\end{eqnarray}
\begin{eqnarray}
\label{ZI13}
b=- \frac{\partial^2}{\partial(\beta h)^2}
{\left. {\tilde{\chi}_0\left(\beta J_0^{(I)},\beta h\right)}
\right|_{_{\beta h =0} }
\frac{\left(\beta J^{(I)}\right)^4}{3!}},
\end{eqnarray}
\begin{eqnarray}
\label{ZI14}
c=- \frac{\partial^4}{\partial(\beta h)^4}
{\left. {\tilde{\chi}_0\left(\beta J_0^{(I)},\beta h\right)}
\right|_{_{\beta h =0} }
\frac{\left(\beta J^{(I)}\right)^6}{5!}},
\end{eqnarray}
\begin{eqnarray}
\label{ZI15}
\tilde{h}=m_0\left(\beta J_0^{(I)},0\right)J^{(I)}+
\tilde{\chi}_0\left(\beta J_0^{(I)},0\right)\beta ^{(I)} J^{(I)}\beta h,
\end{eqnarray}
finally, the last term $\Delta \left(\beta f_0\right) \left(\beta J_0^{(I)},\beta J^{(I)}m\right)$
is defined implicitly to render Eqs. (\ref{ZI10}) and (\ref{ZI11}) exact, 
but terms $\mathop{O}(h^2)$ and $\mathop{O}(m^3 h)$; explicitly
\begin{eqnarray}
\label{ZI16}
&&\Delta \left(\beta f_0\right) \left(\beta J_0^{(I)},\beta J^{(I)}m\right)=
\nonumber \\
&-&\sum_{k=4}^\infty \frac{\partial^{2k-2}}{\partial(\beta h)^{2k-2}}
{\left. {\tilde{\chi}_0\left(\beta J_0^{(I)},\beta h\right)}
\right|_{_{\beta h =0} }
\frac{\left(\beta J^{(I)}\right)^{2k}}{(2k)!}}.
\end{eqnarray}
 
Finally, to come back to the original random model, we have just
to perform the reversed mapping transformations (\ref{MT3}) and (\ref{MT4}) 
in Eqs. (\ref{ZI2})-(\ref{ZI16}). As a result we get 
immediately Eqs. (\ref{THEOa})-(\ref{THEOf1}), but Eq. (\ref{THEOl}).

\section{Derivation of Eq. (\ref{THEOl}) and  Eqs. (\ref{THEOrule})-(\ref{THEOrule9})}
Concerning Eq. (\ref{THEOl}) for the full expression of the free energy
density, it can be obtained by using 
Eqs. (\ref{0logZ2}), (\ref{varphi}), (\ref{mapp2g}) and (\ref{mapp3g}).
Here $\varphi_I$ is the high temperature part of the free energy
density of the related Ising model we have just solved:
\begin{eqnarray}
\label{ZI17} 
&& - \beta f_I = 
 \lim_{N\to\infty}\frac{1}{N}\sum_{(i,j)\in\Gamma_0}\log\left[\cosh(\beta J_0^{(I)})\right]
+\frac{N-1}{2}\nonumber \\
&& \times \log\left[\cosh\left(\beta J^{(I)}/N\right)\right]+
\log\left[2\cosh(\beta h)\right] + \varphi_I
\end{eqnarray}
where we have taken into account the fact that our related Ising model
has $|\Gamma_0|$ connections with coupling $J_0^{(I)}$ and 
$N(N-1)/2$ connections with the coupling $J^{(I)}/N$.
By using Eq. (\ref{ZI7}) calculated in $m_I$ and Eq. (\ref{ZI17}),
for large $N$ we get
\begin{eqnarray}
\label{ZI18}
&& \varphi_I = - \frac{\beta}{2} J^{(I)}{{m_I}}^2
- \beta f_0 \left(\beta J_0^{(I)},\beta J^{(I)}{{m_I}} +\beta h\right)\nonumber \\
&& - \lim_{N\to\infty}\frac{1}{N}\sum_{(i,j)\in\Gamma_0}
\log\left[\cosh(\beta J_0^{(I)})\right] \nonumber \\
&& -\log\left[2\cosh(\beta h)\right] + \mathop{O}\left(\frac{1}{N}\right).
\end{eqnarray}
Therefore, on using Eq. (\ref{mapp2g}), for the non trivial part
$\varphi^{(\Sigma)}$ of the random system, 
up to corrections $\mathop{O}\left(1/N\right)$, we arrive at 
\begin{eqnarray}
\label{ZI19}
&& \varphi^{(\Sigma)} = - \frac{\beta}{2}
J^{(\Sigma)}\left(m^{(\Sigma)}\right)^2
-\log\left[2\cosh(\beta h)\right]
\nonumber \\
&& - \lim_{N\to\infty}\frac{1}{N}\sum_{(i,j)\in\Gamma_0}
\log\left[\cosh(\beta J_0^{(\Sigma)})\right]
\nonumber \\
&& - \beta f_0 \left(\beta J_0^{(\Sigma)},\beta J^{(\Sigma)}m^{(\Sigma)} +\beta h\right) 
\end{eqnarray}
In terms of the function $L^{(\Sigma)}(m)$ Eq. (\ref{ZI19}) reads as
\begin{eqnarray}
\label{ZI20}
&& \varphi^{(\Sigma)} = - L^{(\Sigma)}\left(m^{(\Sigma)}\right)
\nonumber \\
&& - \lim_{N\to\infty}\frac{1}{N}\sum_{(i,j)\in\Gamma_0}
\log\left[\cosh(\beta J_0^{(\Sigma)})\right]
\nonumber \\
&&
-\log\left[2\cosh(\beta h)\right].
\end{eqnarray}
By using Eqs. (\ref{0logZ2}), (\ref{ZI20}), (\ref{mapp2g}) and (\ref{mapp3g}),
with $l=1$ or $2$ for $\Sigma$=F or $\Sigma$=SG, respectively,
we get Eq. (\ref{THEOl}).
 
For $h=0$ Eq. (\ref{ZI20}) can conveniently be rewritten also as
\begin{eqnarray}
\label{ZI21}
&& \varphi^{(\Sigma)} = \varphi_0\left(\beta J_0^{(\Sigma)},0\right)
\nonumber \\
&& + \left[L^{(\Sigma)}\left(0\right)
-L^{(\Sigma)}\left(m^{(\Sigma)}\right)\right],
\end{eqnarray}
where 
\begin{eqnarray}
&& \varphi_0(\beta J_0,\beta h)=
-\beta f_0 \left(\beta J_0^{(\Sigma)},\beta h\right)
\nonumber \\
&& - \lim_{N\to\infty}\frac{1}{N}\sum_{(i,j)\in\Gamma_0}
\log\left[\cosh(\beta J_0^{(\Sigma)})\right]
\nonumber \\
&& -\log\left[2\cosh(\beta h)\right],
\end{eqnarray}
is the high temperature part of the free energy
density of the unperturbed model with coupling $J_0^{(\Sigma)}$ and external field $h$. 
There are some important properties for the function $\varphi_0(\beta J_0,0)$:
it is a monotonic increasing function of $\beta J_0$;
if the lattice $\mathcal{L}_0$ has only loops of even length, 
$\varphi_0(\beta J_0,0)$ is an even function of $\beta J_0$;
furthermore,
if $d_0<2$ and the coupling-range is finite, 
or if $d_0=\infty$ at least in a wide sense
\cite{MOIII}, in the thermodynamic limit we have $\varphi_0(\beta J_0,0)=0$;
if instead $2\leq d_0<\infty$, $\varphi_0(\beta J_0,0)\neq 0$. 
We see here therefore what anticipated in Sec. VIIA: when $J_0\neq 0$,
the symmetry among the random couplings is broken and for $d_0$
sufficiently high this reflects in a non zero $\varphi^{(\Sigma)}$ 
also in the P region.

Next we prove Eqs. (\ref{THEOrule})-(\ref{THEOrule9}).
To this aim we have to calculate Eq. (\ref{ZI21}) at the
leading solution $\bar{m}^{(\Sigma)}$ and to compare $\varphi^{\mathrm{(F)}}$ and $\varphi^{\mathrm{(SG)}}$.
Note that the term in the square parenthesis of Eq. (\ref{ZI21})
is non negative since $\bar{m}^{(\Sigma)}$ is the absolute minimum of $L^{(\Sigma)}$.
We recall that for critical temperature we mean here any temperature
lying on the boundary P-F or P-SG, 
so that $\bar{m}^{(\Sigma)}|_\beta=0$ for any $\beta$ in the P region.

\subsection{$J_0\geq 0$}
If $J_0\geq 0$, for both the solution with label F and SG, 
we have only one second-order phase transition so that
$\bar{m}^{\mathrm{(F)}}=0$ and $\bar{m}^{\mathrm{(SG)}}=0$, respectively, are 
the stable and leading solutions even on the boundary with the P region.

Let us suppose $\beta_c^{\mathrm{(F)}}<\beta_c^{\mathrm{(SG)}}$.
Let be $\varphi_0(\cdot,0)\neq 0$.
From Eq. (\ref{ZI21}) and by using 
$J_0^{\mathrm{(F)}}>J_0^{\mathrm{(SG)}}$, we see that
\begin{eqnarray}
\label{ZI22}
&& \varphi^{(\mathrm{F})}|_{\beta_c^{\mathrm{(F)}}} 
= \varphi_0\left(\beta_c^{\mathrm{(F)}} J_0^{\mathrm{(F)}},0\right)
\nonumber \\
&&> \varphi^{(\mathrm{SG})}|_{\beta_c^{\mathrm{(F)}}} 
= \varphi_0\left(\beta_c^{\mathrm{(F)}} J_0^{\mathrm{(SG)}},0\right).
\end{eqnarray}
Finally, by using this result and the general rule given by Eqs. (\ref{mapp2g}) and
(\ref{mapp3g}), we see 
(and with a stronger reason, due to the factor 1/2 appearing in these equations for the SG solution) that 
the stable phase transition is the P-F one:
$\beta_c=\beta_c^{\mathrm{(F)}}$. 
Similarly, by using Eq. (\ref{ZI21})
for $\beta_c^{\mathrm{(F)}}<\beta<\beta_c^{\mathrm{(SG)}}$, 
we see that even for any $\beta$ in the interval 
$(\beta_c^{\mathrm{(F)}},\beta_c^{\mathrm{(SG)}})$
the stable solution is that with label F. This last observation makes also clear that if
$\varphi_0(\cdot,0)=0$ we reach the same conclusion: F is the stable
phase in all the region
$\beta_c^{\mathrm{(F)}}<\beta<\beta_c^{\mathrm{(SG)}}$ and in particular this 
implies also that the stable phase transition is the P-F one: 
$\beta_c=\beta_c^{\mathrm{(F)}}$.

Let us suppose $\beta_c^{\mathrm{(F)}}>\beta_c^{\mathrm{(SG)}}$.
If $\varphi_0(\cdot,0)\neq 0$, we arrive at 
\begin{eqnarray}
\label{ZI23}
&& \varphi^{(\mathrm{F})}|_{\beta_c^{\mathrm{(SG)}}} 
= \varphi_0\left(\beta_c^{\mathrm{(SG)}} J_0^{\mathrm{(F)}},0\right)
\nonumber \\
&&> \varphi^{(\mathrm{SG})}|_{\beta_c^{\mathrm{(SG)}}} 
= \varphi_0\left(\beta_c^{\mathrm{(SG)}} J_0^{\mathrm{(SG)}},0\right).
\end{eqnarray}
Finally, by using this result and the general rule given by Eqs. (\ref{mapp2g}) and
(\ref{mapp3g}), we see (and with a stronger reason) that 
the stable phase on the boundary is that predicted by the F solution which
has zero magnetization at $\beta_c^{\mathrm{(SG)}}$. This does not
imply that $\beta_c=\beta_c^{\mathrm{(F)}}$, but only that
$\beta_c^{\mathrm{(SG)}}<\beta_c\leq \beta_c^{\mathrm{(F)}}$.
If instead $\varphi_0(\cdot,0)=0$,
by using Eq. (\ref{ZI21})
for $\beta_c^{\mathrm{(SG)}}<\beta<\beta_c^{\mathrm{(F)}}$, 
we see that for any $\beta$ in the interval 
$(\beta_c^{\mathrm{(SG)}},\beta_c^{\mathrm{(F)}})$
the stable solution is SG and then, in particular, the stable boundary is P-SG:
$\beta_c=\beta_c^{\mathrm{(SG)}}$.

\subsection{$J_0<0$}
If $J_0< 0$, for the solution with label F, we may have both first and second-order phase transitions.
In the first case we cannot in general assume that 0 is the stable and leading
solution on the boundary with the P region:
$m^\mathrm{(F)}|_{\beta_c^\mathrm{(F)}}\neq 0$ in general.
As a consequence, for a first-order transition the term in square
parenthesis of Eq. (\ref{ZI21}) may be non zero even on the critical surface. 
Furthermore, as $J_0<0$, for the solution F we have at least two critical temperatures 
that we order as $\beta_{c1}^\mathrm{(F)}\geq \beta_{c2}^\mathrm{(F)}$.
However, despite of these complications, 
if we assume that $\mathcal{L}_0$ has only loops of even length,
$\varphi_0(\cdot,0)$ turns out to be an even function and,
due to the inequality $|J_0^{\mathrm{(F)}}|>J_0^{\mathrm{(SG)}}$,
almost nothing changes in the arguments we have used in the previous case $J_0\geq 0$.

Let us consider first the surfaces $\beta_{c2}^{\mathrm{(F)}}$
and $\beta_{c}^{\mathrm{(SG)}}$. Independently of the kind of phase
transition, first or second-order, we arrive again at Eqs. (\ref{ZI22})
and (\ref{ZI23}), for $\beta_{c2}^{\mathrm{(F)}}<\beta_{c}^{\mathrm{(SG)}}$ 
and $\beta_{c2}^{\mathrm{(F)}}>\beta_{c}^{\mathrm{(SG)}}$, respectively,
with the same prescription for the cases $\varphi_0(\cdot,0)\neq 0$,
or $\varphi_0(\cdot,0)= 0$. 

Let us now consider the surfaces $\beta_{c1}^{\mathrm{(F)}}$
and $\beta_{c}^{\mathrm{(SG)}}$. If $\beta_{c1}^{\mathrm{(F)}}<\beta_{c}^{\mathrm{(SG)}}$
and $\varphi_0(\cdot,0)\neq 0$,
for any $\beta$ in the interval $[\beta_{c1}^{\mathrm{(F)}},\beta_{c}^{\mathrm{(SG)}}]$ we have, 
\begin{eqnarray}
\label{ZI24}
&& \varphi^{(\mathrm{F})}|_{\beta} = \varphi_0\left(\beta J_0^{\mathrm{(F)}},0\right)
\nonumber \\
&&> \varphi^{(\mathrm{SG})}|_{\beta} 
= \varphi_0\left(\beta J_0^{\mathrm{(SG)}},0\right),
\end{eqnarray}
so that the interval $[\beta_{c1}^{\mathrm{(F)}},\beta_{c}^{\mathrm{(SG)}}]$
is a stable P region corresponding to the solution with label F.
Similarly, we arrive at the same conclusion if $\varphi_0(\cdot,0)=0$.
However, the interval of temperatures where the P region 
$[\beta_{c1}^{\mathrm{(F)}},\beta_{c}^{\mathrm{(SG)}}]$ is stable
can be larger when $d_0\geq 2$. In fact (exactly as we have seen for $J_0\geq 0$) 
in this case the P-SG stable boundary may stay at lower temperatures.
Finally, let us consider the case $\beta_{c1}^{\mathrm{(F)}}>\beta_{c}^{\mathrm{(SG)}}$.
If $\varphi_0(\cdot,0)= 0$, by using 
Eq. (\ref{ZI21}) we see that for any $\beta> \beta_{c}^{\mathrm{(SG)}}$
we have that the stable solution corresponds to the SG one, 
so that there is no stable boundary with the P region. 
If instead $\varphi_0(\cdot,0)\neq 0$, 
due to the fact that $\varphi^{(\mathrm{SG})}|_{\beta}$ 
and $\varphi^{(\mathrm{F})}|_{\beta}$ grow in a different way with $\beta$,
we are not able to make an exact comparison, and it is possible that 
the P-F boundary becomes stable starting from some $\beta_{c1}$ with 
$\beta_{c1}\geq \beta_{c1}^{\mathrm{(F)}}$. In general, as in the case $J_0<0$, we could have 
one (or even more) sectors where the P region corresponding to the solution
with label F is stable. 

\section{Conclusions}
In this paper we have presented a novel and general method to
analytically face random Ising models defined over small-world
networks.  The key point of our method is the fact that, at least in
the P region, any such a model can be exactly mapped to a suitable
fully connected model, whose resolvability is in general non trivial
for $d_0>1$, but still as feasible as a non random model. As a main
result we then derive a general self-consistent equation,
Eq. (\ref{THEOa}), which allows to describe effectively the model once
the magnetization of the unperturbed model in the presence of a
uniform external field, $m_0(\beta J_0,\beta h)$, is known.

The physical interpretation of this general result is straightforward.
From Eqs. (\ref{THEOa}) we see that, concerning the magnetization
$m^{(\mathrm{F})}$, the effect of adding long range Poisson
distributed bonds implies that the system - now perturbed - feels,
besides the coupling $J_0$, also an effective external field
$J^{(\mathrm{F})}$ shrunk by $m^{(\mathrm{F})}$ itself.
Concerning $m^{(\mathrm{SG})}$, the effect is that the
system now feels a modified effective coupling $J_0^{(\mathrm{SG})}$
and an effective external field
$J^{(\mathrm{SG})}$ shrunk by $m^{(\mathrm{SG})}$ itself.

We are therefore in the presence of an \textit{effective field theory}
which, as opposed to a simpler \textit{mean-field theory}, describes
$m^{(\mathrm{F})}$ and $m^{(\mathrm{SG})}$ in terms of, not only an
effective external field, but also through the non trivial function $m_0(\beta
J_0,\beta h)$ which, in turn, takes into account the correlations due
to the non zero short-range coupling $J_0$ or $J_0^{(\mathrm{SG})}$
felt by the unperturbed system. The combination of these two effects
gives rise to the typical behavior of models defined over small-world
networks: the presence of a non zero effective external field causes
the existence of a phase transition also at low $d_0$
dimension.  However, the precise determination of both the critical
surface and the correlation functions is obtained in a non trivial way
via the unperturbed magnetization $m_0(\beta J_0,\beta h)$.

We have used the method introduced to analyze the critical behavior of
generic models with $J_0\geq 0$ and $J_0<0$ and showed that they give
rise to two strictly different phase transition scenarios. In the
first case, we have a mean-field second-order phase transition with a
finite correlation length. Whereas, in the second case, we obtain
multiple first- and second-order phase transitions.  Furthermore, we
have shown that the combination of the F and SG solutions results in a
total of four possible kinds of phase diagrams according to the cases
\textbf{(1)} $(J_0\geq 0;~d_0<2,~\mathrm{or}~d_0=\infty)$,
\textbf{(2)} $(J_0\geq 0;~2\leq d_0<\infty)$, \textbf{(3)} $(J_0<
0;~d_0<2,~\mathrm{or}~d_0=\infty)$, and \textbf{(4)} $(J_0< 0;~2\leq
d_0<\infty)$.  One remarkable difference between systems with
$d_0<2,~\mathrm{or}~d_0=\infty$ and those with $2<d_0<\infty$, is that
in the latter case we have, in principle, also first-order P-SG phase transitions
and, moreover, re-entrance phenomena are in principle possible even
for $J_0\geq 0$.

In Secs. IV, V, and VI we have applied the method to solve
analytically those models for which the unperturbed magnetization
$m_0(\beta J_0,\beta h)$ is known analytically, \textit{i.e.}, the
small-world models in dimension $d_0=0,1,\infty$, corresponding to an
ensemble of non interacting units (spins, dimers, etc...), the one
dimensional chain, and the spherical model, respectively.  In
particular, we have studied in detail the small-world model defined over the
one dimensional chain with positive and negative short-range couplings
showing explicitly how, in the second case, multicritical points with
first- and second-order phase transition arise.  Finally, the
small-world spherical model - an exact solvable model (in our
approach) with continuous spin variables - has provided us with an
interesting case study to explore what happens as $d_0$ changes
continuously from 0 to $\infty$. As expected from general grounds,
unlike the non random version of the model, the small-world model
presents always a finite temperature phase transition, even in the
limit $d_0\to 0^+$.  This latter result - besides to be consistent
with what we have found in the $d_0=0$-dimensional discrete models -
has a simple physical explanation in our approach.  In fact, it
consists on mapping the small-world model (a random model) to a
corresponding non random model (no long-range bonds) but immersed in
an effective uniform external field which is active as soon as the
added random connectivity $c$ is not zero (see
Eqs. (\ref{THEOa})-(\ref{THEOc})).

Many interesting variants of the above models can be considered and
are still analytically solvable by our approach (see also the
generalizations considered in Sec. IIIE).  However, our approach can
be also applied numerically to study more complex small-world
models for which the corresponding unperturbed model is not
analytically available \cite{ML}.  In fact, the numerical complexity in
solving such small-world models is comparable to that required in
solving a non random model immersed in a uniform external field.

Models defined on complex small-world networks \cite{ReviewNote} are
an interesting subject of future work.


\begin{acknowledgments}
This work was supported by the FCT (Portugal) grants
SFRH/BPD/24214/2005, pocTI/FAT/46241/2002 and
pocTI/FAT/46176/2003, and the Dysonet Project.
We thank A. L. Ferreira, A. Goltsev and C. Presilla for useful discussions.
\end{acknowledgments}

\appendix
\section{Generalization to non homogeneous external field}
In this appendix we prove Eq. (\ref{THEOC2b}) calculating 
the $\mathop{O}(1/N)$ correction responsible for
the divergence of the susceptibility of the random system at $T_c$.
To this aim we firstly need to generalize our method to
an arbitrary external field. 
Let us consider again a fully connected model having - as done in Sec. VIII -
long-range couplings $J$ (for brevity we will here omit the label $I$)
and short-range couplings $J_0$ but now
immersed in an arbitrary (non homogeneous) external field $\{h_n\}$, 
where $n=1,\ldots,N$. 
After using the Gaussian transformation 
we have the following partition function: 
\begin{eqnarray}
\label{App1}
Z&=& c_N \int_{-\infty}^{\infty} d{{m}} ~ e^{-N L({{m}})},
\end{eqnarray}
where we have introduced the function 
\begin{eqnarray}
\label{App2}
L({{m}})= \frac{\beta}{2} J{{m}}^2 
+\beta f_0 \left(\beta J_0,\{\beta J{{m}} + \beta h_n\}\right),
\end{eqnarray}
$f_0 \left(\beta J_0,\{\beta h_n\}\right)$
being the free energy density of the unperturbed model
in the presence of an arbitrary external field $\{\beta h_n\}$. 
By using 
\begin{eqnarray}
\label{App3}
\partial_{\beta h_i}~\beta f_0(\beta J_0,\{\beta h_n\})=
-m_{0i}(\beta J_0,\{\beta h_n\}), 
\end{eqnarray}
and
\begin{eqnarray}
\label{App4}
\tilde{\chi}_{0;i,j}(\beta J_0,\{\beta h_n\})&\defi&
\media{\sigma_i\sigma_j}_0-\media{\sigma_i}_0\media{\sigma_j}_0 \nonumber \\ 
&=& \partial_{\beta h_j}~m_{0i}(\beta J_0,\{\beta h_n\})
\end{eqnarray}
we get
\begin{eqnarray}
\label{App5}
L'({{m}})= \beta J
\left[{{m}}-\frac{1}{N}\sum_i 
m_{0i}\left(\beta J_0,\{\beta J{{m}} + \beta h_n\}\right)\right], 
\end{eqnarray}
\begin{eqnarray}
\label{App6}
L''({{m}})&=& \beta J 
\left[1- \beta J \frac{1}{N} \right. \nonumber \\ 
&\times& \left. \sum_{i,j} 
\tilde{\chi}_{0ij} \left(\beta J_0,\{\beta J{{m}} + \beta h_n\}\right)\right].
\end{eqnarray}
By performing the saddle point integration we see that the saddle point
${{m}}^{\mathrm{sp}}$ is solution of the equation
\begin{eqnarray}
\label{App7}
{{m}}^{\mathrm{sp}}=\frac{1}{N}\sum_i 
m_{0i}\left(\beta J_0,\{\beta J{{m}}^{\mathrm{sp}} + \beta h_n\}\right)
\end{eqnarray}
hence, by using 
\begin{eqnarray}
\label{App7a}
&& \tilde{\chi}_{0} \left(\beta J_0,\{\beta J{{m}} + \beta h_n\}\right) = 
\nonumber \\ &&\frac{1}{N}  
\sum_{i,j}\tilde{\chi}_{0ij} \left(\beta J_0,\{\beta J{{m}} + \beta h_n\}\right),
\end{eqnarray}
we see that if the stability condition
\begin{eqnarray}
\label{App8}
1- \beta J
\tilde{\chi}_0 \left(\beta J_0,\{\beta J{{m}}^{\mathrm{sp}} + \beta h_n\}\right)>0, 
\end{eqnarray}
is satisfied, in the thermodynamic limit we arrive at the following
expression for the free energy density $f$ of the related Ising model 
immersed in an arbitrary external field
\begin{eqnarray}
\label{App9}
\beta f = \left[\frac{\beta}{2} J{{m}}^2
+\beta f_0 \left(\beta J_0,\{\beta J{{m}}+ \beta h_n\}\right)\right]_{{{m}}={{m}}^{\mathrm{sp}}}.
\end{eqnarray}
On the other hand, by derivation with respect to $\beta h_i$ and 
by using Eq. (\ref{App7}), it is immediate to verify that
\begin{eqnarray}
\label{App10}
m_i\defi \media{\sigma_i}=m_{0i}(\beta J_0,\{\beta J{{m}}^{\mathrm{sp}} + \beta h_n\}),
\end{eqnarray}
and then also (from now on for brevity on we omit the symbol $^{\mathrm{sp}}$)
\begin{eqnarray}
\label{App11}
m=\frac{1}{N}\sum_i m_i.
\end{eqnarray}

We want now to calculate the correlation functions.
From Eq. (\ref{App10}), by deriving with respect to $\beta h_j$ we have
\begin{eqnarray}
\label{App12}
\tilde{\chi}_{ij} &\defi& \frac{\partial m_i}{\partial (\beta h_j)}=
\sum_l \tilde{\chi}_{0;i,l} (\beta J_0,\{\beta J{{m}}+ \beta h_n\})
\nonumber \\ 
&\times & \left(\beta J\frac{\partial m}{\partial (\beta h_j)} + \delta_{l,j}\right),
\end{eqnarray}
which, by summing over the index $i$ and using (\ref{App11}), gives
\begin{eqnarray}
\label{App13}
\frac{\partial m}{\partial (\beta h_j)}=
\frac{\frac{1}{N}\sum_i \tilde{\chi}_{0;i,j} (\beta J_0,\{\beta J{{m}}+ \beta h_n\})}
{1-\beta J \tilde{\chi}_{0} (\beta J_0,\{\beta J{{m}}+ \beta h_n\})}.
\end{eqnarray}
We can now insert Eq. (\ref{App13}) in the rhs of Eq. (\ref{App12}) to get
\begin{eqnarray}
\label{App14}
\tilde{\chi}_{ij}&=&
\tilde{\chi}_{0;i,j} (\beta J_0,\{\beta J{{m}}+ \beta h_n\}) \nonumber \\
&+& \frac{\beta J}{N}\frac{
\sum_l \tilde{\chi}_{0;l,j} 
\sum_k \tilde{\chi}_{0;i,k}} 
{1-\beta J \tilde{\chi}_{0} (\beta J_0,\{\beta J{{m}}+ \beta h_n\})},
\end{eqnarray}
where for brevity we have omitted the argument in $\tilde{\chi}_{0;l,j}$ 
and $\tilde{\chi}_{0;i,k}$, which is the same of $\tilde{\chi}_{0}$ appearing
in the denominator.
If we now come back to choice a uniform external field $h_n=h$, $n=1,\ldots,N$, 
we can use translational invariance and for the related Ising model (fully connected) 
we obtain the following correlation function
\begin{eqnarray}
\label{App15}
\tilde{\chi}_{ij}&=&\frac{\beta J}{N}
\frac{\left[\tilde{\chi}_{0} (\beta J_0,\beta J{{m}}+ \beta h)\right]^2}
{1-\beta J \tilde{\chi}_{0} (\beta J_0,\beta J{{m}}+ \beta h)}
\nonumber \\
&+& \tilde{\chi}_{0;i,j} (\beta J_0,\beta J{{m}}+ \beta h).
\end{eqnarray}
Finally, by performing the mapping substitutions (\ref{MT3}) and (\ref{MT4})
we arrive at Eq. (\ref{THEOC2b}). 

Similarly, any correlation function $C$ of the related Ising model 
will be given by a similar formula with the leading term $C_0$ plus
a correction $\mathop{O}(1/N)$ becoming important only near $T_c$.


\end{document}